\documentclass[3p,authoryear]{elsarticle}
\usepackage{epstopdf}
\usepackage{subfigure,graphicx}
\usepackage{float}
\usepackage{amsmath, amsfonts, amssymb,mathrsfs}
\usepackage{amsthm}
\usepackage{epsf}
\usepackage{amsfonts}
\usepackage{stmaryrd}
\usepackage{amssymb}
\usepackage{leftidx}
\usepackage{color}
\usepackage{mathtools}
\usepackage{placeins}
\usepackage{booktabs}
\usepackage{enumitem}
\usepackage{caption}
\usepackage{multirow}
\usepackage{stackengine}
\usepackage[utf8]{inputenc}
\usepackage[english]{babel}
\usepackage{color}

\theoremstyle{plain}
\newtheorem{theorem}{Theorem}

\newtheorem{remark}[theorem]{Remark}

\def\bfu{{\bf u}}

\def\bfx{{\bf x}}
\def\bfy{{\bf y}}

\def\bfC{{\bf C}}

\def\bfI{{\bf I}}

\def\bfN{{\bf N}}

\def\bfS{{\bf S}}

\def\bfX{{\bf X}}
\def\bfF{{\bf F}}

\def\bfZ{{\bf Z}}

\def\obfC{\overline{\bfC}}

\def\obfF{\overline{\bfF}}

\def\obfS{\overline{\bfS}}

\def\Atan{\mbox{\boldmath$\mathcal{A}$}}

\def\Ctan{\mbox{\boldmath$\mathcal{C}$}}

\def\Atan{\mbox{\boldmath$\mathcal{A}$}}

\def\bfrho{\mbox{\boldmath $\rho$}}

\newcommand\sts{\mathfrak{s}_{\texttt{ts}}}

\newcommand\shs{\mathfrak{s}_{\texttt{hs}}}

\long\def\symbolfootnote[#1]#2{\begingroup%
\def\thefootnote{\fnsymbol{footnote}}\footnote[#1]{#2}\endgroup}

\makeatletter
\renewcommand\@biblabel[1]{}

\makeatother

\begin{document}
\begin{frontmatter}

\title{Nucleation and propagation of fracture in viscoelastic elastomers: \\ A complete phase-field theory\vspace{0.1cm}}

\author[Illinois]{Farhad Kamarei}
\ead{kamarei2@illinois.edu}

\author[3M]{Evan Breedlove}
\ead{elbreedlove@mmm.com}

\author[Illinois]{Oscar Lopez-Pamies}
\ead{pamies@illinois.edu}

\address[Illinois]{Department of Civil and Environmental Engineering, University of Illinois, Urbana--Champaign, IL 61801, USA  \vspace{0.05cm}}

\address[3M]{3M Corporate Research Laboratory,  St. Paul, MN 55144, USA}

\vspace{-0.0cm}

\begin{abstract}

This paper presents a macroscopic theory, alongside its numerical implementation, aimed at describing, explaining, and predicting the nucleation and propagation of fracture in viscoelastic materials subjected to quasistatic loading conditions. The focus is on polymers, in particular, on elastomers. To this end, the starting point of this work is devoted to summarizing the large body of experimental results on how elastomers deform, nucleate cracks, and propagate cracks when subjected to mechanical loads. When viewed collectively, the experiments make it plain that there are three basic ingredients that any attempt at a complete macroscopic theory of fracture in elastomers ought to account for: \emph{i}) the viscoelasticity of the elastomer; \emph{ii}) its strength; and \emph{iii}) its fracture energy. A theory is then introduced that accounts for all these three basic ingredients by extending the phase-field theory initiated by Kumar, Francfort, and Lopez-Pamies (\emph{J. Mech. Phys. Solids} 112 (2018), 523--551) for elastic brittle materials to seamlessly incorporate viscous energy dissipation by deformation, a generalized strength surface that is a hypersurface in stress-deformation space (and not just in stress space as for elastic brittle materials), and the pertinent Griffith criticality condition for materials that dissipate energy not just by the creation of surface but also by deformation, in this case, by viscous deformation (Shrimali and Lopez-Pamies (2023) \emph{Extreme Mech. Lett.} 58, 101944). From an applications point of view, the proposed theory amounts to solving an initial-boundary-value problem comprised of two nonlinear PDEs coupled with a nonlinear ODE for the deformation field $\bfy(\bfX,t)$, a tensorial internal variable $\bfC^v(\bfX,t)$, and the phase field $z(\bfX,t)$. A robust scheme is presented to generate solutions for these equations that makes use of a non-conforming Crouzeix-Raviart finite-element discretization of space and a high-order accurate explicit Runge-Kutta finite-difference discretization of time. To illustrate the descriptive and predictive capabilities of the theory, the last part of this paper presents simulations of prototypical experiments dealing with nucleation of fracture in the bulk, nucleation of fracture from a pre-existing crack, and propagation of fracture in different types of elastomers under various types of loading conditions. 

\endkeyword

\end{abstract}

\end{frontmatter}

\vspace{-0.0cm}

\section{Introduction}\label{Sec: Intro}

Because of their unique properties, polymers have become indispensable materials in modern engineering design. In spite of their pervasive use --- frequently grounded in empirical observations --- numerous fundamental questions about their distinctive behaviors remain open. Arguably, the main source of difficulty at the center of all such questions is one and the same: the inherent dissipative and time-dependent nature of polymers' mechanical and physical behavior. 

In this context, the primary objective of this paper is to tackle one of the most consequential open problems pertaining to their mechanical behavior: \emph{where and when cracks nucleate and propagate in polymers subjected to mechanical forces}. The focus is on viscoelastic elastomers, to wit, soft polymers that in addition to being capable to store energy by elastic deformation, they are capable to dissipate energy by viscous deformation and by the creation of surface, that is, by fracture. Dissipation mechanisms other than viscous and fracture (e.g., strain-induced crystallization, Mullins effect) are considered to be negligible or absent altogether. 

Specifically, in this paper, we introduce a macroscopic theory, alongside its numerical implementation, aimed at describing, explaining, and predicting the nucleation and propagation of fracture in viscoelastic elastomers subjected to mechanical loads that are applied quasistatically, that is, loads for which inertial effects are negligible. We do so by extending the phase-field theory initiated by \cite*{KFLP18} for elastic brittle materials. 

Over the past seven years, a series of validation results \citep{KFLP18,KRLP18,KBFLP20,KLP20,KLP21,KRLP22,KLDLP24,KKLP24,KDLP25b} for a wide range of materials, specimen geometries, and loading conditions have shown that the phase-field fracture theory initiated by \cite*{KFLP18} provides a complete theory of nucleation and propagation of fracture in nominally elastic brittle materials subjected to quasistatic loading conditions, thus its use here as the base theory to build upon the additional complexities of the dissipative and time-dependent deformation, strength, and toughness in viscoelastic elastomers.

We begin in Section \ref{Sec: Ingredients} by summarizing the substantial macroscopic experimental knowledge on how elastomers deform, nucleate cracks, and propagate cracks when subjected to mechanical loads applied quasistatically. In a nutshell, the existing experimental results indicate that \emph{nucleation of fracture}
\renewcommand{\labelitemi}{$\bullet$}
\begin{itemize}

\vspace{-0.1cm}

\item{in a body under a spatially uniform state of stress is governed by the \emph{strength}\footnote{The precise definition of strength for viscoelastic elastomers is introduced in Subsection \ref{Sec: Strength ingredient} below.} of the elastomer;}
    
\vspace{-0.1cm}

\item{from large pre-existing cracks is governed by a \emph{Griffith competition}\footnote{The precise Griffith competition for viscoelastic elastomers is introduced in Subsection \ref{Sec:Nucleation-crack} below.} between part of its viscoelastic energy and fracture energy;}

\vspace{-0.1cm}

\item{under any other circumstance of spatially non-uniform state of stress (e.g., from notches, small pre-existing cracks, or any other subregion in the body under a non-uniform state of stress) is governed by a mediation between the strength and the Griffith competition of energies in the elastomer;}
    
\vspace{-0.1cm}

\end{itemize}
while \emph{propagation of fracture}
\begin{itemize}

\item{is, akin to nucleation from large pre-existing cracks, also governed by the Griffith competition between the viscoelastic and fracture energies.}

\end{itemize}

Having pinpointed in Section \ref{Sec: Ingredients} all the basic features dictated by experimental observations that any theory that aims at providing a complete macroscopic description of nucleation and propagation of fracture in viscoelastic elastomers must possess, we then present in Section \ref{Sec: Theory} one such theory. As noted above, we do so by extending to the realm of  viscoelastic materials the phase-field theory initiated by \cite*{KFLP18} for elastic brittle materials. The extension involves accounting for: $i$) the energy dissipation of elastomers by viscous deformation; $ii$) their generalized strength surface that happens to be describable by a hypersurface in stress-deformation space (and not just in stress space as for elastic brittle materials); and $iii$) the pertinent Griffith competition for materials that dissipate energy not just by the creation of surface but also by deformation, in this case, by viscous deformation.

The proposed theory amounts to solving an initial-boundary-value problem --- given by equations (\ref{BVP-y-theory-reg})-(\ref{BVP-z-theory-reg}) --- comprised of two nonlinear partial differential equations (PDEs) coupled with a nonlinear ordinary differential equation (ODE) for the deformation field $\bfy(\bfX,t)$, a tensorial internal variable $\bfC^v(\bfX,t)$, and the phase field $z(\bfX,t)$. In Section \ref{Sec: Numerical Implementation}, we present a numerical scheme to generate solutions for these equations that makes use of a non-conforming Crouzeix-Raviart finite-element (FE) discretization of space and a high-order accurate explicit Runge-Kutta finite-difference (FD) discretization of time. The combination of these two types of discretizations results in a robust scheme that is capable of handling finite deformations, the near incompressibility of typical elastomers, as well as general time-dependent loading conditions, irrespectively of whether they are applied slowly, fast, or span a large time range. 

In Section \ref{Sec: Simulations}, we showcase the descriptive and predictive capabilities of the proposed theory via sample simulations of three prototypical experiments dealing with nucleation of fracture in the bulk, nucleation of fracture from a pre-existing crack, and propagation of fracture in different types of elastomers under various types of loading conditions. In particular, we present results and direct comparisons with experiments for the nucleation of fracture in polyurethane rubber bands under uniaxial tension \citep{Mueller1968}, the nucleation of fracture from a pre-existing crack in VHB 4905 sheets in the so-called ``pure-shear'' test \citep{Pharretal2012}, and the propagation of fracture in SBR sheets in the so-called trousers test \citep{Greensmith55}. We conclude this work by recording a number of final comments in Section \ref{Sec: Final Comments}.

\subsection{Existing phase-field models of fracture for viscoelastic materials} 

At this stage, it is appropriate to make explicit mention of the handful of phase-field theories of fracture for viscoelastic materials --- not necessarily elastomers --- that have recently appeared in the literature, both in the asymptotic setting of linear viscoelasticity \citep{Waisman19,Dammas21} as well as in the more general setting of finite viscoelasticity \citep{Loew19,Kaliske20,Brighenti21,Dammas23,Lancioni25}.

The first of these seems to be that of \cite{Waisman19}, who modified the classical\footnote{The shortcomings associated with the use of the so-called \texttt{AT}$_2$ regularization for the surface area of cracks --- as opposed to other regularizations such as \texttt{AT}$_1$ --- have been summarized in several works; see, e.g., \cite{Marigo11}.} \texttt{AT}$_2$ phase-field model for isotropic linear elastic brittle materials \citep{Bourdin00} by replacing the linear elastic behavior of the bulk with a linear viscoelastic behavior (in particular, with a generalized Maxwell model) in which the entire stored elastic energy, including the part of the stored elastic energy that is \emph{not} in thermodynamic equilibrium, and a weighted part of the dissipated viscous energy enter the Griffith competition for the propagation of fracture. Their work includes a number of 3D simulations based on material parameters for polar ice and asphalt. Subsequently, \cite{Dammas21} examined the same modification with the added alteration of having different degradation functions for the stored elastic energy and the dissipated viscous energy.

\cite{Loew19} proposed a similar modification to that in \citep{Waisman19} but in the context of finite deformations. In particular, these authors modified the classical \texttt{AT}$_2$ phase-field model for isotropic nonlinear elastic brittle materials \citep{Bourdin08} by replacing the nonlinear elastic behavior of the bulk with a finite linear viscoelastic behavior \citep{ColemanNoll61} in which the entire stored elastic energy and the dissipated viscous energy enter the Griffith competition for the propagation of fracture. Their theory also includes a ``viscosity'' term ($\kappa_1$ in their notation) for the phase field, which likely improves the numerical stability of the governing equations but is of unclear physical meaning and severs the connection with sharp cracks; see also \cite{Lancioni25} for a more recent use of such a ``viscosity'' term. Their work includes a number of 2D simulations of experiments that the authors themselves carried out on a certain EPDM elastomer filled with carbon black.\footnote{It has been long established that fracture in filled elastomers may be notably more complex than fracture in unfilled elastomers \citep{DeGent1996,Hamed2002}. The focus of this work in on unfilled elastomers or, more precisely, on filled elastomers wherein the fillers are \emph{not} much larger than the characteristic size of the underlying polymeric chains.} Soon thereafter, \cite{Kaliske20} proposed a similar alteration with the key differences that the viscoelasticity in the bulk is described by a model within the class introduced by \cite{RG98}, which is more representative of elastomers than the finite linear viscoelastic model used in \citep{Loew19}, and that the entire stored elastic energy, but not the dissipated viscous energy, is chosen to enter the Griffith competition for the propagation of fracture. \cite{Kaliske20} presented as well a number of 2D and 3D simulations centered on the nucleation of fracture from pre-existing cracks and subsequent propagation in various types of elastomers. Later, \cite{Brighenti21} also made use of the same alteration with the difference that the viscoelasticity in the bulk is described by a model based on statistical mechanics. More recently, \cite{Dammas23} have extended the approach in \citep{Dammas21} to finite deformations with the added proposal of making the fracture energy dependent on the rate of deformation.

A common feature in all the above-mentioned phase-field formulations is that they describe the evolution of the phase field as a Griffith competition between the \emph{entire} stored elastic energy, plus possibly the dissipated viscous energy, and the regularized surface fracture energy. Accordingly, they are inconsistent with experimental observations which, as summarized in Subsections \ref{Sec:Nucleation-crack} and \ref{Sec:Propagation-crack} below, indicate that the nucleation of fracture from a large pre-existing crack as well as the propagation of fracture in viscoelastic elastomers follows a Griffith competition between only certain part of the stored elastic energy --- in particular, the ``equilibrium'' part of the stored elastic energy --- and the fracture energy.

\begin{remark}\label{Remark1}
{\rm For the basic case of elastic brittle materials, when the only mechanism of energy dissipation is the creation of surface, there is a unique way of setting up the Griffith energy competition that describes the growth of cracks: the stored elastic energy competes with the fracture energy. By contrast, when dealing with materials that dissipate energy by deformation, and not just by the creation of surface, there are many ways in which a Griffith-type competition among the various energies that are stored and dissipated can be potentially set up. In these more general cases, the correct Griffith energy competition, if there is one, needs to be determined from experimental observations. For viscoelastic elastomers, experiments indicate that the correct energy competition appears to be that between the ``equilibrium'' part of the stored elastic energy --- and \emph{not} the entire stored elastic energy --- and the fracture energy \citep{SLP23}.}
\end{remark}

What is more, none of the above-mentioned formulations are capable of describing fracture nucleation in elastomers in general. As summarized in Subsections \ref{Sec: Strength ingredient} and \ref{Sec:Nucleation-transition} below, and as previously discussed at length in the simpler setting of elastic brittle materials \citep{LPDFL25,KDLP25}, this is because these formulations are direct extensions of the classical variational phase-field theory \citep{Bourdin00,Bourdin08} and hence are fundamentally incomplete as they cannot account for the strength of elastomers as an independent macroscopic material property.

\section{A summary of experimental observations of fracture nucleation and propagation in viscoelastic elastomers}\label{Sec: Ingredients}

The history of elastomers, beginning with that of natural rubber, is long and fascinating \citep{deGennes96}. Fundamental investigations of their mechanics of deformation and fracture started, in earnest, during the 1930s and 1940s \citep{Busse34,Busse38,Tobolsky42}. By now, a large body of macroscopic experimental knowledge exists that provides a fairly comprehensive picture of how elastomers deform, nucleate fracture, and propagate fracture when subjected to mechanical loads applied quasistatically. In the next three subsections, we summarize such a knowledge and thereby identify the three basic ingredients that any attempt at a comprehensive macroscopic theory of fracture in viscoelastic elastomers ought to account for, to wit:
\begin{enumerate}[label=\Roman*.]

\vspace{-0.1cm}

\item{The viscoelasticity of the elastomer;}

\vspace{-0.1cm}

\item{Its strength; and}

\vspace{-0.1cm}

\item{Its fracture energy or intrinsic fracture toughness.}

\vspace{-0.1cm}

\end{enumerate}

\subsection{Deformation: The viscoelasticity}\label{Sec: Ingredient1-viscoelasticity}

That the deformation of elastomers is well described by viscoelasticity has been a well settled matter for decades; for classical accounts and reviews, see, for instance, \cite{Findley76,deGennes79,Ferry80,Doi98,Wineman2009,KLP16} and references therein.

From a modeling point of view, by virtue of their superior numerical tractability and demonstrated capabilities to describe quantitatively the viscoelastic behavior of a wide range of elastomers, constitutive models based on internal variables are typically preferred over those based on hereditary integrals, especially when dealing with finite deformations. Within the former, the vast majority belong to the so-called family of two-potential constitutive models, or generalized standard materials \citep{Coleman67,Halphen75,Ziegler87,KLP16}. The key idea behind this class of models is that they expediently describe how the elastomer of interest stores and dissipates energy by means of two scalar-valued functions: a free energy function $\psi$ and a dissipation potential $\phi$. In this work, we will make use of the two-potential formulation introduced by \cite{KLP16}, which accounts for the typical non-Gaussian elasticity of elastomers as well as for their deformation-dependent shear-thinning viscosity. We emphasize, nonetheless, that the theory of fracture proposed in this work applies to any constitutive formulation of choice based on internal variables.

\subsection{Nucleation of fracture}

Much like for the simpler class of nominally elastic brittle materials \citep{KLP20,KBFLP20,LPDFL25}, experimental investigations of the nucleation of fracture in viscoelastic elastomers can be broadly divided into three distinct types: nucleation in the bulk under spatially uniform states of stress, from large pre-existing cracks, and under any other circumstance where the stress field is spatially non-uniform. Below, we summarize each of these three types, one at a time.

\subsubsection{Nucleation of fracture under states of spatially uniform stress: The strength}\label{Sec: Strength ingredient}

Soon after Charles Goodyear discovered the process of rubber vulcanization, Stephen Perry invented the rubber band, receiving a patent for it in 1845. That invention triggered the first studies into fracture of viscoelastic elastomers. In particular --- not unlike we playfully do today --- early experiments involved stretching rubber bands at various loading rates. This led to a spatially uniform uniaxial stress in the regions between the stretched ends until a crack suddenly appeared in such regions and severed the band. The critical values of the stretch and stress at which the crack nucleated identified the \emph{uniaxial tensile strength} of the rubber for the applied loading rate. 

Despite having kickstarted the field over 180 years ago, a complete definition of the strength of viscoelastic elastomers has remained absent from the literature. Here, we propose a precise and complete definition of strength that generalizes that recently introduced by \cite{KLP20} and \cite{KBFLP20} for the simpler case of elastic brittle materials. When a specimen of the elastomer of interest is subjected to a state of \emph{spatially uniform} --- arbitrary, not just uniaxial --- \emph{history of stress} $\{\bfS(t),t\in[0,T]\}$, fracture will nucleate from one or more of its inherent defects at a critical value along the loading path. The set of all such critical values defines a hypersurface in stress-deformation-time space. In terms of the first Piola-Kirchhoff stress tensor $\bfS(t)$, the deformation gradient tensor $\bfF(t)$, and time $t\in[0,T]$, we formally write
\begin{equation}
\mathcal{F}\left(\bfS(t),\bfF(t),t\right)=0.\label{Strength-Surface-General}
\end{equation}
Because of the vast separation of length scales between the size of a macroscopic piece of elastomer and its underlying defects\footnote{Typical elastomers contain inherent defects of mostly submicron size \citep{Gent90,Valentinetal2010,Poulain17}.}, such a hypersurface is an intrinsic, albeit stochastic, macroscopic material property. We refer to it as the \emph{strength surface} of the elastomer.

\begin{remark}\label{Remark2}
\emph{The above definition of strength based on the response of the material under spatially uniform states of stress is precisely that introduced by \cite{KLP20} and \cite{KBFLP20} for elastic brittle materials. Sure, the strength of elastic brittle materials is described by a surface $\mathcal{F}\left(\bfS\right)=0$ solely in stress space. The form (\ref{Strength-Surface-General}) makes it plain that the strength of materials that dissipate energy by deformation, such as viscoelastic elastomers, requires, in general, a much richer mathematical representation.}
\end{remark}

The majority of the experimental strength data available in the literature pertains to nominally isotropic incompressible elastomers under uniaxial tension, when
$\bfS(t)={\rm diag}(0,0,s(t))$ and $\bfF(t)\approx{\rm diag}(\lambda^{-1/2}(t),$ $\lambda^{-1/2}(t),$ $\lambda(t))$, under equi-biaxial tension, when $\bfS(t)={\rm diag}(s(t),s(t),0)$ and $\bfF(t)\approx$ ${\rm diag}(\lambda(t),$ $\lambda(t),\lambda^{-2}(t))$, and under equi-triaxial or hydrostatic tension, when $\bfS(t)={\rm diag}(s(t),s(t),$ $s(t))$ and $\bfF(t)\approx{\rm diag}(1,1,1)$. The uniaxial and equi-biaxial data is mostly due to an experimental campaign led by Thor L. Smith 
some 70 years ago \citep{Smith58,Smith60,Smith63,Smith64,Smith64b,Smith65,Smith69}; see also \cite{Greensmith60a} and \cite{Mueller1968}. The first systematic experimental campaign on hydrostatic strength data can be traced back to the celebrated poker-chip experiments of \cite{GL59} some 70 years ago as well; see the recent review by \cite{BCLLP24}. Remarkably, the data shows that, to a first degree of approximation, the strength of elastomers is independent of the history of the loading and hence that it can be represented as a hypersurface in stress-deformation space. This implies that we can remove the explicit time dependence in (\ref{Strength-Surface-General}) and, with a slight abuse of notation, rewrite
\begin{equation}
\mathcal{F}\left(\bfS,\bfF\right)=0.\label{Strength-Surface-T-independent}
\end{equation}
\paragraph{Isotropic viscoelastic elastomers} For elastomers whose viscoelastic response is isotropic, which is the class of elastomers of interest in this work, their strength is expected to also be isotropic. Standard arguments of material frame indifference and material symmetry then imply that for such elastomers the strength surface (\ref{Strength-Surface-T-independent}) admits a representation\footnote{Of course, many other representations that make use of other stress and deformation measures are possible \citep{KLP20,LP23}. Based on the various sets of experimental data that we have examined, the representation in terms of $\bfS^{(2)}$ and $\bfC$ utilized here appears to be an expedient one.} in terms of the ten invariants ${\rm tr}\,\bfS^{(2)}$, ${\rm tr}\,{\bfS^{(2)}}^2$, ${\rm tr}\,{\bfS^{(2)}}^3$, ${\rm tr}\,\bfC$, ${\rm tr}\,\bfC^2$, ${\rm tr}\,\bfC^3$, ${\rm tr}\,\bfS^{(2)}\bfC$, ${\rm tr}\,{\bfS^{(2)}}^2\bfC$, ${\rm tr}\,\bfS^{(2)}\bfC^2$, ${\rm tr}\,{\bfS^{(2)}}^2\bfC^2$ associated with the second Piola-Kirchhoff stress tensor $\bfS^{(2)}=\bfF^{-1}\bfS$ and the right Cauchy-Green deformation tensor $\bfC=\bfF^T\bfF$ \citep{Boehler1987}. 

In this work, we will make use of the specific class of strength surfaces
\vspace{-0.15cm}
\begin{equation}
\mathcal{F}\left(\bfS,\bfF\right)=
\sqrt{\dfrac{\mathfrak{I}^2_1}{3}-\mathfrak{I}_2}+g_1(I_1)\mathfrak{I}_1+g_0(I_1)=0\quad {\rm with}\quad\left\{\begin{array}{l}g_0(I_1)=-\dfrac{\sqrt{3}\shs\sts(I_1)}{3\shs-\sts(I_1)} \vspace{0.2cm}\\
g_1(I_1)=\dfrac{\sts(I_1)}{\sqrt{3}\left(3\shs-\sts(I_1)\right)}\end{array}\right.\label{Strength-Surface-DP}
\end{equation}
\vspace{-0.15cm}
where $\mathfrak{I}_1$, $\mathfrak{I}_2$, $I_1$ stand for the stress and deformation invariants
\begin{align}
\left\{\hspace{-0.1cm}\begin{array}{l}
\mathfrak{I}_1={\rm tr}\,\bfS^{(2)}=\mathfrak{s}_1+\mathfrak{s}_2+\mathfrak{s}_3\vspace{0.15cm}\\
\mathfrak{I}_2=\dfrac{1}{2}\left(\left({\rm tr}\,\bfS^{(2)}\right)^2-{\rm tr}\,{\bfS^{(2)}}^2\right)=\dfrac{1}{2}\left((\mathfrak{s}_1+\mathfrak{s}_2+\mathfrak{s}_3)^2-\mathfrak{s}_1^2-\mathfrak{s}_2^2-\mathfrak{s}_3^2\right)\end{array}\right. \hspace{-0.1cm},\quad
I_1={\rm tr}\,\bfC=\lambda_1^2+\lambda_2^2+\lambda_3^2,\label{Invariants}
\end{align}
and where $\sts(I_1)$ and $\shs$ are the material function and material constant that describe the strength of the elastomer at hand under uniaxial tension and hydrostatic loading when $\bfS^{(2)}={\rm diag}(\mathfrak{s}>0,0,0)$ and $\bfS^{(2)}={\rm diag}(\mathfrak{s}>0,\mathfrak{s}>0,\mathfrak{s}>0)$, respectively. In the above expressions, $\mathfrak{s}_1$, $\mathfrak{s}_2$, $\mathfrak{s}_3$ stand for the eigenvalues of $\bfS^{(2)}$, that is, the principal second Piola-Kirchhoff stresses, and $\lambda_1$, $\lambda_2$, $\lambda_3$ stand for the eigenvalues of $\sqrt{\bfC}$, that is, the principal stretches. 

The strength surface (\ref{Strength-Surface-DP}) is a generalization of the Drucker-Prager-type strength surface introduced in \citep{KLP20} for elastic brittle materials --- which, in turn, is an adaptation of the surface originally introduced by \cite{DruckerPrager1952} to model the yielding of soils --- in that the material parameters $g_0(I_1)$ and $g_1(I_1)$ are functions of deformation as opposed to constants. Having $g_0(I_1)$ and $g_1(I_1)$ depend only on the first principal invariant $I_1$ of the right Cauchy-Green deformation tensor, as opposed to also on its second and third principal invariants, is motivated both by the fact that viscoelastic elastomers are by and large nearly incompressible, so that $J=\det\sqrt{\bfC}\approx 1$, and by simplicity. Indeed, the strength surface (\ref{Strength-Surface-DP}) is arguably the simplest model that allows to fit and interpolate in a reasonable manner the strength data that is typically available from experiments for viscoelastic elastomers, thus its use here.

\paragraph{The material function $\mathfrak{s}_{\emph{\texttt{ts}}}(I_1)$ and material constant $\mathfrak{s}_{\emph{\texttt{hs}}}$} In general, the material function $\sts(I_1)$ and material constant $\shs$ need to be determined directly from experiments by fitting uniaxial and hydrostatic strength data for the elastomer of interest.

\begin{table}[t!]\centering
\caption{Constants in the formula (\ref{Strength-Solithane}) for the uniaxial tensile strength $\sts(I_1)$ and value of the hydrostatic strength $\shs$ obtained by fitting the experimental data of \cite{Knauss67} for Solithane 113 (50/50).}
\begin{tabular}{ccccccc|c}
\hline
$a_0$ & $a_1$ (MPa) & $b_1$ & $a_2$  (MPa)  & $b_2$ & $c_0$ & $c_1$ &  $\shs$ (MPa) \\
\hline
$0$ & $3.7159$ & $0.8543$ & $1.0538\times10^{-6}$ & $8.8912$ & $3.103$ & $10$ & $2.27$ \\
\hline
\end{tabular} \label{Table1}
\end{table}
%
%
\begin{figure}[t!]
\centering
\begin{subfigure}
\centering
\includegraphics[width=0.74\linewidth]{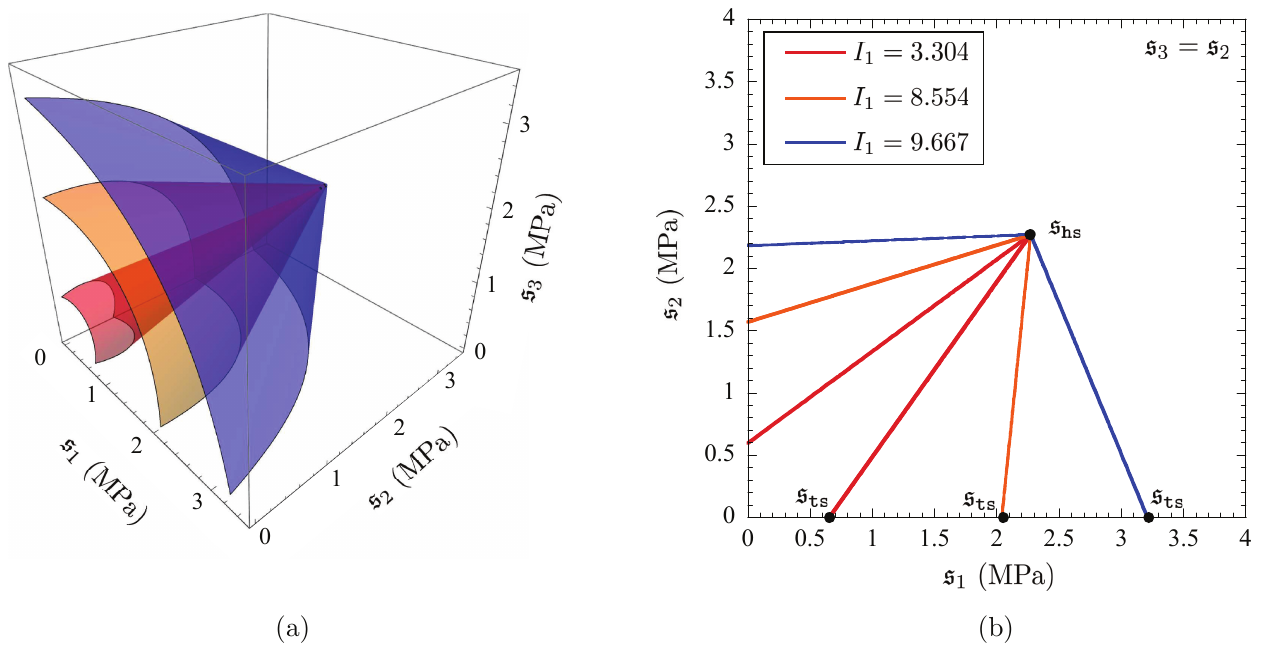}
\end{subfigure}
\begin{subfigure}
\centering
\includegraphics[width=0.74\linewidth]{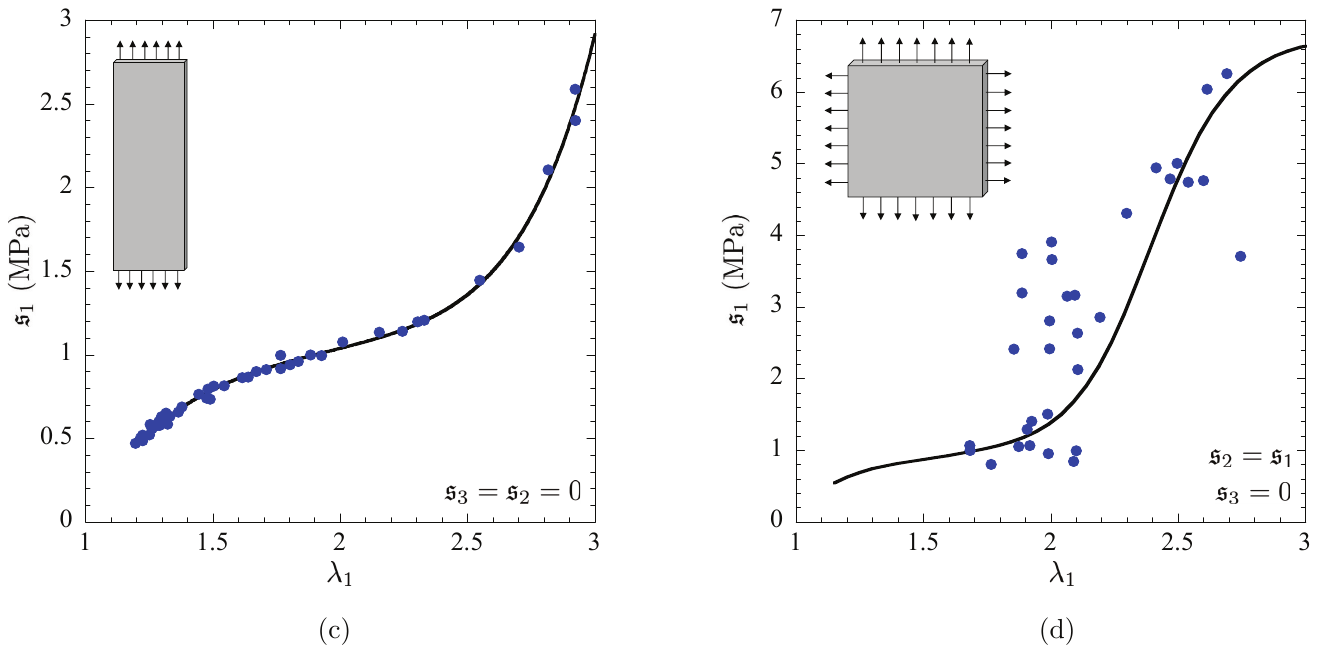}
\end{subfigure}
\caption{{\small Strength surface (\ref{Strength-Surface-DP}) with  (\ref{Strength-Solithane}) and the material constants given in Table \ref{Table1} for one of the polyurethane elastomers studied by \cite{Knauss67}, Solithane 113 with 50/50 composition. (a) Plot of the surface in the space of principal second Piola-Kirchhoff stresses $(\mathfrak{s}_1,\mathfrak{s}_2,\mathfrak{s}_3)$ for three fixed values of the deformation measure $I_1$. (b) Plot of the surface in the space of principal stresses $(\mathfrak{s}_1,\mathfrak{s}_2)$ with $\mathfrak{s}_3=\mathfrak{s}_2$ for the same three fixed values of $I_1$. (c) Plot of the surface in the space of principal-stress-principal-stretch $(\mathfrak{s}_1,\lambda_1)$ with $\mathfrak{s}_3=\mathfrak{s}_2=0$ and $\lambda_3=\lambda_2=\lambda_1^{-1/2}$. (d) Plot of the surface in the space of principal-stress-principal-stretch $(\mathfrak{s}_1,\lambda_1)$ with $\mathfrak{s}_2=\mathfrak{s}_1$, $\mathfrak{s}_3=0$,  $\lambda_2=\lambda_1$, and $\lambda_3=\lambda_1^{-2}$. Parts (c) and (d) include the experimental data (circles) reported by \cite{Knauss67} for direct comparison, alongside simplified schematics of the associated tests.}}\label{Fig1}
\end{figure}

Experiments show that the material function $\sts(I_1)$ is qualitatively similar to the ``equilibrium'' stress-stretch response of the elastomer at hand under uniaxial tension applied at a very slow loading rate when viscous effects are negligible; see, e.g., \cite{Smith65}. This implies that one can use the stress-stretch relation predicted for uniaxial tension by any sufficiently general nonlinear elastic model of choice as a formula for $\sts(I_1)$. In this work, we make use of the formula \citep{LP10}
\begin{equation}
\sts(I_1)=\left\{\hspace{-0.1cm}\begin{array}{ll}
+\infty  & 3\leq I_1<c_0 \vspace{0.2cm}\\
a_0+\displaystyle\sum_{r=1}^2a_r \sqrt{\frac{I_1}{3}-1} \left(2+\left(1+\sqrt{\frac{I_1}{3}-1}\right)^3\right)^{b_r-1}
   \left(1+\sqrt{\frac{I_1}{3}-1}\right)^{-b_r-\frac{1}{2}} \quad & {\rm if}\quad  c_0\leq I_1<c_1 \vspace{0.2cm}\\
a_0+\displaystyle\sum_{r=1}^2a_r \sqrt{\frac{c_1}{3}-1} \left(2+\left(1+\sqrt{\frac{c_1}{3}-1}\right)^3\right)^{b_r-1}
   \left(1+\sqrt{\frac{c_1}{3}-1}\right)^{-b_r-\frac{1}{2}}   &  c_1\leq I_1
\end{array}\right.,\label{Strength-Solithane}
\end{equation}
where $a_0,a_1,b_1,a_2,b_2,c_0,c_1$ are material constants. The last two of these constants, $c_0$ and $c_1$, stand for the threshold values of the deformation invariant $I_1$ below which the elastomer cannot possibly fracture under uniaxial tension, irrespective of how the load is applied in time, and above which the uniaxial tensile strength saturates. The rest of the constants serve to modulate the nonlinear relation between the critical values of the stress $\sts$ and the critical values of the deformation $c_0\leq I_1\leq c_1$ at which the elastomer fractures under uniaxial tension.

Experiments also indicate that the value of the material constant $\shs$ is typically within the narrow range
\begin{equation*}
\shs\in[0.5,5]\, {\rm MPa}
\end{equation*}
for most elastomers, even when their uniaxial tensile strength can be drastically different; see, e.g., Fig.~15 in \citep{BCLLP24}.

By way of an example, Fig.~\ref{Fig1} shows the comparison between the strength surface (\ref{Strength-Surface-DP}), with the constitutive choice (\ref{Strength-Solithane}), where the values of the constants $a_0,a_1,b_1,a_2,b_2,c_0,c_1$, $\shs$ are given in Table \ref{Table1}, and the experimental data of Knauss (1967) for Solithane 113 (50/50 composition); see Fig.~6 in that reference. 

Three comments are in order. First, the strength surface (\ref{Strength-Surface-DP}) with the formula  (\ref{Strength-Solithane}) for the uniaxial tensile strength $\sts(I_1)$ is able to describe accurately the available experimental data. Second, as expected, the experimental data exhibits significant scatter (in this case, the biaxial data more so than the uniaxial data) indicating that the value of $\shs$ and the material constants in (\ref{Strength-Solithane})  should not be taken as deterministic but rather as stochastic parameters; we will come back to this important point in Subsection \ref{Sec: rubber bands} below. We note, in particular, that the precise stochastic spatial distribution of strength in viscoelastic elastomers is itself a property of the material, one that reflects the intrinsic distribution of its microscopic flaws, much like in glasses and ceramics. Finally, it should be emphasized that more experimental data (beyond uniaxial and equi-biaxial tension) is needed to further probe the accuracy of the Drucker-Prager-type strength surface (\ref{Strength-Surface-DP}) for viscoelastic elastomers or guide its generalization; see, e.g., \cite{Chockalingam25}.

\subsubsection{Nucleation of fracture from large pre-existing cracks: The intrinsic fracture toughness}\label{Sec:Nucleation-crack}

When a specimen of the elastomer of interest contains a large\footnote{``Large'' refers to large relative to the characteristic size of the underlying heterogeneities in the elastomer under investigation. By the same token, ``small'' refers to sizes that are of the same order or just moderately larger than the sizes of the heterogeneities.} pre-existing crack, fracture may nucleate from the crack front, in other words, the crack may grow. The first systematic experimental campaign on this type of fracture nucleation appears to date back to the work of \cite{Busse34}, who carried out single edge notch fracture tests in specimens made of natural rubber. Those experiments made clear that a criterion based on strength applied pointwise could not possibly explain the observations. Indeed, because of the sharp geometry of the cracks, the stresses around the crack fronts greatly exceeded the strength of the elastomer as soon as the specimens were loaded, and yet the cracks did not grow. It was the later works of \cite{RT53} and \cite{Greensmith55} in the 1950s, following in the footsteps of \cite{Griffith21}, and the flurry of subsequent experimental investigations \citep{Mullins59,LakeandThomas67,Gent96,Thomas00,Knauss15,Creton20} that revealed that it is according to a Griffith-type competition between bulk deformation energy and surface fracture energy that fracture nucleates from a large pre-existing crack in an elastomer. Precisely, these investigations established that fracture nucleates from a large pre-existing crack in an elastomer whenever the change in \emph{total} deformation (stored and dissipated) energy $\mathcal{W}$ in the bulk with respect to an added surface area to the pre-existing crack $\Gamma_0$ reaches a certain critical tearing energy:
\begin{equation}
-\dfrac{\partial \mathcal{W}}{\partial \Gamma_0}=T_{c}.\label{Tc-0}
\end{equation}
In this expression, the added surface area refers to the undeformed configuration, the derivative is to be carried out under fixed boundary conditions on the parts of the boundary which are not traction-free, so that no work is done by the external forces, and the critical tearing energy $T_c$ is a material property that is loading-history dependent or, as otherwise described, ``rate dependent''. 

Traditionally, the critical tearing energy $T_c$ has been written in the form \citep{GentSchultz72,Gent96}
\begin{equation*}
T_{c}=G_c(1+f_c),
\end{equation*}
where the material constant $G_c$ denotes the fracture energy, or intrinsic fracture toughness, of the elastomer, while the non-negative material function $f_c$ stands for the part of $T_c$ that describes its dependence on the loading history. Experiments carried out at extremely low loading rates, as well as at high temperatures and on elastomers in a swollen state, show that $f_c$ vanishes when viscous dissipation is minimized \citep{LakeandThomas67,Knauss71,Ahagon-Gent75,GT82,GentLai94}. They also show that $f_c$ scales with the viscosity of the elastomer \citep{Mullins59,Knauss73,GentLai94,Gent96}. Save for these qualitative characteristics, how the material function $f_c$ depends on the loading history for arbitrary loading conditions remained an open problem for decades rendering the Griffith criticality condition in its ordinary form (\ref{Tc-0}) of limited utility. 

%
\begin{figure}[b!]
\centering
\includegraphics[width=0.99\linewidth]{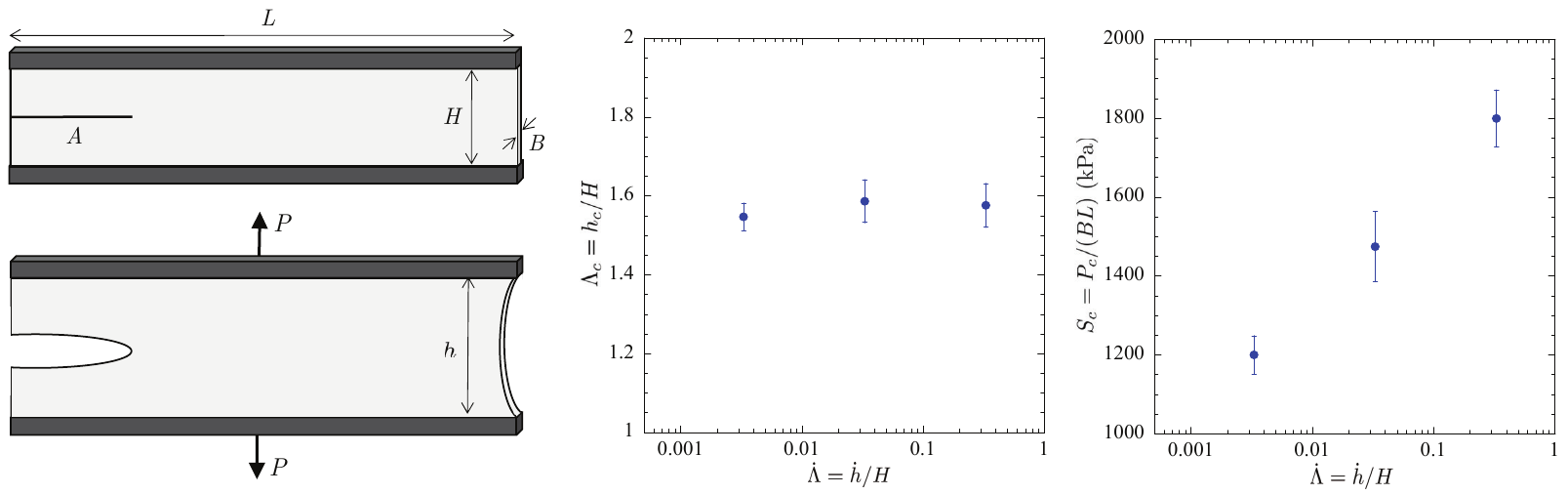}
\caption{{\small Schematic of the specimens ($L=60$ mm, $H=6$ mm, $B=1.5$ mm, $A=10$ mm) and results for the critical global stretch $\Lambda_c=h_c/H$ and the corresponding critical global stress $S_c=P_c/(B L)$, plotted as functions of the loading rate $\dot{\Lambda}=\dot{h}/H$, at which fracture nucleates from the large pre-existing cracks in the ``pure-shear'' fracture tests on SBR reported by \cite{Cai17}.}}\label{Fig2}
\end{figure}
%

In a recent contribution, as a direct consequence of two elementary observations, \cite{SLP23} have uncovered a general formula for $f_c$. This formula shows that equation (\ref{Tc-0}) can be simplified to a fundamental form that involves \emph{not} the historically elusive critical tearing energy $T_c$, but only the fracture energy $G_c$ of the elastomer. The first of these two observations is that, for any viscoelastic elastomer, the total deformation energy $\mathcal{W}$ in (\ref{Tc-0}) can be partitioned into the three different contributions
\begin{equation*}
\mathcal{W}=\underbrace{\mathcal{W}^{{\rm Eq}}+\mathcal{W}^{{\rm NEq}}}_\text{stored}+\underbrace{\mathcal{W}^{v}}_\text{dissipated}. 
\end{equation*}
The second observation is that the so-called ``pure-shear'' fracture tests of elastomers consistently show that fracture occurs from the pre-existing crack in the specimens at a critical stretch that is roughly independent of the loading rate at which the test is carried out; see, e.g., \cite{Major10,Pharretal2012,Cai17,Kangetal2020}. By combining these two elementary observations, \cite{SLP23} have shown that the Griffith criticality condition (\ref{Tc-0}) reduces to the fundamental form
\begin{equation}
-\dfrac{\partial \mathcal{W}^{{\rm Eq}}}{\partial \Gamma_0}=G_{c}.\label{Gc-0}
\end{equation}
By the same token, they have also shown that the critical tearing energy is given by the formula
\begin{equation*}
T_{c}=G_c-\dfrac{\partial \mathcal{W}^{{\rm NEq}}}{\partial \mathrm{\Gamma}_0}-\dfrac{\partial \mathcal{W}^{v}}{\partial \mathrm{\Gamma}_0},
\end{equation*}
where the right-hand side is to be evaluated at the time at which (\ref{Gc-0}) is attained along the loading path of interest. In these last expressions, $\mathcal{W}^{v}$ represents the part of the total energy that has been dissipated via viscous deformation, while the combination $\mathcal{W}^{{\rm Eq}}+\mathcal{W}^{{\rm NEq}}$ represents the part of the total energy that is stored by the elastomer via elastic deformation. In this combination, $\mathcal{W}^{{\rm NEq}}$ stands for the part of the stored energy that can be dissipated eventually via viscous deformation as the elastomer reaches a state of thermodynamic equilibrium. On the other hand, $\mathcal{W}^{{\rm Eq}}$ denotes the part of the stored energy that the elastomer will retain at thermodynamic equilibrium.

The following two comments are in order. First, from a fundamental point of view,  as noted in Remark \ref{Remark1} above, the criticality condition (\ref{Gc-0}) makes it plain that when dealing with materials that dissipate energy by deformation, the Griffith energy competition that describes the growth of cracks may indeed involve only a certain part --- as opposed to the entirety --- of the deformation energy. Second, from a practical point of view, the criticality condition (\ref{Gc-0}) is straightforward to employ for arbitrary specimen geometries and arbitrary loading conditions \citep{SLP23b,SLP23c,SKLP24}. This is because its evaluation requires having knowledge of the finite viscoelastic behavior of the elastomer --- from which $\mathcal{W}^{{\rm Eq}}$, $\mathcal{W}^{{\rm NEq}}$, $\mathcal{W}^v$ can be computed --- and of its fracture energy $G_c$. Both of these material properties can be measured once and for all.

By way of an example, Fig.~\ref{Fig2} reproduces the experimental results in \citep{Cai17} for the critical global stretch $\Lambda_c=h_c/H$ and the corresponding critical global stress $S_c=P_c/(B L)$ at which fracture nucleates from the large pre-existing cracks in ``pure-shear'' fracture tests on SBR. In accordance with the Griffith criticality condition (\ref{Gc-0}), as elaborated in \citep{SLP23}, the results show that $\Lambda_c$ is independent of the loading rate $\dot{\Lambda}=\dot{h}/H$, whereas $S_c$ increases monotonically with increasing $\dot{\Lambda}$.

%
\begin{figure}[t!]
\centering
\includegraphics[width=0.8\linewidth]{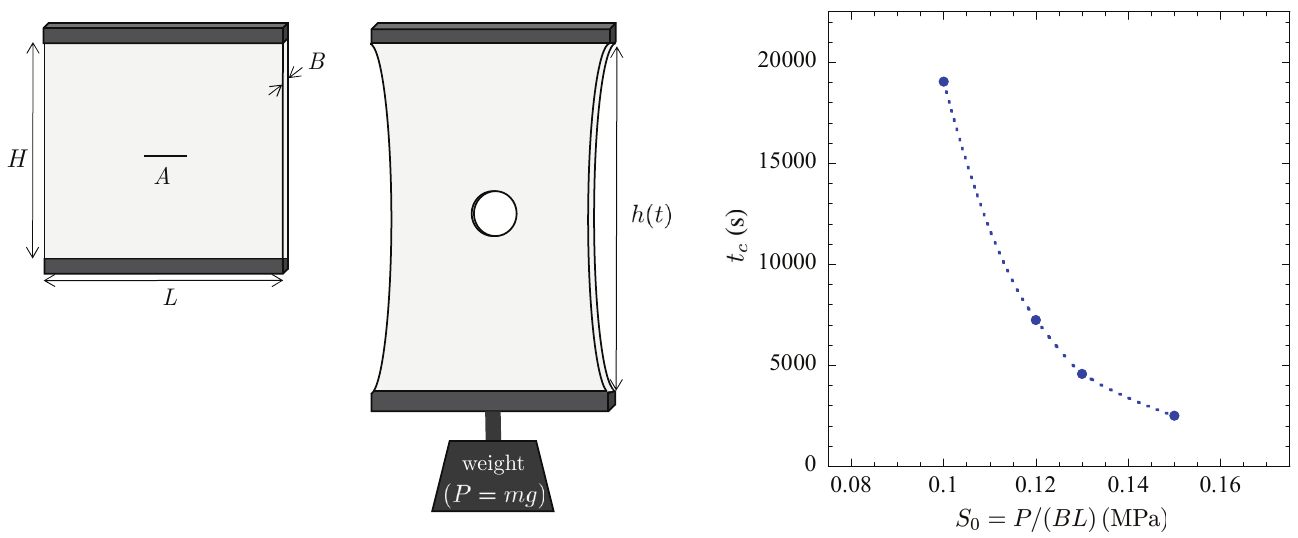}
\caption{{\small Schematic of the specimens ($L=101.6$ mm, $H=L=101.6$ mm, $B=0.7938$ mm, $A=5.08$ mm) and results for the critical time $t_c$, plotted as a function of the applied global stress $S_0=P/(B L)$, at which fracture nucleates from the large pre-existing cracks in the delayed fracture tests on a polyurethane elastomer reported by \cite{Knauss70}.}}\label{Fig3}
\end{figure}
%

As elaborated in \citep{SLP23b}, another telltale test that illustrates nucleation of fracture from large pre-existing cracks in viscoelastic elastomers according to the Griffith criticality condition (\ref{Gc-0}) is the so-called delayed fracture test. In a typical delayed fracture test, a sheet of the elastomer of interest containing a large pre-existing crack is subjected to a load that is applied rapidly over a very short time interval $[0, t_0]$ and then held constant. Nucleation of fracture from the pre-existing crack occurs at a critical time $t_c \gg t_0$, hence the name of the test. Fig.~\ref{Fig3} reproduces the experimental results in \citep{Knauss70}, where the pre-existing crack is located in the center of the specimen and the load is applied in a uniaxial fashion. The results show that nucleation of fracture can occur hours after the load has been applied. In particular, they show that smaller applied global stresses $S_0=P/(B L)$ lead to significantly larger critical times $t_c$ at fracture.

\subsubsection{Nucleation of fracture under states of non-uniform stress: Mediation between the strength and the Griffith energy competition}\label{Sec:Nucleation-transition}

%
\begin{figure}[b!]
\centering
\includegraphics[width=0.75\linewidth]{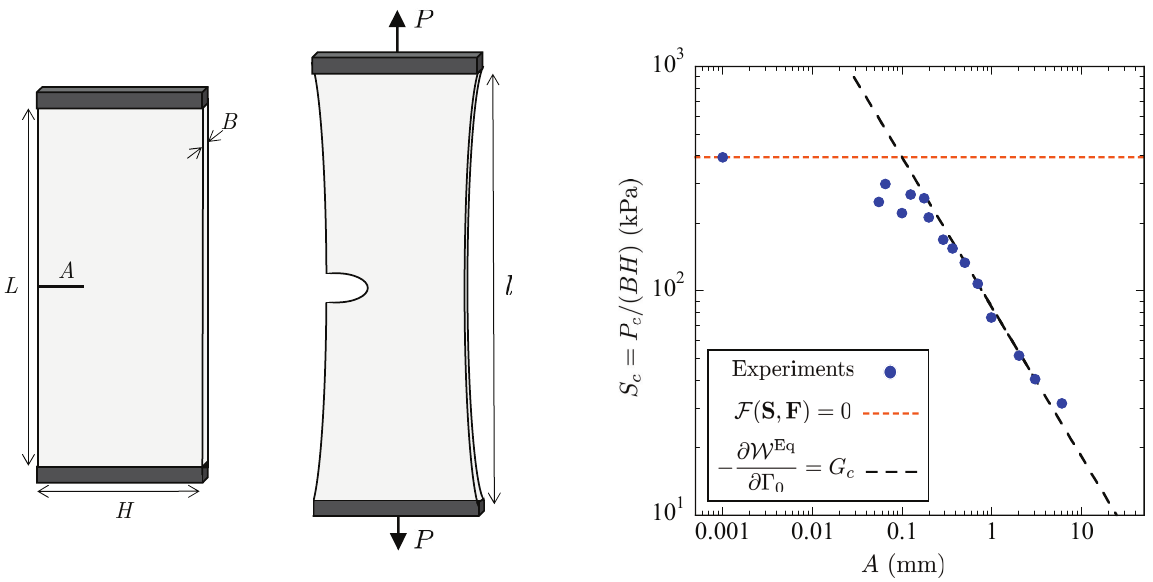}
\caption{{\small  
Schematic of the specimens ($L\leq100$ mm, $H\leq50$ mm, $B=0.5$ mm) and results for the critical global stress $S_c=P_c/(B H)$, plotted as a function of the size $A$ of the pre-existing crack, at which fracture nucleates from the pre-existing cracks in the single edge notch fracture tests on the acrylic elastomer VHB 4905 reported by \cite{Chen17}. The results illustrate that fracture nucleation: $i$) from large cracks ($A>0.3$ mm) can be described by the Griffith criticality condition (\ref{Gc-0}); $ii$) from very small cracks ($A<1$ $\mu$m) is described by the strength (\ref{Strength-Surface-T-independent}); while $iii$) from cracks of intermediate length is described by an interpolation between the Griffith and strength limits.}}\label{Fig4}
\end{figure}
%

The two preceding subsections have focused on how fracture nucleates in elastomers under two opposite limiting conditions: when the stress field is spatially uniform and when the stress field is most non-uniform, in fact, singular. They have shown that two different criteria govern nucleation of fracture under each such condition. Namely, the strength criterion (\ref{Strength-Surface-T-independent}) governs fracture nucleation under spatially uniform states of stress, while the Griffith energy criterion (\ref{Gc-0}) governs fracture nucleation from large pre-existing cracks. In this subsection, we summarize the existing experimental knowledge on fracture nucleation in elastomers between these two opposite limiting conditions, when the stress field is not spatially uniform, but not as non-uniform as around large pre-existing cracks. These include nucleation of fracture from notches, smooth and sharp, small pre-existing cracks, and from any other subregion in the body under a non-uniform state of stress.

Experiments on specimens featuring U- and V-notches \citep{Greensmith60,Andrews63,Berto18}, as well as specimens featuring small pre-existing edge cracks \citep{Thomas1970,Hamed2016,Chen17}  have shown that nucleation of fracture from the front of the notch or crack is  governed neither by the strength nor by the Griffith competition between bulk deformation energy and surface fracture energy in the elastomer but by a mediation between these two extremal criteria. The same is true for nucleation within the bulk in subregions where the stress in non-uniform, such as for instance, in the so-called poker-chip and two-particle experiments; see, e.g., \cite{GL59,GuoRavi23,KLP21,KKLP24} and \cite{GentPark84,Poulain17,KFLP18,KRLP18}. More specifically, all these experiments have shown that the violation of the strength criterion (\ref{Strength-Surface-T-independent}) is a necessary but \emph{not} sufficient condition for fracture to nucleate in subregions within elastomers where the stress field is not uniform. Sufficiency appears to be reached when the violation of the strength criterion (\ref{Strength-Surface-T-independent}) is over a large enough subregion, the size of such regions being characterized by a family of length scales that is material specific.

By way of an example, Fig.~\ref{Fig4} shows the experimental data in \citep{Chen17} for the critical global stress $S_c=P_c/(B H)$ at which fracture nucleates from the pre-existing cracks in single edge notch fracture tests on VHB 4905, a homogeneous  acrylic elastomer from the 3M Company. For cracks larger than about $0.3$ mm, fracture nucleation is well described by the Griffith criticality condition (\ref{Gc-0}). For cracks smaller than $0.3$ mm, the critical global stress $S_c$ transitions from being Griffith-dominated to being strength-dominated. The tests were carried out at an approximately constant stretch rate of $1.7\times 10^{-2}$ s$^{-1}$, thus the unique value of uniaxial tensile strength (\emph{circa} $400$ kPa in nominal stress) in the data. The intersection between the results for the prediction for the critical global stress $S_c$ by the Griffith criticality condition and by the strength serves to roughly identify the material length scale (in this case, about $0.1$ mm) of the subregion over which the violation of the strength criterion (\ref{Strength-Surface-T-independent}) must occur for fracture nucleation to take place. This particular length scale associated with uniaxial tensile strength has been typically referred to as the Irwin length scale \citep{Irwin60}, or more recently, as the fracto-cohesive length scale \citep{Chen17}.

\subsection{Propagation of fracture: The intrinsic fracture toughness}\label{Sec:Propagation-crack}

Finally, we turn to the experimental investigations of the propagation of fracture in viscoelastic elastomers. Traditionally, these have mostly revolved around  ``pure-shear'', single edge notch, and double cantilever beam fracture tests, where the crack front is tracked in time with help of a camera \citep{Mueller1968,Knauss71,Thomas2000a,Kinloch2003,Corre20}, and more so around the so-called trousers fracture test, where, as elaborated next, the use of a camera is not essential.

\begin{figure}[t!]
\centering
\includegraphics[width=0.9\linewidth]{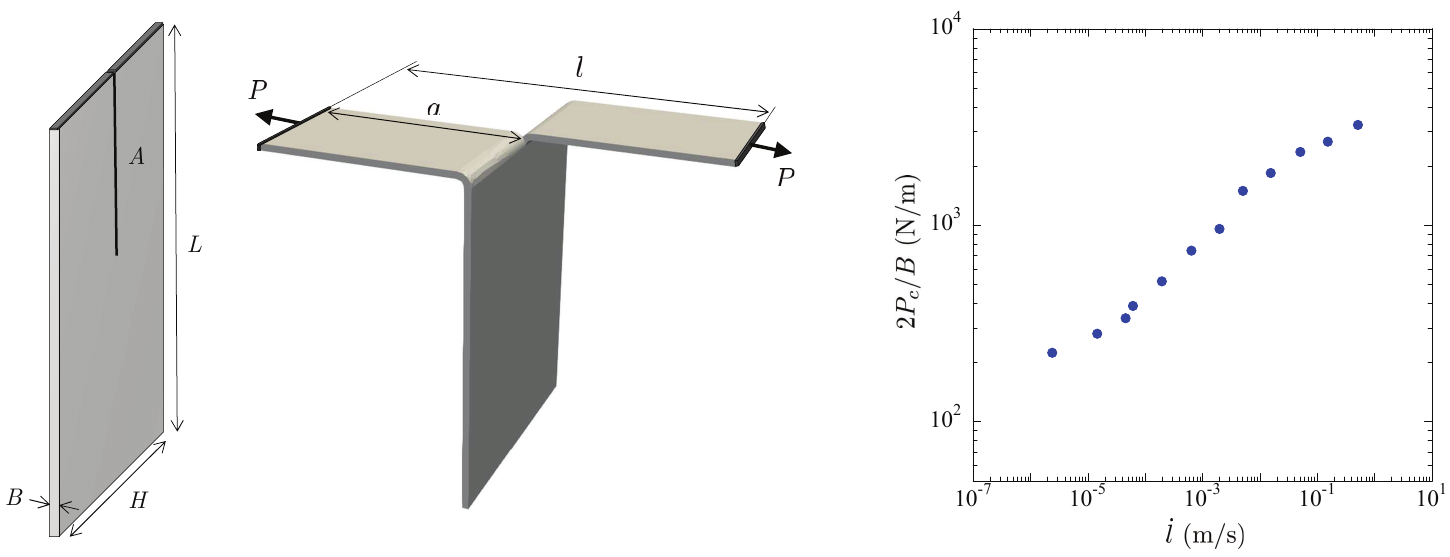}
\caption{{\small Schematic of the specimens ($L=150$ mm, $H=40$ mm, $B=1$ mm, $A=50$ mm) and results for the normalized critical force $2P_c/B$ at which the crack grows at a constant speed $\dot{a}\approx\dot{l}/2$ in the trousers tests on a SBR (vulcanizate $D$) reported by \cite{Greensmith55}. The results are shown as a function of the applied constant rate $\dot{l}$ of separation between the grips.}}\label{Fig5}
\end{figure}

While \cite{RT53} introduced the trousers test as a convenient test to study nucleation of fracture from a large pre-existing crack in elastomers, it was later recognized by \cite{Greensmith55} that it is also a particularly convenient test to study propagation of fracture. The main reason for this is that the crack in these tests can be made to propagate at a constant speed $\dot{a}$ when the test is carried out either at a constant rate $\dot{l}$ of separation between the grips or, roughly equivalently, at a constant force $P$; see Fig.~\ref{Fig5}. An additional reason behind the convenience of trousers tests is that, for the case of specimens of appropriate dimensions, they allow to estimate in closed form the speed at which the crack propagates, as well as the associated critical energy release rate. It bears emphasizing, however, that the specimens commonly used for trousers tests are \emph{not} of such appropriate dimensions; see the recent analysis by \cite{KLP25}.

%
%

By way of an example, Fig.~\ref{Fig5} reproduces one of the results from the pioneering trousers tests carried out by \cite{Greensmith55} on SBR (vulcanizate $D$); see Fig.~6
in their work. The results for the normalized critical force $2P_c/B$ at which the crack grows at a constant speed $\dot{a}\approx\dot{l}/2$ are plotted as a function of the applied constant rate $\dot{l}$ of separation between the grips. These results show that the Griffith criticality condition (\ref{Tc-0}) is a necessary condition for propagation of fracture in viscoelastic elastomers for crack growth occurs when (\ref{Tc-0}) is continuously satisfied \citep{KLP25}. The results also reveal how the value of $P_c$ --- and hence of $T_c$ --- scales with the viscous dissipation in the elastomer. In particular, $P_c$ is bounded from below and from above and increases monotonically from its smallest to largest value with increasing $\dot{l}$. In the limit of vanishingly small loading rate ($\dot{l}\searrow 0$), when viscous effects are minimized, $P_c$ approaches its smallest value. As the loading rate increases, when viscous effects increase, so does $P_c$ monotonically. In the limit of ``infinitely'' fast\footnote{Since inertia is not taken into account, ``infinitely'' fast should be understood in the usual sense of a loading condition that is applied in a time scale that is much smaller than the characteristic relaxation time of the elastomer, but still large enough that inertial effects can be neglected.} loading rates ($\dot{l}\nearrow +\infty$), when viscous effects are maximized, $P_c$ approaches its largest value.

Much like for fracture nucleation from large pre-existing cracks, the analysis presented in \citep{SLP23c} of the trousers fracture test for viscoelastic elastomers has shown that the Griffith criticality condition (\ref{Tc-0}) can be reduced to the fundamental form (\ref{Gc-0}), which we repeat here for symmetry of presentation
\begin{equation}
-\dfrac{\partial \mathcal{W}^{{\rm Eq}}}{\partial \Gamma_0}=G_{c},\label{Gc-0-Propagation}
\end{equation}
to describe the propagation of fracture in viscoelastic elastomers. It bears re-emphasizing that the Griffith criticality condition (\ref{Gc-0-Propagation}) is applicable for arbitrary specimen geometries and arbitrary loading conditions and that its use is straightforward, as its evaluation requires having knowledge of the finite viscoelastic behavior of the elastomer and of its fracture energy.

\section{The proposed theory}\label{Sec: Theory}

In the sequel, we present a macroscopic theory of deformation and fracture nucleation and propagation in viscoelastic elastomers subjected to quasistatic loading conditions that directly accounts for all the three basic ingredients identified by the existing body of experimental observations summarized in the preceding section. As announced in the Introduction, we do so by extending the phase-field theory initiated by \cite*{KFLP18} for elastic brittle materials to viscoelastic materials.

\subsection{Initial configuration and kinematics}

Consider a body made of an elastomer that in its initial configuration, at time $t=0$, occupies an open bounded domain\footnote{For notational simplicity, we restrict the presentation to 3D domains. Nevertheless, the corresponding formulations for 1D and 2D domains should be apparent.} $\Omega_0\subset \mathbb{R}^3$, with boundary $\partial \Omega_0$ and outward unit normal $\bfN$. Identify material points by their initial position vector $\bfX\in\Omega_0$.

In response to externally applied mechanical stimuli to be described below, the position vector $\bfX$ of a material point moves to a new position specified by
\begin{equation*}
\bfx=\bfy(\bfX, t),
\end{equation*}
where $\bfy$ is an invertible mapping from $\Omega_0$ to the current configuration $\Omega(t)$, also contained in $\mathbb{R}^3$. Making use of standard notation, we write the deformation gradient and velocity fields at $\bfX$ and $t$ as
\begin{equation*}
\bfF(\bfX, t)=\nabla\bfy(\bfX,t)=\frac{\partial \bfy}{\partial \bfX}(\bfX,t) \qquad {\rm and}\qquad \textbf{V}(\bfX,t)=\dot{\bfy}(\bfX, t)= \frac{\partial \bfy}{\partial t}(\bfX,t).
\end{equation*}
Here and subsequently, the ``dot'' notation is employed to denote the Lagrangian time derivative (i.e., with $\bfX$ held fixed) of any field quantity.

In addition to the above-described deformation of the elastomer, the externally applied mechanical stimuli may also result in the nucleation and subsequent propagation of cracks in the elastomer. We describe such cracks in a regularized fashion via an order parameter or phase field
\begin{equation*}
z=z(\bfX, t)
\end{equation*}
taking values in the range $[0,1]$. The value $z=1$ identifies the intact regions of the elastomer and $z=0$ those that have been fractured, while the transition from $z=1$ to $z=0$ is set to occur smoothly over regions of small thickness of regularization length scale $\varepsilon>0$.

\subsection{Constitutive behavior of the elastomer: Its viscoelasticity, strength, and fracture energy}

\subsubsection{Viscoelasticity}

The elastomer is taken to be homogeneous and isotropic. Making use of the two-potential formalism \citep{Halphen75,Ziegler87,KLP16,Yavari24}, we describe its viscoelastic behavior by two thermodynamic potentials, a free energy function, or free energy density per unit undeformed volume, of the form
\begin{equation}
\psi(\bfF,\bfC^v)=\psi^{{\rm Eq}}(I_1,J)+\psi^{{\rm NEq}}(I_1^e,J)\label{psi}
\end{equation}
that describes how the elastomer stores energy through elastic deformation and a dissipation potential of the form
\begin{equation}
\phi(\bfF,\bfC^v,\dot{\bfC}^v)=\left\{\begin{array}{ll}
\hspace{-0.1cm}\dfrac{1}{2}\dot{\bfC}^v\cdot\Atan(\bfF,\bfC^v)\dot{\bfC}^v& \, {\rm if}\, \,  \det\bfC^v=1\\ \\
\hspace{-0.1cm}+\infty &\, {\rm else}\end{array}\right.\quad {\rm with}\quad \mathcal{A}_{ijkl}(\bfF,\bfC^v)=\dfrac{\eta(I_1^e,I_2^e,I_1^v)}{2}{C^{v}}^{-1}_{ik}{C^{v}}^{-1}_{jl} \label{phi}
\end{equation}
that describes how the elastomer dissipates energy through viscous deformation. In these expressions, the symmetric second-order tensor $\bfC^v$ is an internal variable of state that stands for a measure of the ``viscous part'' of the deformation gradient $\bfF$,
\begin{align*}
I_1={\rm tr}\,\bfC, \quad  J=\sqrt{\det\bfC}, \quad I^v_1={\rm tr}\,\bfC^v,\quad I_1^e={\rm tr}(\bfC{\bfC^{v}}^{-1}),\quad I_2^e=\dfrac{1}{2}\left({I_1^e}^2-{\rm tr}\, (\bfC{\bfC^{v}}^{-1}\bfC{\bfC^{v}}^{-1})\right), 
\end{align*}
where, again, $\bfC=\bfF^T\bfF$ denotes the right Cauchy-Green deformation tensor, and $\psi^{{\rm Eq}}$, $\psi^{{\rm NEq}}$, $\eta$ are any suitably well-behaved non-negative material functions of their arguments.

Granted the two thermodynamic potentials (\ref{psi}) and (\ref{phi}), it follows that the first Piola-Kirchhoff stress tensor $\bfS$  is given by the relation \citep{KLP16}
\begin{equation*}
\bfS(\bfX,t)=\frac{\partial \psi}{\partial\bfF}(\bfF,\bfC^v),
\end{equation*}
where $\bfC^v$ is implicitly defined by the evolution equation
\begin{equation*}
\left\{\begin{array}{l}\dfrac{\partial \psi}{\partial \bfC^v}(\bfF,\bfC^v)+\dfrac{\partial \phi}{\partial \dot{\bfC}^v}(\bfF,\bfC^v,\dot{\bfC}^v)={\bf0}\\ \\
\bfC^v(\bfX,0)=\bfI\end{array}\right. .
\end{equation*}
Making use of the specific isotropic forms (\ref{psi}) and (\ref{phi}), this relation can be rewritten more explicitly as
\begin{equation}
\bfS(\bfX,t)=2\psi^{{\rm Eq}}_{I_1}\bfF+2\psi^{{\rm NEq}}_{I^e_1}\bfF{\bfC^v}^{-1}+\left(\psi^{{\rm Eq}}_{J}+\psi^{{\rm NEq}}_{J}\right)J\bfF^{-T},\label{S-I1-J}
\end{equation}
where $\bfC^v$ is solution of the evolution equation
\begin{equation}
\left\{\begin{array}{l}\dot{\bfC}^v(\bfX,t)=\dfrac{2\psi^{{\rm NEq}}_{I^e_1}}{\eta(I_1^e,I_2^e,I_1^v)}\left(\bfC-\dfrac{1}{3}\left(\bfC\cdot{\bfC^v}^{-1}\right)\bfC^v\right)\\ \\
\bfC^v(\bfX,0)=\bfI\end{array}\right. ,\label{Evolution-I1-J}
\end{equation}
and where we have made use of the notation $\psi^{{\rm Eq}}_{I_1}=\partial \psi^{{\rm Eq}}(I_1,J)/\partial I_1$, $\psi^{{\rm Eq}}_{J}=\partial \psi^{{\rm Eq}}(I_1,J)/\partial J$, $\psi^{{\rm NEq}}_{I^e_1}=\partial \psi^{{\rm NEq}}$ $(I^e_1,J)/\partial I^e_1$, $\psi^{{\rm NEq}}_{J}=\partial \psi^{{\rm NEq}}(I^e_1,J)/\partial J$.

For a detailed account of the constitutive relation (\ref{S-I1-J})-(\ref{Evolution-I1-J}) and its physical interpretation, the interested reader is referred to \cite{KLP16}. Here, we do make the following two remarks.

\begin{remark}
\emph{The constitutive relation (\ref{S-I1-J})-(\ref{Evolution-I1-J}) is admittedly general as it includes four basic models as limiting cases. The first one, which corresponds to setting the viscosity function either to $\eta=0$ or $\eta=+\infty$, is that of a \emph{non-Gaussian elastic solid}. This model, in turn, includes that of an \emph{isotropic linear elastic solid} as a special case. The third one, which corresponds to setting the equilibrium and non-equilibrium energies to $\psi^{{\rm Eq}}=0$ and $\psi^{{\rm NEq}}=+\infty$, is the basic model of an incompressible \emph{non-Newtonian fluid}, which can be further simplified to an incompressible \emph{Newtonian fluid} by setting the viscosity function $\eta$ to be a constant.
}
\end{remark}

\begin{remark}
\emph{The constitutive relation (\ref{S-I1-J})-(\ref{Evolution-I1-J}) is nothing more than a generalization of the classical \cite{Zener48} or standard solid model  to the setting of finite deformations. To see this, note that the rheological representation of the constitutive relation (\ref{S-I1-J})-(\ref{Evolution-I1-J}) is given by the Zener rheological representation shown in Fig.~\ref{Fig6}. Note as well that in the limit of small deformations, as $\bfF\rightarrow \bfI$, the constitutive relation (\ref{S-I1-J})-(\ref{Evolution-I1-J}) reduces to the classical linear viscoelastic Zener model
\begin{equation*}
\bfS(\bfX,t)=2\,\mu^{{\rm Eq}}\,\boldsymbol{\varepsilon}+\Lambda^{{\rm Eq}} \, {\rm tr}(\boldsymbol{\varepsilon})\,\bfI+2\,\mu^{{\rm NEq}}(\boldsymbol{\varepsilon}-\boldsymbol{\varepsilon}^v)+\Lambda^{{\rm NEq}} \, {\rm tr}(\boldsymbol{\varepsilon}-\boldsymbol{\varepsilon}^v)\,\bfI,
\end{equation*}
where $\boldsymbol{\varepsilon}=1/2(\bfF+\bfF^T-2\bfI)$ stands for the infinitesimal strain tensor and $\boldsymbol{\varepsilon}^v=1/2(\bfC^v-\bfI)$ is solution of the evolution equation
\begin{equation*}
\left\{\begin{array}{l}\dot{\boldsymbol{\varepsilon}}^v(\bfX,t)=\dfrac{\mu^{{\rm NEq}}}{\eta_0}\left(\boldsymbol{\varepsilon}-\boldsymbol{\varepsilon}^v-\dfrac{1}{3}{\rm tr}(\boldsymbol{\varepsilon}-\boldsymbol{\varepsilon}^v)\bfI\right) \\ \\
\boldsymbol{\varepsilon}^v(\bfX,0)=\textbf{0}\end{array}\right. .  
\end{equation*}
In these expressions, $\mu^{{\rm Eq}}=2\psi^{{\rm Eq}}_{I_1}(3,1)=-\psi^{{\rm Eq}}_{J}(3,1)$ and $\Lambda^{{\rm Eq}}=\psi^{{\rm Eq}}_{J J}(3,1)+4\psi^{{\rm Eq}}_{I_1 I_1}(3,1)+4\psi^{{\rm Eq}}_{I_1 J}(3,1)-\mu^{{\rm Eq}}$ stand for the initial shear modulus and first Lam\'e constant associated with the elastic energy in equilibrium, similarly,  $\mu^{{\rm NEq}}=2\psi^{{\rm NEq}}_{I^e_1}(3,1)=-\psi^{{\rm NEq}}_{J}(3,1)$ and $\Lambda^{{\rm NEq}}=\psi^{{\rm NEq}}_{J J}(3,1)+4\psi^{{\rm NEq}}_{I^e_1 I^e_1}(3,1)+4\psi^{{\rm NEq}}_{I^e_1 J}(3,1)-\mu^{{\rm NEq}}$ stand for the initial shear modulus and first Lam\'e constant associated with the non-equilibrium elastic energy, while $\eta_0=\eta(3,3,3)$ stands for the initial viscosity.
}
\begin{figure}[H]
   \centering \includegraphics[width=2.5in]{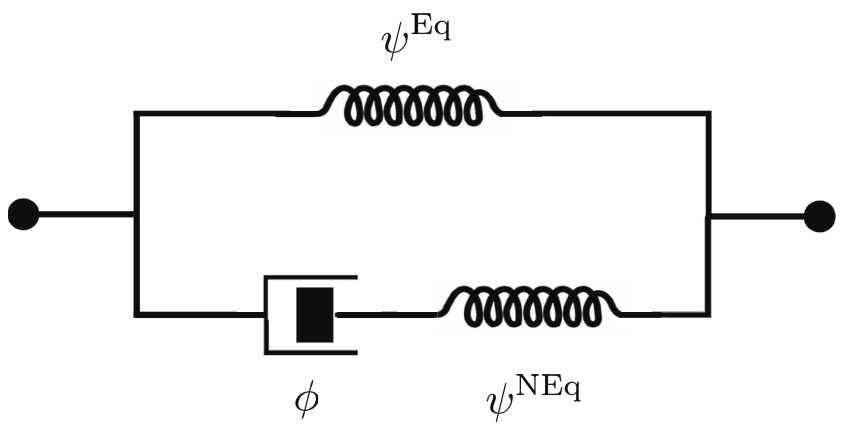}
   \caption{\small Rheological representation of the two-potential viscoelastic model (\ref{S-I1-J})-(\ref{Evolution-I1-J}).}
   \label{Fig6}
\end{figure}
\end{remark}

In the comparisons with experiments that are presented in Section \ref{Sec: Simulations} below, we will make use of the following specific forms for the equilibrium and non-equilibrium free energy functions in (\ref{psi}) and viscosity function in (\ref{phi}):

\begin{equation}\label{PsiLP-vis}
\left\{\hspace{-0.1cm}\begin{array}{l}
\psi^{{\rm Eq}}(I_1,J)=\displaystyle\sum\limits_{r=1}^2\dfrac{3^{1-\alpha_r}}{2 \alpha_r}\mu_r\left(\,I_1^{\alpha_r}-3^{\alpha_r}\right)-\displaystyle\sum\limits_{r=1}^2\mu_r\ln J+\dfrac{\kappa}{2}(J-1)^2\vspace{0.2cm}\\
\psi^{{\rm NEq}}(I^e_1,J)=\displaystyle\sum\limits_{r=1}^2\dfrac{3^{1-\beta_r}}{2 \beta_r}\nu_r\left(\,{I^e_1}^{\beta_r}-3^{\beta_r}\right)-\displaystyle\sum\limits_{r=1}^2\nu_r\ln J\vspace{0.2cm}\\
\eta(I_1^e,I_2^e,I_1^v)=\widetilde{\eta}(I_1^v,\mathcal{J}_2^{{\rm NEq}})=\eta_{\infty}+\dfrac{\eta_0-\eta_{\infty}+K_1\left(I_1^{v \gamma_1}-3^{\gamma_1}\right)}{1+\left(K_2 \mathcal{J}_2^{{\rm NEq}}\right)^{\gamma_2}}
\end{array}\right. ,
\end{equation}
with $\mathcal{J}_2^{{\rm NEq}}=\left(\dfrac{{I_1^e}^2}{3}-I_2^e\right)\left(\displaystyle\sum\limits_{r=1}^2 3^{1-\beta_r}\nu_r {I_1^e}^{\beta_r-1}\right)^2$, which result in the constitutive relation

\begin{equation}
\bfS(\bfX,t)=\displaystyle\sum\limits_{r=1}^2 3^{1-\alpha_r} \mu_r I_1^{\alpha_r-1} \bfF+\displaystyle\sum\limits_{r=1}^2 3^{1-\beta_r}\nu_r {I^e_1}^{\beta_r-1}\bfF{\bfC^v}^{-1}-\left(\displaystyle\sum\limits_{r=1}^2\dfrac{\mu_r+\nu_r}{J}-\kappa(J-1)\right)J\bfF^{-T}\label{S-KLP}
\end{equation}
with evolution equation
\begin{equation}
\left\{\begin{array}{l}
\dot{\bfC}^v(\bfX,t)=\dfrac{\sum\limits_{r=1}^2 3^{1-\beta_r}\nu_r {I^e_1}^{\beta_r-1}}{\widetilde{\eta}(I_1^v,\mathcal{J}_2^{{\rm NEq}})}\left(\bfC-\dfrac{1}{3}\left(\bfC\cdot{\bfC^v}^{-1}\right)\bfC^v\right)\\ \\
\bfC^v(\bfX,0)=\bfI\end{array}\right. \label{Evolution-KLP}
\end{equation}
for $\bfC^v$. 

The constitutive prescription (\ref{S-KLP})-(\ref{Evolution-KLP}) has been shown to be accurately descriptive and predictive of a wide range of elastomers, which typically exhibit non-Gaussian elasticity \citep{Treloar2005,LP10}, as well as nonlinear deformation-dependent shear-thinning viscosity \citep{Doi98,BB98,Lion2006,KLP16,GSKLP21,Cohen21,Ricker23}, thus its use here.

In total, the constitutive prescription (\ref{S-KLP})-(\ref{Evolution-KLP}) contains fifteen material constants. Five of them, $\mu_1$, $\mu_2$, $\kappa$, $\alpha_1$, $\alpha_2$, serve to characterize the initial and the large-deformation non-Gaussian elasticity of the elastomer at states of thermodynamic equilibrium. Another four, $\nu_1$, $\nu_2$, $\beta_1$, $\beta_2$,  characterize the initial and the large-deformation non-Gaussian elasticity of the elastomer at non-equilibrium states. Finally, the last six constants, $\eta_0$, $\eta_{\infty}$, $K_1$, $\gamma_1$, $K_2$, $\gamma_2$, serve to characterize the nonlinear deformation-dependent shear-thinning viscosity. These parameters can be determined by fitting the model simultaneously to experimental data from a few standard tests, such as uniaxial relaxation/creep data and/or uniaxial tension/compression at different constant deformation rates. It is important, however, that the fitted data span the entire range of finite deformations and deformation rates prior to fracture, as deformations and deformation rates are typically large around regions where fracture nucleates and propagates and hence they need to be accurately described by the model \citep{SKLP24}.

\subsubsection{Strength}

We take that the strength of the elastomer is characterized by the Drucker-Prager-type strength surface (\ref{Strength-Surface-DP}), with the formula (\ref{Strength-Solithane}) for the uniaxial tensile strength. In practice, it is difficult to carry out experiments that probe the entire space of uniform stresses and uniform deformation gradients in order to measure the entire strength surface (\ref{Strength-Surface-T-independent}) of a given elastomer of interest. In fact, as noted in Subsection \ref{Sec: Strength ingredient}, only uniaxial, equi-biaxial, and hydrostatic strength data is typically available from experiments. The Drucker-Prager-type strength surface (\ref{Strength-Surface-DP}) with (\ref{Strength-Solithane}) is able to describe all such available data, interpolate in a reasonable manner between them, and has the additional merit of being simple, thus its use here. At any rate, we emphasize that the central idea behind the proposed theory is independent of the specific choice of strength surface (\ref{Strength-Surface-T-independent}).

In total, the strength surface (\ref{Strength-Surface-DP}) with (\ref{Strength-Solithane}) contains eight material constants. Seven of them, $a_0$, $a_1$, $b_1$, $a_2$, $b_2$, $c_0$, $c_1$, serve to characterize the stress-stretch pairs at which the elastomer fractures under uniaxial tension. The last constant, $\shs$, characterizes the stress at which the elastomer fractures under pure hydrostatic loading. As illustrated in Fig.~\ref{Fig1} above, these parameters can be determined by fitting experimental data for uniaxial tensile strength and hydrostatic or equi-biaxial tensile strength. It is important that the fitted data is sufficiently complete so that it describes the inherent stochasticity of the strength of the elastomer at hand.

\subsubsection{Fracture energy}

Finally, we take that the fracture energy of the elastomer is characterized by the constant
\begin{align*}
G_c. 
\end{align*}
As discussed in Subsections \ref{Sec:Nucleation-crack} and \ref{Sec:Propagation-crack}, tests aimed at measuring $G_c$ can be readily carried out with conventional equipment. However, it is important to note that these tests might not entirely eliminate viscous effects. If that is the case, the actual value of $G_c$ may have to be inferred from a combination of tests.

\begin{remark}
\emph{For the constitutive viscoelastic behavior (\ref{S-I1-J})-(\ref{Evolution-I1-J}) used in this work, the equilibrium elastic energy $\mathcal{W}^{{\rm Eq}}$ in the Griffith criticality condition (\ref{Gc-0}) is given by
\begin{equation}\label{Int-WEq}
\mathcal{W}^{{\rm Eq}}=\displaystyle\int_{\mathrm{\Omega}_0}\psi^{{\rm Eq}}(I_1,J)\,{\rm d}\bfX.
\end{equation}
Here, recalling that $I_1(\bfX,t)={\rm tr}\,(\nabla \bfy^T\nabla \bfy)$ and $J(\bfX,t)=\det\nabla\bfy$, it is important to emphasize that the equilibrium energy (\ref{Int-WEq}) --- and hence its derivative $-\partial \mathcal{W}^{{\rm Eq}}/\partial \Gamma_0$ and, consequently, how large pre-existing cracks nucleate, as well as how cracks propagate, in viscoelastic elastomers according to the Griffith criticality condition (\ref{Gc-0}) --- depends on the \emph{entire} viscoelastic behavior of the elastomer --- as characterized by $\psi^{{\rm Eq}}$, $\psi^{{\rm NEq}}$, $\eta$ --- through the solution of the governing equations (to be spelled out below) for the deformation field $\bfy(\bfX,t)$.
}
\end{remark}

\subsection{Initial conditions, boundary conditions, and source terms}

In its initial configuration, we consider that the body is undeformed and stress-free. We also take the phase field to be initially equal to one throughout, that is, the elastomer is intact. Precisely, we have that
\begin{align*}
\bfy(\bfX,0)=\bfX,\quad \bfX\in \Omega_0,\qquad \bfC^v(\bfX,0)=\bfI,\quad \bfX\in \Omega_0, \qquad{\rm and}\qquad z(\bfX,0)=1,\quad \bfX\in \Omega_0.
\end{align*}
The last of these initial conditions may be replaced along the boundary of sharp corners and pre-existing cracks by the initial condition $z=0$.

On a portion $\partial  \Omega_0^{\mathcal{D}}$ of the boundary $\partial \Omega_0$ of the body, the deformation field $\bfy$ is taken to be given by a known function $\overline{\bfy}$, while the complementary part of the boundary $\partial  \Omega_0^{\mathcal{N}}=\partial \Omega_0\setminus\partial \Omega_0^{\mathcal{D}}$ is subjected to a prescribed nominal traction $\bfS\bfN=\overline{\textbf{s}}$. Moreover, the fracturing of the elastomer is taken to be solely driven by the mechanical loads applied to it, and not by the direct imposition of a phase-transition stimulus on its boundary. For this reason, the Neumann boundary condition $\nabla z\cdot\bfN=0$ is assumed to hold on the entire boundary $\partial \Omega_0$. All in all, we have that
\begin{align*}
\left\{\begin{array}{ll}
\bfy(\bfX,t)=\overline{\bfy}(\bfX,t),&\;(\bfX,t)\in \partial  \Omega_0^{\mathcal{D}}\times [0,T]\vspace{0.15cm}\\
\bfS\bfN=\overline{\textbf{s}}(\bfX,t),&\; (\bfX,t)\in \partial  \Omega_0^{\mathcal{N}}\times [0,T]\end{array}\right.
\qquad{\rm and}\qquad \nabla z\cdot\bfN=0,\quad (\bfX,t)\in \partial\Omega_0\times [0,T]. 
\end{align*}
The last of these boundary conditions may be replaced along the boundary of sharp corners and pre-existing cracks by the Dirichlet boundary condition $z=0$.

We assume that body forces are negligible.

\subsection{The governing equations of deformation and fracture in their general form}

Following \cite*{KFLP18}, there are two governing principles that the mechanical forces and energy applied to the body ought to satisfy, those are the balance of linear momentum\footnote{Balance of angular momentum is automatically satisfied by virtue of the material frame indifference of the thermodynamic potentials (\ref{psi}) and (\ref{phi}); see, e.g., \cite{KLP16}.} and a certain balance of configurational forces; see also \cite{KRLP18}. When combined with the constitutive behavior for elastomers introduced above, they yield the system of coupled PDEs and ODE

\begin{equation}
\left\{\begin{array}{ll}
\hspace{-0.15cm} {\rm Div}\left[z^2\left(2\psi^{{\rm Eq}}_{I_1}\,\nabla\bfy+2\psi^{{\rm NEq}}_{I^e_1}\,\nabla\bfy \,{\bfC^v}^{-1}+\left(\psi^{{\rm Eq}}_{J}+\psi^{{\rm NEq}}_{J}\right)( \det\nabla\bfy)\,\nabla\bfy^{-T}\right)\right]={\bf0}, \quad(\bfX,t)\in \Omega_0\times[0,T]\\[12pt]
\hspace{-0.15cm} \det\nabla\bfy>0, \quad(\bfX,t)\in\Omega_0\times[0,T]\\[12pt]
\hspace{-0.15cm}\bfy(\bfX,t)=\overline{\bfy}(\bfX,t), \quad(\bfX,t)\in\partial  \Omega_0^{\mathcal{D}}\times[0,T]\\[12pt]
\hspace{-0.15cm} \left(z^2\left(2\psi^{{\rm Eq}}_{I_1}\,\nabla\bfy+2\psi^{{\rm NEq}}_{I^e_1}\,\nabla\bfy \,{\bfC^v}^{-1}+\left(\psi^{{\rm Eq}}_{J}+\psi^{{\rm NEq}}_{J}\right)( \det\nabla\bfy)\,\nabla\bfy^{-T}\right)\right)\bfN=\overline{\textbf{s}}(\bfX,t), \quad(\bfX,t)\in\partial \Omega_0^{\mathcal{N}}\times[0,T]\\[12pt]
\hspace{-0.15cm} \bfy(\bfX,0)=\bfX, \quad \bfX\in\Omega_0\end{array}\right. , \label{BVP-y-theory}
\end{equation}
\begin{equation}
\hspace{-1.3cm}\left\{\begin{array}{l}\hspace{-0.15cm}\dot{\bfC}^v(\bfX,t)=\dfrac{2\psi^{{\rm NEq}}_{I^e_1}}{\eta(I_1^e,I_2^e,I_1^v)}\left(\nabla\bfy^T\nabla\bfy-\dfrac{1}{3}\left(\nabla\bfy^T\nabla\bfy\cdot{\bfC^v}^{-1}\right)\bfC^v\right), \quad(\bfX,t)\in\Omega_0\times[0,T] \\[12pt]
\hspace{-0.15cm}\bfC^v(\bfX,0)=\bfI, \quad \bfX\in\Omega_0\end{array}\right., \label{BVP-Cv-theory}
\end{equation}
and

\begin{equation}
\left\{\begin{array}{l}\hspace{-0.15cm}{\rm Div}\left[\varepsilon\, G_c \nabla z\right]=\dfrac{8}{3}z\, \psi^{{\rm Eq}}(I_1,J)-\dfrac{4}{3}c_\texttt{e}(\bfX,t)-\dfrac{G_c}{2\varepsilon}
\quad\mbox{ if } \dot{z}(\bfX,t)< 0, \quad(\bfX,t)\in \Omega_0\times[0,T]
\\[12pt]
\hspace{-0.15cm}
{\rm Div}\left[\varepsilon\, G_c \nabla z\right]\ge \dfrac{8}{3}z\, \psi^{{\rm Eq}}(I_1,J)-\dfrac{4}{3}c_\texttt{e}(\bfX,t)-\dfrac{G_c}{2\varepsilon}
\quad\mbox{ if }  z(\bfX,t)=1 \mbox{ or } \dot{z}(\bfX,t)>0, \quad(\bfX,t)\in \Omega_0\times [0,T]\\[12pt]
\hspace{-0.15cm}
{\rm Div}\left[\varepsilon\, G_c \nabla z\right]\le \dfrac{8}{3}z\, \psi^{{\rm Eq}}(I_1,J)-\dfrac{4}{3}c_\texttt{e}(\bfX,t)-\dfrac{G_c}{2\varepsilon}
\quad\mbox{ if }  z(\bfX,t)=0, \quad(\bfX,t)\in \Omega_0\times [0,T]\\[12pt]
\hspace{-0.15cm}\nabla z\cdot\bfN=0, \quad (\bfX,t)\in \partial\Omega_0\times [0,T]\\[12pt]
\hspace{-0.15cm} z(\bfX,0)=1,\quad \bfX \in \Omega_0
\end{array}\right.  \label{BVP-z-theory}
\end{equation}
as the governing equations for the deformation field $\bfy(\bfX,t)$, the internal variable $\bfC^v(\bfX,t)$, and the phase field $z(\bfX,t)$ that describe how an arbitrary body made of a viscoelastic elastomer deforms, nucleates fracture, and propagates fracture in response to arbitrary quasistatic mechanical loads. 

\begin{remark}\label{Remark: Correct Griffith}
{\rm Equation (\ref{BVP-y-theory})$_1$ is nothing more than the balance of linear momentum 
\begin{equation*}
{\rm Div}\,\bfS={\bf0}\qquad {\rm with}\qquad \bfS=\dfrac{\partial\Psi}{\partial \bfF}(\bfF,\bfC^v,z), \quad {\rm where}\quad \Psi(\bfF,\bfC^v,z)=z^2\psi(\bfF,\bfC^v),
\end{equation*}
governing the evolution of the deformation field, while equations (\ref{BVP-z-theory})$_{1-3}$ stand for the balance of configurational forces 
\begin{equation*}
\left\{\begin{array}{l}\hspace{-0.15cm}{\rm Div}\,\Ctan+c_{\texttt{i}}+c_{\texttt{e}}=0
\quad\mbox{ if } \dot{z}< 0
\\[6pt]
\hspace{-0.15cm}
{\rm Div}\,\Ctan+c_{\texttt{i}}+c_{\texttt{e}} \geq 0
\quad\mbox{ if }  z=1 \mbox{ or } \dot{z}>0\\[6pt]
\hspace{-0.15cm}
{\rm Div}\,\Ctan+c_{\texttt{i}}+c_{\texttt{e}}\leq 0
\quad\mbox{ if }  z=0 
\end{array}\right.  \quad {\rm with}\qquad 
\left\{\begin{array}{l}\hspace{-0.15cm}
\Ctan=\dfrac{\partial\Psi}{\partial\bfZ}(\bfF,\bfC^v,z)+\dfrac{\partial\Phi}{\partial\dot\bfZ}(\bfF,\bfC^v,\bfZ,\dot{\bfC}^v,\dot z,\dot\bfZ;\varepsilon)
\\[12pt]
\hspace{-0.15cm}
c_{\texttt{i}}=-\dfrac{\partial\Psi}{\partial z}(\bfF,\bfC^v,z)-\dfrac{\partial\Phi}{\partial\dot z}(\bfF,\bfC^v,\bfZ,\dot{\bfC}^v,\dot z,\dot\bfZ;\varepsilon)
\end{array}\right.,
\end{equation*}
where 
\begin{equation*}
\Phi(\bfF,\bfC^v,\bfZ,\dot{\bfC}^v,\dot z,\dot\bfZ;\varepsilon)=\phi(\bfF,\bfC^v,\dot{\bfC}^v)-2\dot z\,\psi^{{\rm NEq}}(I_1^e,J)+\dfrac{3G_c}{8}\left(-\dfrac{\dot z}{\varepsilon}+2\varepsilon \bfZ\cdot\dot\bfZ\right),
\end{equation*}
governing the evolution of the phase field. In these last equations, consistent with the Griffith criticality condition (\ref{Gc-0}), only the equilibrium part $\psi^{{\rm Eq}}$ of the free energy function ends up entering the competition that describes the nucleation and propagation of fracture. In the same equations, $c_\texttt{e}(\bfX,t)$ is the driving force through which the strength surface of the elastomer enters the theory. As elaborated in the next subsection, its constitutive prescription depends on the particular form of the strength surface. 
}
\end{remark}

\begin{remark}\label{Remark: irreversibility}
{\rm The inequalities in (\ref{BVP-z-theory}) stem from the facts that, by definition, the phase field is bounded according to $0\leq z\leq 1$ and, by constitutive assumption, fracture is an irreversible process, in other words, healing is not allowed. Experimental evidence has revealed that internally nucleated cracks in some elastomers may self-heal \citep{Poulain17,Poulain18}. The inequalities (\ref{BVP-z-theory}) can be augmented to describe such a healing process, but we shall not consider healing in this work; see Subsection 3.2 in \cite*{KFLP18} for the relevant details and \cite{FGLP19} for the corresponding ``sharp-theory'' perspective.
}
\end{remark}

\begin{remark}{\rm The parameter $\varepsilon$ in (\ref{BVP-z-theory}), with units of $length$, regularizes sharp cracks. Accordingly, by definition, it can be arbitrarily small. In practice, $\varepsilon$ should be selected to be smaller than the smallest material length scale built in (\ref{BVP-y-theory})-(\ref{BVP-z-theory}),  which comes about because of the different units of the equilibrium free energy function $\psi^{{\rm Eq}}(I_1,J)$ ($force/length^2$), the strength function $\mathcal{F}(\bfS,\bfF)$ ($force/length^2$), and the fracture energy $G_c$ ($force/length$).
}\label{Remark_length}
\end{remark}

\begin{remark}{\rm In the absence of viscous dissipation, when $\psi^{{\rm NEq}}=0$ and $\eta=0$, the equation (\ref{BVP-Cv-theory}) is satisfied trivially, while equations (\ref{BVP-y-theory}) and (\ref{BVP-z-theory}) reduce to the governing equations introduced in \citep{KFLP18} for elastic brittle materials; see Subsection 3.3 in that work. The governing equations (\ref{BVP-y-theory})-(\ref{BVP-z-theory}) provide thus a seamless extension of the phase-field theory initiated by \cite*{KFLP18} for elastic brittle materials to account for the additional complexities of the dissipative and time-dependent deformation, strength, and toughness in viscoelastic elastomers.
}\label{Remark_Elastic}
\end{remark}

\subsection{The constitutive prescription for the driving force $c_{\texttt{\emph{e}}}(\bfX,t)$}

For the basic case of purely elastic brittle materials, given a strength surface $\mathcal{F}(\bfS)=0$, \cite{KLP20} and \cite{KBFLP20} provided a blueprint to construct constitutive prescriptions for the driving force $c_\texttt{e}(\bfX,t)$. In this subsection, we show that the same blueprint can be utilized \emph{mutatis mutandis} to construct constitutive prescriptions for the driving force $c_\texttt{e}(\bfX,t)$ for strength surfaces of the more general form $\mathcal{F}(\bfS,\bfF)=0$. In particular, analogous to the prescription recently put forth by \cite{KKLP24}, we work out a fully explicit constitutive prescription for $c_\texttt{e}(\bfX,t)$ for the case when the strength surface is of the Drucker-Prager type (\ref{Strength-Surface-DP}).

\subsubsection{The general functional form of $c_{\texttt{\emph{e}}}(\bfX,t)$}

Given an elastomer with strength surface of the Drucker-Prager type (\ref{Strength-Surface-DP}), we begin by considering driving forces of the form
\begin{align}
c_{\texttt{e}}(\bfX,t)=\beta_2^\varepsilon(I_1)\sqrt{\dfrac{\mathfrak{I}^2_1}{3}-\mathfrak{I}_2}+\beta_1^\varepsilon(I_1)  \mathfrak{I}_1+\beta_0^\varepsilon(I_1)+z\left(1-\dfrac{\sqrt{\mathfrak{I}^2_1}}{\mathfrak{I}_1}\right)\psi^{{\rm Eq}}(I_1,J)+\dfrac{3}{4}{\rm Div}\left[\varepsilon(\delta^{\varepsilon}-1) G_c\nabla z \right],\label{cehat}
\end{align}
where $\beta_0^\varepsilon(I_1)$, $\beta_1^\varepsilon(I_1)$, $\beta_2^\varepsilon(I_1)$, and $\delta^{\varepsilon}$, are $\varepsilon$-dependent coefficients, to be spelled out below, and  $\mathfrak{I}_1$ and $\mathfrak{I}_2$ stand for the principal invariants (\ref{Invariants})$_1$ of the second Piola-Kirchhoff stress tensor
\begin{align*}
\bfS^{(2)}(\bfX,t)=\bfF^{-1}(\bfX,t)\bfS(\bfX,t)=z^2\left(2\psi^{{\rm Eq}}_{I_1}\bfI+2\psi^{{\rm NEq}}_{I^e_1}{\bfC^v}^{-1}+\left(\psi^{{\rm Eq}}_{J}+\psi^{{\rm NEq}}_{J}\right)J\bfC^{-1}\right),
\end{align*}
to wit,
\begin{align}\label{I1I2-S-Invariants}
\left\{\begin{array}{l}
\mathfrak{I}_1=6z^2\psi^{{\rm Eq}}_{I_1}+2z^2\psi^{{\rm NEq}}_{I^e_1}\,{\rm tr}\,{\bfC^v}^{-1}+z^2\left(\psi^{{\rm Eq}}_{J}+\psi^{{\rm NEq}}_{J}\right)J^{-1} I_2\\ \\
\mathfrak{I}_2=\dfrac{z^4}{2}\mathfrak{I}^2_1-6z^4(\psi^{{\rm Eq}}_{I_1})^2 -2z^4(\psi^{{\rm NEq}}_{I^e_1})^2\,{\rm tr}\,{\bfC^v}^{-2}
-z^4\dfrac{J^2}{2}\left(\psi^{{\rm Eq}}_{J}+\psi^{{\rm NEq}}_{J}\right)^2\,{\rm tr}\,\bfC^{-2}-\vspace{0.15cm}\\
\hspace{0.8cm}4z^4\psi^{{\rm Eq}}_{I_1}\psi^{{\rm NEq}}_{I^e_1}\,{\rm tr}\,{\bfC^v}^{-1}-2z^4\psi^{{\rm Eq}}_{I_1}\left(\psi^{{\rm Eq}}_{J}+\psi^{{\rm NEq}}_{J}\right)J^{-1}I_2-2z^4J\psi^{{\rm NEq}}_{I^e_1}\left(\psi^{{\rm Eq}}_{J}+\psi^{{\rm NEq}}_{J}\right)\,{\rm tr}\,(\bfC^{-1}{\bfC^v}^{-1})\end{array}\right.,
\end{align}
where $I_2=(I_1^2-{\rm tr}\,\bfC^2)/2$ and we recall that $I_1={\rm tr}\,\bfC$ and $\bfC=\bfF^T\bfF$. Rather conveniently, note that these relations depend explicitly (without the need to having to resort to polar decompositions) on the deformation field $\bfy(\bfX,t)$, in particular on its gradient $\nabla\bfy(\bfX,t)$, and on the internal variable $\bfC^v(\bfX,t)$ and the phase field $z(\bfX,t)$. 

\subsubsection{The asymptotic form of the coefficients $\beta_0^\varepsilon(I_1)$, $\beta_1^\varepsilon(I_1)$, $\beta_2^\varepsilon(I_1)$ in the limit as $\varepsilon\searrow 0$}

The coefficients $\beta_0^\varepsilon(I_1)$, $\beta_1^\varepsilon(I_1)$, $\beta_2^\varepsilon(I_1)$ in the driving force (\ref{cehat}) must be selected so that for spatially uniform deformation gradients $\nabla\bfy(\bfX,t)=\overline{\bfF}(t)$ and spatially uniform phase field $z(\bfX,t)=1$ throughout the entire domain $\Omega_0$ occupied by the body, the evolution equation (\ref{BVP-z-theory})$_1$ for the phase field is satisfied when the given strength surface $\mathcal{F}(\overline{\bfS}(t),\overline{\bfF}(t))=0$, with $\overline{\bfS}(t)=\partial \psi(\overline{\bfF}(t),\obfC^v(t))/\partial\bfF$, is first violated along the given loading path in the limit as the regularization length $\varepsilon\searrow 0$. In this last expression, $\obfC^v(t)$ stands for the solution of the evolution equation (\ref{BVP-Cv-theory}) for the internal variable $\bfC^v(\bfX,t)$ corresponding to the spatially uniform deformation gradient $\nabla\bfy(\bfX,t)=\overline{\bfF}(t)$. Precisely, noting that the left- and right-hand sides in (\ref{BVP-z-theory})$_1$ reduce to
\begin{align*}
{\rm Div}\left[\varepsilon\, G_c \nabla z\right]=0\quad{\rm and} \quad \dfrac{8}{3}z\, \psi^{{\rm Eq}}(I_1,J)-\dfrac{4}{3}c_\texttt{e}(\bfX,t)-\dfrac{G_c}{2\varepsilon}=-\dfrac{4}{3}c_\texttt{e}(\bfX,t)-\dfrac{G_c}{2\varepsilon}+O(\varepsilon^0),
\end{align*}
and that the driving force (\ref{cehat}) reduces to 
\begin{align*}
c_{\texttt{e}}(\bfX,t)=\beta_2^\varepsilon(I_1)\sqrt{\dfrac{\mathfrak{I}^2_1}{3}-\mathfrak{I}_2}+\beta_1^\varepsilon(I_1)  \mathfrak{I}_1+\beta_0^\varepsilon(I_1),
\end{align*}
when $\nabla\bfy(\bfX,t)=\overline{\bfF}(t)$, $\bfC^v(\bfX,t)=\overline{\bfC}^v(t)$, $z(\bfX,t)=1$, and $\varepsilon\searrow 0$, we require that
\begin{align}
-\dfrac{4}{3}\left(\beta_2^\varepsilon(I_1)\sqrt{\dfrac{\mathfrak{I}^2_1}{3}-\mathfrak{I}_2}+\beta_1^\varepsilon(I_1)  \mathfrak{I}_1+\beta_0^\varepsilon(I_1)\right)-\dfrac{G_c}{2\varepsilon}=&C\mathcal{F}(\overline{\bfS}(t),\overline{\bfF}(t))\nonumber\\
=&C\left(\sqrt{\dfrac{\mathfrak{I}^2_1}{3}-\mathfrak{I}_2}+\dfrac{\sts(I_1)}{\sqrt{3}\left(3\shs-\sts(I_1)\right)} \mathfrak{I}_1-\dfrac{\sqrt{3}\shs\sts(I_1)}{3\shs-\sts(I_1)} \right)\label{Cond-F-0}
\end{align}
for some constant $C$ in the limit as $\varepsilon\searrow 0$. This requirement can be easily satisfied by setting
\begin{equation}
\left\{\begin{array}{l}\beta_0^\varepsilon(I_1)=(\delta^\varepsilon-1)\dfrac{3 G_c}{8 \varepsilon}\vspace{0.2cm}\\
\beta^\varepsilon_1(I_1)=-\dfrac{1}{\shs}\delta^\varepsilon\dfrac{G_c}{8\varepsilon}\vspace{0.2cm}\\
\beta^\varepsilon_2(I_1)=-\dfrac{\sqrt{3}(3\shs-\sts(I_1))}{\shs\sts(I_1)}\delta^\varepsilon\dfrac{G_c}{8\varepsilon}\end{array}\right. , \label{betas-0}
\end{equation}
where $\delta^\varepsilon$ is, for now, an arbitrary coefficient. Note that the choice (\ref{betas-0}) leads to the constant $C=\delta^\varepsilon G_c(3\shs-\sts(I_1))/(2\sqrt{3}\shs \sts(I_1)\varepsilon)$ in (\ref{Cond-F-0}).

\subsubsection{The coefficients $\beta_0^\varepsilon(I_1)$, $\beta_1^\varepsilon(I_1)$, $\beta_2^\varepsilon(I_1)$ for finite values of the regularization length $\varepsilon$}

In practice, however small, the value of the regularization length $\varepsilon$ is always finite. It is thus desirable to choose coefficients in the constitutive prescription (\ref{cehat}) for the driving force that result in a phase-field theory that is capable of describing accurately the strength surface of the elastomer  not only asymptotically in the limit as $\varepsilon\searrow 0$ but also at finite values of $\varepsilon$. This can be accomplished by adding corrections of $O(\varepsilon^0)$ and higher orders to the expressions for $\beta_1^\varepsilon(I_1)$ and $\beta_2^\varepsilon(I_1)$ in (\ref{betas-0}).

In order to determine the corrections in $\beta_1^\varepsilon(I_1)$ and $\beta_2^\varepsilon(I_1)$, as a natural choice among other possibilities, we require that when the entire domain $\Omega_0$ occupied by the body is subjected to spatially uniform uniaxial tension $\obfS(t)={\rm diag}(s(t)>0,0,0)$ with $\obfF(t)={\rm diag}(\lambda(t),\lambda_{l}(t),\lambda_{l}(t))$ and spatially uniform hydrostatic loading $\obfS(t)={\rm diag}(s(t)>0,s(t)>0,s(t)>0)$ with $\obfF(t)={\rm diag}(\lambda(t),\lambda(t),\lambda(t))$, when the phase field is also uniformly $z(\bfX,t)=1$, the evolution equation (\ref{BVP-z-theory})$_1$ is satisfied when the given strength surface $\mathcal{F}(\obfS(t),\obfF(t))=0$ is violated, this for any value of $\varepsilon$. Precisely, we require that
\begin{equation}
\left\{\begin{array}{l}
\dfrac{8}{3}\psi^{{\rm Eq}}_{\texttt{ts}}(I_1)-\dfrac{4}{3}\left(\beta_2^\varepsilon(I_1)\dfrac{\sts(I_1)}{\sqrt{3}}+\beta_1^\varepsilon(I_1)  \sts(I_1)+\beta_0^\varepsilon(I_1)\right)-\dfrac{G_c}{2\varepsilon}=0\vspace{0.2cm}\\
\dfrac{8}{3}\psi^{{\rm Eq}}_{\texttt{hs}}-\dfrac{4}{3}\left(3\beta_1^\varepsilon(I_1)  \shs+\beta_0^\varepsilon(I_1)\right)-\dfrac{G_c}{2\varepsilon}=0\end{array}\right.,\label{Cond-F-1}
\end{equation}
where $\psi^{{\rm Eq}}_{\texttt{ts}}(I_1)$ and $\psi^{{\rm Eq}}_{\texttt{hs}}$ stand for the values 
\begin{equation*}
\psi^{{\rm Eq}}_{\texttt{ts}}(I_1)=\psi^{{\rm Eq}}(\l_{\texttt{ts}}^2+2\l^2_l,\l_{\texttt{ts}}\l^2_l)\qquad {\rm and}\qquad \psi^{{\rm Eq}}_{\texttt{hs}}=\psi^{{\rm Eq}}(3\l^2_{\texttt{hs}},\l^3_{\texttt{hs}})
\end{equation*}
of the equilibrium free energy function $\psi^{{\rm Eq}}(I_1,J)$ at the deformations at which the strength surface is violated under states of spatially uniform uniaxial tension and hydrostatic stress when

\begin{equation*}
\left\{\begin{array}{l}\mathfrak{s}_1=\sts(I_1)\vspace{0.2cm}\\
\mathfrak{s}_2=\mathfrak{s}_3=0\vspace{0.2cm}\\
\lambda_1=\l_{\texttt{ts}}\vspace{0.2cm}\\
\lambda_2=\lambda_3=\l_l\vspace{0.2cm}\\
\bfC^v={\rm diag}(\l_v^2,\l_v^{-1},\l_v^{-1})\end{array}\right.\qquad {\rm and}\qquad \left\{\begin{array}{l}\mathfrak{s}_1=\mathfrak{s}_2=\mathfrak{s}_3=\shs\vspace{0.2cm}\\
\lambda_1=\lambda_2=\lambda_3=\l_{\texttt{hs}}\vspace{0.2cm}\\
\bfC^v=\bfI\end{array}\right.,
\end{equation*}
respectively. In these expressions, the stretches $\l_{\texttt{ts}}$ and  $\l_l$, together with the variable $\l_v$, are defined implicitly in terms of the uniaxial tensile strength $\sts(I_1)$ as the root closest to $(\l_{\texttt{ts}},\l_l,\l_v)=(1,1,1)$ of the system of nonlinear algebraic equations
\begin{align*}
&\left\{\begin{array}{l}
I_1=\l^2_{\texttt{ts}}+2\l^2_l\vspace{0.2cm}\\
\sts(I_1)=2 \psi^{{\rm Eq}}_{I_1}(I_1, \l_{\texttt{ts}}\l_l^2)+\dfrac{2}{\l^{2}_v} \psi^{{\rm NEq}}_{I^e_1}(\l^2_{\texttt{ts}}\l_v^{-2}+2\l^2_l\l_v, \l_{\texttt{ts}}\l_l^2)+\dfrac{\l^2_l}{\l_{\texttt{ts}}}\left(\psi^{{\rm Eq}}_{J}(I_1, \l_{\texttt{ts}}\l_l^2)+\psi^{{\rm NEq}}_{J}(\l^2_{\texttt{ts}}\l_v^{-2}+2\l^2_l\l_v, \l_{\texttt{ts}}\l_l^2)\right)\vspace{0.2cm}\\
0=2 \psi^{{\rm Eq}}_{I_1}(I_1, \l_{\texttt{ts}}\l_l^2)+2\l_v \psi^{{\rm NEq}}_{I^e_1}(\l^2_{\texttt{ts}}\l_v^{-2}+2\l^2_l\l_v, \l_{\texttt{ts}}\l_l^2)+\l_{\texttt{ts}}\left(\psi^{{\rm Eq}}_{J}(I_1, \l_{\texttt{ts}}\l_l^2)+\psi^{{\rm NEq}}_{J}(\l^2_{\texttt{ts}}\l_v^{-2}+2\l^2_l\l_v, \l_{\texttt{ts}}\l_l^2)\right)\end{array}\right.,  
\end{align*}
while the stretch $\l_{\texttt{hs}}$ is defined implicitly in terms of the hydrostatic strength $\shs$ as the root closest to $\l_{\texttt{hs}}=1$ of the nonlinear algebraic equation
\begin{equation*}
\shs=2\left(\psi^{{\rm Eq}}_{I_1}(3\l^2_{\texttt{hs}}, \l^3_{\texttt{hs}})+\psi^{{\rm NEq}}_{I^e_1}(3\l^2_{\texttt{hs}}, \l^3_{\texttt{hs}})\right)+\l_{\texttt{hs}}\left(\psi^{{\rm Eq}}_{J}(3\l^2_{\texttt{hs}}, \l^3_{\texttt{hs}})+\psi^{{\rm NEq}}_{J}(3\l^2_{\texttt{hs}}, \l^3_{\texttt{hs}})\right).
\end{equation*}

At this point, it is straightforward to deduce that the choice
\begin{equation}
\left\{\begin{array}{l}\beta_0^\varepsilon(I_1)=(\delta^\varepsilon-1)\dfrac{3 G_c}{8 \varepsilon}\vspace{0.2cm}\\
\beta^\varepsilon_1(I_1)=-\dfrac{1}{\shs}\delta^\varepsilon\dfrac{G_c}{8\varepsilon}+\dfrac{2\psi^{{\rm Eq}}_{\texttt{hs}}}{3\shs}\vspace{0.2cm}\\
\beta^\varepsilon_2(I_1)=-\dfrac{\sqrt{3}(3\shs-\sts(I_1))}{\shs\sts(I_1)}\delta^\varepsilon\dfrac{G_c}{8\varepsilon}-
\dfrac{2\psi^{{\rm Eq}}_{\texttt{hs}}}{\sqrt{3}\shs}+\dfrac{2\sqrt{3}\psi^{{\rm Eq}}_{\texttt{ts}}(I_1)}{\sts(I_1)}\end{array}\right. , \label{betas}
\end{equation}
where $\delta^\varepsilon$ is still an arbitrary coefficient, satisfies the asymptotic condition (\ref{Cond-F-0}) in the limit as $\varepsilon\searrow 0$, as well as the conditions (\ref{Cond-F-1}) for any value of $\varepsilon$. Expressions (\ref{betas}) are hence the coefficients that we are after for the constitutive prescription (\ref{cehat}).

\subsubsection{The constitutive prescription for the coefficient $\delta^\varepsilon$}\label{Sec: delta}

By construction, the constitutive prescription (\ref{cehat}) with coefficients (\ref{betas}) for the driving force $c_{\texttt{e}}(\bfX,t)$ leads to a phase-field fracture theory that predicts nucleation of fracture in a body that is subjected to spatially uniform stress according to the strength surface (\ref{Strength-Surface-DP}) of the elastomer, irrespective of the value of the coefficient $\delta^\varepsilon$. However, the value of $\delta^\varepsilon$ \emph{does} affect when the theory predicts nucleation of fracture from a large pre-existing crack, as well as when it predicts propagation of fracture, and hence must be prescribed accordingly. ﻿Given that the prescription (\ref{cehat}) for $c_{\texttt{e}}(\bfX,t)$, as well as the coefficients (\ref{betas}), are analogous to those utilized by \cite{KKLP24} for purely elastic elastomers, the same type of formula derived for $\delta^\varepsilon$ by these authors applies here. We thus set
\begin{equation}
\delta^\varepsilon=\left(\dfrac{\sts(I_1)+(1+2\sqrt{3})\,\shs}{(8+3\sqrt{3})\,\shs}\right)\dfrac{3 G_c}{16\psi^{{\rm Eq}}_{\texttt{ts}}(I_1)\varepsilon}+\dfrac{2}{5}.
\label{delta-eps-final}
\end{equation}
Note that the coefficient (\ref{delta-eps-final}) depends on the deformation invariant $I_1$ via its dependence on the uniaxial tensile strength $\sts(I_1)$ and associated equilibrium free energy function $\psi^{{\rm Eq}}_{\texttt{ts}}(I_1)$. For notational simplicity, we do \emph{not} indicate this dependence by including explicitly an argument and instead continue to simply write $\delta^\varepsilon$.

\begin{remark}
\emph{In the sequel, we will make use of a FE discretization of the space dependence of the governing equations (\ref{BVP-y-theory})-(\ref{BVP-z-theory}). In such a type of discretization an error is incurred that scales with the mesh size, $\texttt{h}$ say. Similar to the correction that can be added to $G_c$ within the setting of the original phase-field fracture theory (see, e.g., Subsection 8.1.1 in \cite{Bourdin08}), here too it is possible to include a similar correction in the formula (\ref{delta-eps-final}) for $\delta^{\varepsilon}$ so that the resulting discretized equations are consistent with the actual value $G_c$ of the critical energy release rate of the elastomer \citep{KKLP24}. For first-order FEs of size $\texttt{h}$, the formula for $\delta^{\varepsilon}$ with the correction reads
\begin{equation*}
\delta^\varepsilon=\left(1+\dfrac{3}{8}\dfrac{\texttt{h}}{\varepsilon}\right)^{-2}\left(\dfrac{\sts(I_1)+(1+2\sqrt{3})\,\shs}{(8+3\sqrt{3})\,\shs}\right)\dfrac{3 G_c}{16\psi^{{\rm Eq}}_{\texttt{ts}}(I_1)\varepsilon}+\left(1+\dfrac{3}{8}\dfrac{\texttt{h}}{\varepsilon}\right)^{-1}\dfrac{2}{5}.
\end{equation*}
}
\end{remark}

\subsubsection{The resulting expression for $c_{\texttt{\emph{e}}}(\bfX,t)$}

We are now in a position to spell out the resulting final expression for the driving force $c_{\texttt{e}}(\bfX,t)$. Substitution of the coefficients (\ref{betas}) in expression (\ref{cehat}) yields
\begin{align}
c_{\texttt{e}}(\bfX,t)=\widehat{c}_{\texttt{e}}(\bfX,t)+(\delta^\varepsilon-1)\dfrac{3 G_c}{8 \varepsilon}+\dfrac{3}{4}{\rm Div}\left[\varepsilon (\delta^{\varepsilon}-1) G_c\nabla z \right]\label{ce-Final}
\end{align}
with
\begin{align}
\widehat{c}_{\texttt{e}}(\bfX,t)=&\left(-\dfrac{\sqrt{3}(3\shs-\sts(I_1))}{\shs\sts(I_1)}\delta^\varepsilon\dfrac{G_c}{8\varepsilon}-
\dfrac{2\psi^{{\rm Eq}}_{\texttt{hs}}}{\sqrt{3}\shs}+\dfrac{2\sqrt{3}\psi^{{\rm Eq}}_{\texttt{ts}}(I_1)}{\sts(I_1)}\right)\sqrt{\dfrac{\mathfrak{I}^2_1}{3}-\mathfrak{I}_2}+\left(-\dfrac{1}{\shs}\delta^\varepsilon\dfrac{G_c}{8\varepsilon}+\dfrac{2\psi^{{\rm Eq}}_{\texttt{hs}}}{3\shs}\right)  \mathfrak{I}_1+\nonumber\\
&z\left(1-\dfrac{\sqrt{\mathfrak{I}^2_1}}{\mathfrak{I}_1}\right)\psi^{{\rm Eq}}(I_1,J),\label{cehat-Final}
\end{align}
where the coefficient $\delta^\varepsilon$ is given by expression (\ref{delta-eps-final}), and where we recall that $\mathfrak{I}_1$ and $\mathfrak{I}_2$ are given by expressions (\ref{I1I2-S-Invariants}) in terms of the deformation field $\bfy(\bfX,t)$, the internal variable $\bfC^v(\bfX,t)$, and the phase field $z(\bfX,t)$. 

In terms of its dependence on the constitutive properties of the elastomer, in addition to its dependence on the regularization length $\varepsilon$, note that the constitutive prescription (\ref{ce-Final}), with (\ref{cehat-Final}) and (\ref{delta-eps-final}), depends on the equilibrium free energy function $\psi^{{\rm Eq}}(I_1,J)$, the non-equilibrium free energy function $\psi^{{\rm NEq}}(I^e_1,J)$, and the viscosity $\eta(I_1^e,I_2^e,I_1^v)$ characterizing the viscoelastic behavior of the elastomer, the material function $\sts(I_1)$ and material constant $\shs$ characterizing its strength, and on its fracture energy $G_c$. That is, the driving force $c_{\texttt{e}}(\bfX,t)$ depends on all three material properties: the viscoelasticity of the elastomer, its strength, and its fracture energy.

\subsection{The governing equations}

Making use of the constitutive prescription (\ref{ce-Final}), with (\ref{cehat-Final}) and (\ref{delta-eps-final}), for the driving force $c_{\texttt{e}}(\bfX,t)$ in the general form (\ref{BVP-y-theory})-(\ref{BVP-z-theory}) of the governing equations leads to the final form of the governing equations

\begin{equation}
\left\{\begin{array}{ll}
\hspace{-0.15cm} {\rm Div}\left[z^2\left(2\psi^{{\rm Eq}}_{I_1}\,\nabla\bfy+2\psi^{{\rm NEq}}_{I^e_1}\,\nabla\bfy \,{\bfC^v}^{-1}+\left(\psi^{{\rm Eq}}_{J}+\psi^{{\rm NEq}}_{J}\right)( \det\nabla\bfy)\,\nabla\bfy^{-T}\right)\right]={\bf0}, \quad(\bfX,t)\in \Omega_0\times[0,T]\\[12pt]
\hspace{-0.15cm} \det\nabla\bfy>0, \quad(\bfX,t)\in\Omega_0\times[0,T]\\[12pt]
\hspace{-0.15cm}\bfy(\bfX,t)=\overline{\bfy}(\bfX,t), \quad(\bfX,t)\in\partial  \Omega_0^{\mathcal{D}}\times[0,T]\\[12pt]
\hspace{-0.15cm} \left(z^2\left(2\psi^{{\rm Eq}}_{I_1}\,\nabla\bfy+2\psi^{{\rm NEq}}_{I^e_1}\,\nabla\bfy \,{\bfC^v}^{-1}+\left(\psi^{{\rm Eq}}_{J}+\psi^{{\rm NEq}}_{J}\right)( \det\nabla\bfy)\,\nabla\bfy^{-T}\right)\right)\bfN=\overline{\textbf{s}}(\bfX,t), \quad(\bfX,t)\in\partial \Omega_0^{\mathcal{N}}\times[0,T]\\[12pt]
\hspace{-0.15cm} \bfy(\bfX,0)=\bfX, \quad \bfX\in\Omega_0\end{array}\right. , \label{BVP-y-theory-reg}
\end{equation}
\begin{equation}
\hspace{-1.3cm}\left\{\begin{array}{l}\hspace{-0.15cm}\dot{\bfC}^v(\bfX,t)=\dfrac{2\psi^{{\rm NEq}}_{I^e_1}}{\eta(I_1^e,I_2^e,I_1^v)}\left(\nabla\bfy^T\nabla\bfy-\dfrac{1}{3}\left(\nabla\bfy^T\nabla\bfy\cdot{\bfC^v}^{-1}\right)\bfC^v\right), \quad(\bfX,t)\in\Omega_0\times[0,T] \\[12pt]
\hspace{-0.15cm}\bfC^v(\bfX,0)=\bfI, \quad \bfX\in\Omega_0\end{array}\right., \label{BVP-Cv-theory-reg}
\end{equation}
and
\begin{equation}
\left\{\begin{array}{l}\hspace{-0.15cm}{\rm Div}\left[\varepsilon\,\delta^\varepsilon G_c \nabla z\right]=\dfrac{8}{3}z\, \psi^{{\rm Eq}}(I_1,J)-\dfrac{4}{3}\widehat{c}_\texttt{e}(\bfX,t)-\dfrac{\delta^\varepsilon G_c}{2\varepsilon},
\quad\mbox{ if } \dot{z}(\bfX,t)< 0, \quad(\bfX,t)\in \Omega_0\times[0,T]
\\[12pt]
\hspace{-0.15cm}
{\rm Div}\left[\varepsilon\, \delta^\varepsilon G_c \nabla z\right]\ge \dfrac{8}{3}z\, \psi^{{\rm Eq}}(I_1,J)-\dfrac{4}{3}\widehat{c}_\texttt{e}(\bfX,t)-\dfrac{\delta^\varepsilon G_c}{2\varepsilon},
\quad\mbox{ if }  z(\bfX,t)=1 \mbox{ or } \dot{z}(\bfX,t)>0, \quad(\bfX,t)\in \Omega_0\times [0,T]\\[12pt]
\hspace{-0.15cm}
{\rm Div}\left[\varepsilon\, \delta^\varepsilon G_c \nabla z\right]\le \dfrac{8}{3}z\, \psi^{{\rm Eq}}(I_1,J)-\dfrac{4}{3}\widehat{c}_\texttt{e}(\bfX,t)-\dfrac{\delta^\varepsilon G_c}{2\varepsilon},
\quad\mbox{ if }  z(\bfX,t)=0, \quad(\bfX,t)\in \Omega_0\times [0,T]\\[12pt]
\hspace{-0.15cm}\nabla z\cdot\bfN=0, \quad (\bfX,t)\in \partial\Omega_0\times [0,T]\\[12pt]
\hspace{-0.15cm} z(\bfX,0)=1,\quad \bfX \in \Omega_0
\end{array}\right.  \label{BVP-z-theory-reg}
\end{equation}
for the deformation field $\bfy(\bfX,t)$, the internal variable $\bfC^v(\bfX,t)$, and the phase field $z(\bfX,t)$.

\section{Numerical implementation}\label{Sec: Numerical Implementation}

In this section, we introduce a numerical scheme to solve the initial-boundary-value problem (\ref{BVP-y-theory-reg})-(\ref{BVP-z-theory-reg}). There are three main challenges in the construction of any such a scheme. The first one is the selection of an appropriate space discretization that is capable of dealing with large deformations and the near incompressibility typical of elastomers. 
The fact that the internal variable $\bfC^v$ defined by the ODE (\ref{BVP-Cv-theory-reg}) satisfies the nonconvex constraint $\det\bfC^v=1$ poses the second main challenge. Indeed, commonly used time integration schemes are known to be unable to deliver solutions that satisfy such a constraint. Finally, the third main challenge is the selection of an appropriate solver for the nonlinear algebraic equations resulting from the space and time discretizations that is capable of dealing with the large rapid changes in the deformation field $\bfy(\bfX,t)$ that can ensue from the nucleation of fracture. To address these challenges, we make use of a space discretization based on Crouzeix-Raviart non-conforming FEs of low order, an explicit high-order accurate Runge-Kutta FD discretization of time, and a nested staggered method of solution that uses an implicit gradient flow solver. 

A few words about the organization of this section are in order. We begin in Subsection \ref{Sec: Weak form} by recasting the PDEs (\ref{BVP-y-theory-reg}) and (\ref{BVP-z-theory-reg})
into a weak form featuring admissible sets for the deformation field $\bfy(\bfX, t)$ and the phase field $z(\bfX, t)$ that are free of the inequality constraints $\det\nabla\bfy(\bfX,t)>0$, $0\leq z(\bfX,t)\leq 1$, and $\dot{z}(\bfX,t)\leq 0$. In Subsection \ref{Sec: Time discretization}, we introduce a partition of the time interval under consideration $[0, T]$ into discrete times and spell out the corresponding time discretization of the governing equations in weak form. In Subsection \ref{Sec: Space discretization}, we further discretize in space the obtained set of time-discretized equations by means of a FE approximation. The last subsection presents the solver utilized to generate numerical solutions for the nonlinear algebraic equations resulting from the discretizations.

\subsection{Weak form of the governing equations}\label{Sec: Weak form}

The inequalities $\det\nabla\bfy(\bfX,t)>0$, $0\leq z(\bfX,t)\leq 1$, and $\dot{z}(\bfX,t)\leq 0$ are difficult to enforce \emph{a priori}. Among other possible approaches, we choose to address this difficulty here by adopting a standard procedure followed in finite elasticity, namely, by including penalizing terms directly in the equations. Specifically, we restrict attention to  equilibrium free energy functions $\psi^{{\rm Eq}}(I_1,J)$ with the asymptotic behavior
\begin{align}
\psi^{{\rm Eq}}(I_1,J)\to\infty\quad\mathrm{as}\quad J\rightarrow 0+,\label{W-condition}
\end{align}
such as the example (\ref{PsiLP-vis})$_1$, and recast the PDEs (\ref{BVP-z-theory-reg})$_{1-3}$ in the form 
\begin{align}
{\rm Div}\left[\varepsilon\,\delta^\varepsilon G_c \nabla z\right]=\dfrac{8}{3}z\, \psi^{{\rm Eq}}(I_1,J)-\dfrac{4}{3}\widehat{c}_\texttt{e}(\bfX,t)-\dfrac{\delta^\varepsilon G_c}{2\varepsilon}+\dfrac{8}{3\zeta}\mathfrak{p}(z,\dot{z}),
\quad(\bfX,t)\in \Omega_0\times[0,T]\label{BVP-z-theory-reg-penalty}
\end{align}
with 
\begin{align}
\mathfrak{p}(z,\dot{z})=|1-z|-(1-z)-|z|+z+\mathcal{H}(z_\alpha-z)\left(|\dot{z}|-\dot{z}\right). \label{z-penaltyfunction}
\end{align}
In these expressions, $\zeta$ is a penalty parameter that should be selected to be small relative to the term $2\varepsilon/(\delta^\varepsilon G_c)$, $\mathcal{H}(\cdot)$ stands for the Heaviside function, and $z_\alpha$ denotes the value below which the phase field $z$ is considered to be non-increasing in time. All the results that are presented below correspond to $z_\alpha=0.05$ and $\zeta^{-1}=10^4 \delta^\varepsilon G_c/(2\varepsilon)$. Note that while the property (\ref{W-condition}) penalizes the violation of the inequality $\det\nabla\mathbf{y}(\mathbf{X},t)>0$, the penalty function (\ref{z-penaltyfunction}) in (\ref{BVP-z-theory-reg-penalty}) enforces compliance with the inequalities $0\leq z(\bfX,t)\leq 1$ and $\dot{z}(\bfX,t)\leq 0$ for $z\leq z_\alpha$.

Granted the above reformulation, upon the definition of the admissible sets
\begin{align*}
\mathcal{Y}=\big\{\mathbf{y}\in W^{1,p}(\Omega_{0};\mathbb{R}^{3}):\mathbf{y}=\overline{\mathbf{y}}\mbox{ on } \partial\Omega_0^\mathcal{D}\big\}\quad {\rm and}\quad \mathcal{Y}_{0}=\big\{\mathbf{y}\in W^{1,p}(\Omega_{0};\mathbb{R}^{3}):\mathbf{y}=\mathbf{0}\mbox{ on } \partial\Omega_0^\mathcal{D}\big\},
\end{align*}
the initial-boundary-value problem (\ref{BVP-y-theory-reg})-(\ref{BVP-z-theory-reg}) can be recast as the problem of finding the deformation field $\mathbf{y} \in \mathcal{Y}$ and phase field $z \in W^{1,2}(\Omega_0)$ such that
\begin{align}
\begin{aligned}
&\int_{\Omega_0}\left(\left(z^2+\xi^\varepsilon_I\right)\left(2\psi^{{\rm Eq}}_{I_1}\,\nabla\bfy+2\psi^{{\rm NEq}}_{I^e_1}\,\nabla\bfy \,{\bfC^v}^{-1}\right)+\left(z^2+\xi^\varepsilon_J\right)\left(\psi^{{\rm Eq}}_{J}+\psi^{{\rm NEq}}_{J}\right)( \det\nabla\bfy)\,\nabla\bfy^{-T}\right)\cdot\nabla\mathbf{u}\,  \mathrm{d}\bfX -\\
&\int_{\partial\Omega_0^\mathcal{N}}\overline{\textbf{s}}\cdot\mathbf{u} \, \mathrm{d}\bfX=0\quad\forall\mathbf{u}\in\mathcal{Y}_0, t\in[0,T],
\end{aligned}\label{Weakform_u}
\end{align}
with $\bfC^v(\bfX,t)$ defined by
\begin{equation}
\left\{\begin{array}{l}\hspace{-0.15cm}\dot{\bfC}^v(\bfX,t)=\mathbf{G}\left(\nabla\bfy,\bfC^v\right)\\[12pt]
\hspace{1.3cm}\equiv\dfrac{2\psi^{{\rm NEq}}_{I^e_1}}{\eta(I_1^e,I_2^e,I_1^v)}\left(\nabla\bfy^T\nabla\bfy-\dfrac{1}{3}\left(\nabla\bfy^T\nabla\bfy\cdot{\bfC^v}^{-1}\right)\bfC^v\right), \quad(\bfX,t)\in\Omega_0\times[0,T] \\[12pt]
\hspace{-0.15cm}\bfC^v(\bfX,0)=\bfI, \quad \bfX\in\Omega_0\end{array}\right., \label{BVP-Cv-theory-reg2}
\end{equation}
and
\begin{align}
\begin{aligned}
&\int_{\Omega_0}\left(\varepsilon\delta^\varepsilon G_c \nabla z\cdot\nabla v+\left(\frac{8}{3}z \psi^{\text{Eq}}(I_1,J)-\frac{4}{3}\widehat{c}_\texttt{e}-\frac{\delta^{\varepsilon}G_c}{2\varepsilon} + \dfrac{8}{3\zeta}\mathfrak{p}(z,\dot{z})\right)v \right)\mathrm{d}\bfX=0\quad\forall v \in W^{1,2}(\Omega_0), t\in[0,T].
\end{aligned}\label{Weakform_z}
\end{align}
Here, we note that the deformation field $\bfy(\bfX,t)$ is expected to belong to some Sobolev space $W^{1,p}(\Omega_0;\mathbb{R}^{3})$,  $\xi^\varepsilon_I\geq \xi^\varepsilon_J>0$ are small-value positive constants of $o(\varepsilon)$ introduced to prevent the numerical instabilities associated with vanishingly small stiffness, and, for later convenience, we have introduced the function $\mathbf{G}$ to denote the right-hand side of the evolution equation (\ref{BVP-Cv-theory-reg2}).

\subsection{Time discretization} \label{Sec: Time discretization}

Consider now a partition of the time interval $\left[0,T\right]$ into discrete times $\{t_{k}\}_{k=0,1,\ldots,M}$ with $t_{0} = 0$ and $t_{M} = T$. With help of the notation $\bfy_{k}(\bfX)=\bfy(\bfX,t_{k})$, $\nabla\bfy_{k}(\bfX)=\nabla\bfy(\bfX,t_{k})$, $\bfC_{k}^{v}(\bfX)=\bfC^{v}(\bfX,t_{k})$, $\dot{\bfC}_{k}^{v}(\bfX)=\dot{\bfC}^{v}(\bfX,t_{k})$, $z_{k}(\bfX)=z(\bfX,t_{k})$, $\nabla z_{k}(\bfX)=\nabla z(\bfX,t_{k})$, $\overline{\bfy}_{k}(\bfX)=\overline{\bfy}(\bfX,t_{k})$, $\overline{\textbf{s}}_{k}(\bfX)=\overline{\textbf{s}}(\bfX,t_{k})$, $\psi^{{\rm Eq}}_{I_{1_k}}=\psi^{{\rm Eq}}_{I_{1}}|_{t=t_k}$, $\psi^{{\rm Eq}}_{J_{k}}=\psi^{{\rm Eq}}_{J}|_{t=t_k}$, $\psi^{{\rm NEq}}_{I^e_{1_k}}=\psi^{{\rm NEq}}_{I^e_{1}}|_{t=t_k}$, $\psi^{{\rm NEq}}_{J_{k}}=\psi^{{\rm NEq}}_{J}|_{t=t_k}$, $\widehat{c}_{\texttt{e}_{k}}=\widehat{c}_{\texttt{e}}(\bfX,t_k)$, $\delta_{k}^\varepsilon=\delta^\varepsilon|_{t=t_k}$, the equations (\ref{Weakform_u})-(\ref{Weakform_z}) at any given discrete time $t_{k}$ can be written in the form
\begin{align}
\begin{aligned}
&\int_{\Omega_0}\left(\left(z_k^2+\xi^\varepsilon_I\right)\left(2\psi^{{\rm Eq}}_{I_{1_k}}\,\nabla\bfy_{k}+2\psi^{{\rm NEq}}_{I^e_{1_k}}\,\nabla\bfy_{k} \,{\bfC_{k}^v}^{-1}\right)+\left(z_k^2+\xi^\varepsilon_J\right)\left(\psi^{{\rm Eq}}_{J_{k}}+\psi^{{\rm NEq}}_{J_{k}}\right)( \det\nabla\bfy_{k})\,\nabla\bfy_{k}^{-T}\right)\cdot\nabla\mathbf{u}\,  \mathrm{d}\bfX -\\
&\int_{\partial\Omega_0^\mathcal{N}}\overline{\textbf{s}}_{k}\cdot\mathbf{u} \, \mathrm{d}\bfX=0\quad\forall\mathbf{u}\in\mathcal{Y}_0,
\label{Weakform_u_discrete}
\end{aligned}
\end{align}
\begin{equation}
\dot{\bfC}_{k}^v(\bfX)=\mathbf{G}\left(\nabla\bfy_k(\bfX),\bfC^v_k(\bfX)\right),\quad \bfX\in\Omega_0, \label{BVP-Cv-theory-discrete}
\end{equation}
and
\begin{align}
\begin{aligned}
&\int_{\Omega_0}\left(\varepsilon\delta_{k}^\varepsilon G_c \nabla z_{k}\cdot\nabla v+\right.\\
&\left.\left(\frac{8}{3}z_{k} \psi^{\text{Eq}}(\nabla\bfy_k\cdot\nabla\bfy_k,\det\nabla\bfy_k)-\frac{4}{3}\widehat{c}_{\texttt{e}_{k}}-\frac{\delta_{k}^{\varepsilon}G_c}{2\varepsilon} + \dfrac{8}{3\zeta}\mathfrak{p}(z_k,z_{k}-z_{k-1})\right)v \right)\,\mathrm{d}\bfX=0 \quad\forall v \in W^{1,2}(\Omega_0),
\label{Weakform_z_discrete}
\end{aligned}
\end{align}
where we emphasize that we are yet to choose a time discretization (explicit or implicit) for $\dot{\bfC}_{k}^v(\bfX)$ in terms of $\bfC^v(\bfX,t)$.

\subsection{Space discretization}\label{Sec: Space discretization}

Next, we further discretize in space the time-discretized equations (\ref{Weakform_u_discrete})-(\ref{Weakform_z_discrete}). To this end, consider a partition $^{\texttt{h}}\Omega_0=\bigcup_{e=1}^{\texttt{N}_e}\mathcal{E}^{(e)}$ with $\mathcal{E}^{(i)}\cap\mathcal{E}^{(j)}=\varnothing\ \forall i\ne j$ of the domain $\Omega_0$ that comprises $\texttt{N}_e$ non-overlapping simplicial elements $\mathcal{E}^{(e)}$; here, $\texttt{h}$ stands for the diameter of the largest element. Specifically, consider the case where each element $e$ possesses $4$ nodes with coordinates $\bfX^{(e,l)}$, $l=1,..., 4$, located at the centroids of its $2$-dimensional faces. In terms of these nodal coordinates, with help of the reference simplicial element $\mathcal{T}=\left\{\bfrho: 0\le \rho_1,...,\rho_3 \le 1,\sum_{l=1}^{3}\rho_l\le 1\right\}$, the domain occupied by each element $e$ is defined parametrically by $\mathcal{E}^{(e)}=\left\{\bfX : \sum_{l=1}^{4} N^{(l)}_{CR}(\bfrho)\bfX^{(e,l)},\bfrho\in\mathcal{T}\right\}$. In this last expression,  $N^{(l)}_{CR}(\bfrho)$ stand for the Crouzeix-Raviart linear shape functions 
\begin{align*}
\left\{\begin{array}{l} N_{CR}^{(1)}(\rho_{1},\rho_{2},\rho_{3})=1-3\rho_{1}\\ [8pt] N_{CR}^{(2)}(\rho_{1},\rho_{2},\rho_{3})=1-3\rho_{2}\\ [8pt] N_{CR}^{(3)}(\rho_{1},\rho_{2},\rho_{3})=1-3\rho_{3}\\ [8pt]
N_{CR}^{(4)}(\rho_{1},\rho_{2},\rho_{3})=3(\rho_{1}+\rho_{2}+\rho_{3})-2\end{array}\right. . 
\end{align*}
Additionally, consider the alternative case where each element $e$ possesses $4$ nodes with coordinates $\widehat{\bfX}^{{(e,l)}}$, $l=1,..., 4$, located at its vertices. In terms of these nodal coordinates, the domain occupied by each element $e$ is defined parametrically by $\mathcal{E}^{(e)}=\left\{\bfX : \sum_{l=1}^{4} N^{(l)}_{L}(\bfrho)\widehat{\bfX}^{(e,l)},\bfrho\in \mathcal{T}\right\}$, where $N^{(l)}_{L}(\bfrho)$ denote the standard Lagrangian linear shape functions

\begin{align*}
\left\{\begin{array}{l} N_{L}^{(1)}(\rho_{1},\rho_{2},\rho_{3})=\rho_{1}\\ [8pt] N_{L}^{(2)}(\rho_{1},\rho_{2},\rho_{3})=\rho_{2}\\ [8pt] N_{L}^{(3)}(\rho_{1},\rho_{2},\rho_{3})=\rho_{3}\\ [8pt]
N_{L}^{(4)}(\rho_{1},\rho_{2},\rho_{3})=1-\rho_{1}-\rho_{2}-\rho_{3}.\end{array}\right. . 
\end{align*}

Given the above-described simplicial discretizations, we look for approximate solutions ${}^{\texttt{h}}\bfy_k(\bfX)$ and ${}^{\texttt{h}}z_k(\bfX)$ of the deformation field $\bfy_k(\bfX)$ and the phase field $z_k(\bfX)$ at time $t_k$ in the non-conforming Crouzeix-Raviart and conforming Lagrangian FE spaces 
\begin{align}
\begin{aligned}{}^{\texttt{h}}\mathcal{Y}=\{{}^{\texttt{h}}\bfy_k\colon ^hy^i_k(\bfX)|_{\mathcal{E}^{(e)}}&=\sum_{l=1}^{4}N_{CR}^{(l)}(\boldsymbol{\rho})y^i_{k(e,l)},\forall e=1,\ldots,\texttt{N}_e,{}^{\texttt{h}}\bfy_k\text{ is continuous on }\mathcal{C},\\{}^{\texttt{h}}\bfy_k(\bfX^{(j)})&=
\overline{\bfy}_k(\bfX^{(j)})\quad\forall\bfX^{(j)}\in\mathcal{C}\cap\partial\Omega_0^\mathcal{D}\}\end{aligned}\label{FEspace_formulation_u}
\end{align}
and
\begin{align}
{}^{\texttt{h}}\mathcal{Z}=\{^hz_k\colon ^hz_k(\bfX)|_{\mathcal{E}^{(e)}}=\sum_{l=1}^{4}N_{L}^{(l)}(\boldsymbol{\rho})z_{k(e,l)},\forall e=1,\ldots,\texttt{N}_e,^{\texttt{h}}z_k\text{ is continuous on }\mathcal{V}\},\label{FEspace_formulation_z}
\end{align}
respectively. Here, $\mathcal{C}$ and $\mathcal{V}$ denote the sets of centroids of the $2$-dimensional faces and vertices of all the elements in ${}^{\texttt{h}}\Omega_0$, while $^hy^i_{k(e,l)}$ and $^hz_{k(e,l)}$ denote the component $i$ of the deformation field and the phase field at node $l$ of element $e$ at time $t_k$.

\begin{remark}
{\rm The class of non-conforming FE space (\ref{FEspace_formulation_u}) were originally introduced by \cite{CR73} in the context of Stokes flow of incompressible viscous fluids under Dirichlet boundary conditions. Since then, they have been successfully utilized for a variety of other types of (linear and nonlinear) PDEs and boundary conditions; see, e.g., \cite{Falk91,Burman05,Boffi-Brezzi13,Henao16,KFLP18}. From a practical point of view, the alluring feature of non-conforming FE discretizations is that they may lead --- sometimes, however, with the compulsory addition of stabilization terms \citep{Falk91} --- to consistent and stable formulations with significantly fewer degrees of freedom than competing conforming dicretizations. This is indeed the case for the problem of interest here, where despite the low-order linear approximation of the FE space (\ref{FEspace_formulation_u}), with help of a suitable stabilization term to be described below, the resulting formulation can be shown (via numerical tests) to be stable irrespectively of the compressibility of the elastomer of interest.
}
\end{remark}

Standard assembly procedures permit to construct global shape functions ${^{\texttt{h}}N^{(n)}_{CR}}(\bfX)$, $n=1,...,\texttt{N}_n$, and ${^{\texttt{h}}N^{(n)}_{L}}(\bfX)$, $n=1,...,\widehat{\texttt{N}}_n$, so that the deformation trial field ${^{\texttt{h}}\bfy_k}(\bfX)$ and the phase trial field ${^{\texttt{h}}z_k}(\bfX)$ in (\ref{FEspace_formulation_u}) and (\ref{FEspace_formulation_z}) can be written in the global form
\begin{equation}
{^{\texttt{h}}y^i_k}(\bfX)=\sum_{n=1}^{\texttt{N}_n}{^{\texttt{h}}N^{(n)}_{CR}}(\bfX)y_{k (n)}^i\qquad {\rm and}\qquad {^{\texttt{h}}z_k}(\bfX)=\sum_{n=1}^{\widehat{\texttt{N}}_n}{^{\texttt{h}}N^{(n)}_{L}}(\bfX)z_{k (n)},\label{trial-fields}
\end{equation}
where $\texttt{N}_n$ and $\widehat{\texttt{N}}_n$ stand, respectively, for the total number of centroid and vertex nodes in $^{\texttt{h}}\Omega_0$, while $y_{k (n)}^i$ and $z_{k (n)}$ correspond physically to the component $i$ of the deformation field ${^{\texttt{h}}\bfy_k}(\bfX)$ and the phase field ${^{\texttt{h}}z_k}(\bfX)$ at node $n$ and time $t_k$. We emphasize that the representation (\ref{trial-fields})$_1$ applies \emph{not} to all $\bfX\in {^{\texttt{h}}\Omega_0}$ but to $\bfX\in ({^{\texttt{h}}\Omega_0}\setminus\mathcal{I})\cup \mathcal{C}$, where  $\mathcal{I}$ stands for the set of all internal $2$-dimensional faces in ${^{\texttt{h}}\Omega_0}$, given that the deformation trial fields in the non-conforming FE space (\ref{FEspace_formulation_u}) are not necessarily bijective, and, in particular, can have discontinuities at $\bfX\in \mathcal{I}\setminus \mathcal{C}$. Indeed, for a given internal face $I\in \mathcal{I}$, denoting the two elements sharing that face by the `left (L)' and the `right (R)' elements and their respective gather matrices\footnote{The definition of the gather matrices $G_{ln}^{(e)}$ ($e=1,...,\texttt{N}_e$) adopted here is such that $y^i_{k(e,l)}=\sum_{n=1}^{\texttt{N}_n}G_{ln}^{(e)}y_{k(n)}^i$.} by $G^{(L)}_{ln}$ and $G^{(R)}_{ln}$, $l=1,...,4$ and $n=1,...,\texttt{N}_n$, the deformation trial field at $\bfX\in I$ is given by the two possibly distinct representations
\begin{align}
{^{\texttt{h}}y_k^i}(\bfX) =\left\{ \begin{array}{l}
                   \sum_{l=1}^{4}\sum_{n=1}^{\texttt{N}_n}N^{(l)}_{CR}(\bfrho)G^{(L)}_{ln}y^i_{k(n)} \\
                   \\
                   \sum_{l=1}^{4}\sum_{n=1}^{\texttt{N}_n}N^{(l)}_{CR}(\bfrho)G^{(R)}_{ln}y^i_{k(n)}
                    \end{array}\right.
            \label{y-jump}
\end{align}
in terms of the global degrees of freedom $y^i_{k(n)}$. Similarly, the corresponding approximations ${^{\texttt{h}}\bfu}$ and ${^{\texttt{h}}v}$ of the test functions $\bfu$ and $v$ admit the global representation
\begin{equation}
{^{\texttt{h}}u^i}(\bfX)=\left\{\hspace{-0.1cm}\begin{array}{l}
\sum_{n=1}^{\texttt{N}_n}{^{\texttt{h}}N^{(n)}_{CR}}(\bfX)u_{(n)}^i,\quad \bfX\in ({^{\texttt{h}}\Omega_0}\setminus\mathcal{I})\cup \mathcal{C}\vspace{0.2cm}\\
\left\{ \hspace{-0.1cm}\begin{array}{l}
\sum_{l=1}^{4}\sum_{n=1}^{\texttt{N}_n}N^{(l)}_{CR}(\bfrho)G^{(L)}_{ln}u^i_{(n)} \vspace{0.2cm}\\
\sum_{l=1}^{4}\sum_{n=1}^{\texttt{N}_n}N^{(l)}_{CR}(\bfrho)G^{(R)}_{ln}u^i_{(n)}
\end{array}\right.,\; \bfX\in I \end{array}\right.\quad {\rm and}\quad {^{\texttt{h}}v}(\bfX)=\sum_n^{\widehat{\texttt{N}}_n}{^{\texttt{h}}N^{(n)}_{L}}(\bfX)v_{(n)},\; \bfX\in {^{\texttt{h}}\Omega_0}\label{test-fields}
\end{equation}

Direct substitution of the representations (\ref{trial-fields})-(\ref{test-fields}) in the governing  equations (\ref{Weakform_u_discrete})-(\ref{Weakform_z_discrete}) leads to a system of nonlinear algebraic equations for the global degrees of freedom $y_{k (n)}^i$ and $z_{k (n)}$ that depend on the values, say ${}^{\texttt{h}}\bfC_{k}^v$, of the internal variable $\bfC^v_k(\bfX)$ at the Gaussian quadrature points employed to carry out the integrals in (\ref{Weakform_u_discrete}) and (\ref{Weakform_z_discrete}). Precisely, substitution of the representations (\ref{trial-fields})-(\ref{test-fields}) in the governing  equations (\ref{Weakform_u_discrete})-(\ref{Weakform_z_discrete}) leads to the system of nonlinear algebraic equations
\begin{align}
\begin{aligned}
\mathcal{G}_1({^{\texttt{h}}\bfy_{k}},{}^{\texttt{h}}\bfC_{k}^v,{^{\texttt{h}}z_{k}})\equiv &\int_{\Omega_0}\left(\left({}^hz_{k}^2+\xi^\varepsilon_I\right)\left(2\psi^{{\rm Eq}}_{I_{1_k}}\,\nabla{}^{\texttt{h}}\bfy_{k}+2\psi^{{\rm NEq}}_{I^e_{1_k}}\,\nabla{}^{\texttt{h}}\bfy_{k} \,{{}^{\texttt{h}}\bfC_{k}^v}^{-1}\right)+\left({}^hz_{k}^2+\xi^\varepsilon_J\right)\left(\psi^{{\rm Eq}}_{J_{k}}+\psi^{{\rm NEq}}_{J_{k}}\right)\times\right.\\
&\left.\left.\,(\det\nabla{}^{\texttt{h}}\bfy_{k})\nabla{}^{\texttt{h}}\bfy_{k}^{-T}\right)\right)\cdot\nabla{}^{\texttt{h}}\mathbf{u}\,  \mathrm{d}\bfX -
\int_{\partial\Omega_0^\mathcal{N}}\overline{\textbf{s}}_{k}\cdot{}^{\texttt{h}}\mathbf{u}\, \mathrm{d}\bfX+\sum_{I\in\mathcal{I}}\dfrac{q}{\texttt{h}_{I}}\int_{I}\llbracket{^{\texttt{h}}\bfy_{k}}\rrbracket\cdot \llbracket{^{\texttt{h}}\bfu}\rrbracket \,{\rm d} \bfX  =0\quad\forall u^i_{(n)},
\label{G1_u_equation}
\end{aligned}
\end{align}
\begin{align}
\begin{aligned}
\mathcal{G}_2({^{\texttt{h}}\bfy_{k}},{}^{\texttt{h}}\bfC_{k}^v,{}^{\texttt{h}}\dot{\bfC}_{k}^v)\equiv {{}^{\texttt{h}}\dot{\bfC}}_{k}^v-\mathbf{G}\left(\nabla{}^{\texttt{h}}\bfy_k(\bfX),{{}^{\texttt{h}}\bfC}^v_k(\bfX)\right)={\bf0} \quad\forall \bfX\in \mathcal{N}_{Gauss},
\label{G2_Cv_equation}
\end{aligned}
\end{align}
with $\mathcal{N}_{Gauss}$ denoting the set of all Gaussian quadrature points, and
\begin{align}
\begin{aligned}
\mathcal{G}_3({^{\texttt{h}}\bfy_{k}},{^{\texttt{h}}z_{k}})\equiv &\int_{\Omega_0}\left(\varepsilon\delta_{k}^\varepsilon G_c \nabla ^{\texttt{h}}z_{k}\cdot\nabla ^{\texttt{h}}v+
\left(\frac{8}{3} {^{\texttt{h}}z_{k}} \psi^{\text{Eq}}(\nabla^{\texttt{h}}\bfy_k\cdot\nabla^{\texttt{h}}\bfy_k,\det\nabla^{\texttt{h}}\bfy_k)-\frac{4}{3}\widehat{c}_{\texttt{e}_{k}}-\frac{\delta_{k}^{\varepsilon}G_c}{2\varepsilon} +\right.\right.\\
& \left.\left.\dfrac{8}{3\zeta}\mathfrak{p}({^{\texttt{h}}z}_k,{^{\texttt{h}}z_{k}}-{^{\texttt{h}}z}_{k-1})\right) {^{\texttt{h}}v} \right)\,\mathrm{d}\bfX=0 \quad\forall v_{(n)},
\label{G3_z_equation}
\end{aligned}
\end{align}
for $\bfy_{k (n)}$, ${}^{\texttt{h}}\bfC_{k}^v$, and $z_{k (n)}$.

\begin{remark}
{\rm The addition of the stabilization term 
\begin{align}
\sum_{I\in\mathcal{I}}\dfrac{q}{\texttt{h}_{I}}\int_{I}\llbracket{^{\texttt{h}}\bfy_{k}}\rrbracket\cdot \llbracket{^{\texttt{h}}\bfu}\rrbracket \,{\rm d} \bfX \label{stability-term}
\end{align}
to equation (\ref{G1_u_equation}) is essential to avoid spurious rigid body rotations when dealing with pure traction or mixed boundary conditions; see \cite{Falk91} for a thorough study of this issue in the simpler context of linear elasticity. The specific stabilization term (\ref{stability-term}) that we employ here --- which penalizes the jump of the deformation field across element faces and where ${\texttt{h}_{I}}$ stands for the diameter of the internal face $I$ in the set $\mathcal{I}$, $\llbracket \cdot\rrbracket$ denotes the jump of a given quantity across the corresponding internal face, and $q$ is a stabilization parameter (with unit \emph{force/length}$^{N-1}$) whose selection is described below --- was chosen based on its performance over a wide range of numerical tests involving various specimen geometries and boundary conditions.
}
\end{remark}

\subsection{The solver}\label{Sec: Solver}

Having discretized the governing equations (\ref{BVP-y-theory-reg})-(\ref{BVP-z-theory-reg}) into the system of coupled nonlinear algebraic equations (\ref{G1_u_equation})-(\ref{G3_z_equation}), the final step is to solve these for the global degrees of freedom $\bfy_{k (n)}$, the internal variables ${}^{\texttt{h}}\bfC_{k}^v$ at the Gaussian quadrature points, and the global degrees of freedom $z_{k (n)}$  at time $t_k$. We do so by following a nested staggered scheme, which involves solving  equations (\ref{G1_u_equation}) and (\ref{G2_Cv_equation}) first iteratively between them and then iteratively with (\ref{G3_z_equation}) at every time step $t_k$ until convergence is reached. Specifically, the pseudo algorithm to solve (\ref{G1_u_equation})-(\ref{G3_z_equation}) for $\mathbf{y}_{k(n)}$, ${}^{\texttt{h}}\bfC_{k}^v$ and $z_{k(n)}$ at $t_{k}$ is as follows:

\begin{itemize}
	
\item{\emph{Step 0.} Set $r=1$ and $s=1$ and define tolerances $TOL_1,TOL_2,TOL_3>0$ and maximum number of iterations $\texttt{I}_1$ and $\texttt{I}_2$. For a given solution  ${}^{\texttt{h}}\mathbf{y}_{k-1}$, ${}^{\texttt{h}}\bfC_{k-1}^v$, and ${}^{\texttt{h}}z_{k-1}$ at $t_{k-1}$, define also ${}^{\texttt{h}}\mathbf{y}^{0,0}_{k}={}^{\texttt{h}}\mathbf{y}_{k-1}$, ${}^{\texttt{h}}\bfC^{v^{0,0}}_{k}={}^{\texttt{h}}\bfC_{k-1}^v$, and $ {}^{\texttt{h}}z^{0}_{k}={}^{\texttt{h}}z_{k-1}$.
}
	
\item{\emph{Step 1.} Given the increments $\overline{\bfy}_k(\bfX)$ and $\overline{\mathbf{s}}_k(\bfX)$ in boundary data, find ${}^{\texttt{h}}\mathbf{y}^{r,s-1}_{k}\in{}^{\texttt{h}}\mathcal{Y}$ such that
\begin{align}
\mathcal{G}_1\left({}^{\texttt{h}}\mathbf{y}^{r,s-1}_{k},{}^{\texttt{h}}\bfC_{k}^{v^{r-1,s-1}},{}^{\texttt{h}}z_{k}^{s-1}\right)=0. \label{subproblem1}
\end{align}
}
	
\item{\emph{Step 2.} Having solved sub-problem (\ref{subproblem1}) for ${}^{\texttt{h}}\mathbf{y}^{r,s-1}_{k}$, find ${}^{\texttt{h}}\bfC_{k}^{v^{r,s-1}}$ such that
\begin{align}
\mathcal{G}_2\left({}^{\texttt{h}}\mathbf{y}^{r,s-1}_{k},{}^{\texttt{h}}\bfC_{k}^{v^{r,s-1}}\right)={\bf 0}. \label{subproblem2}
\end{align}
}
    
\item{\emph{Step 3.} If $\parallel\mathcal{G}_1({}^{\texttt{h}}\mathbf{y}_{k}^{r,s-1},{}^{\texttt{h}}\bfC_{k}^{v^{r,s-1}},{}^{\texttt{h}}z_{k}^{s-1})\parallel/\parallel\mathcal{G}_1({}^{\texttt{h}}\mathbf{y}_{k}^{0,s-1},{}^{\texttt{h}}\bfC_{k}^{v^{0,s-1}},^hz_{k}^{s-1})\parallel\leq TOL_1$ and $\parallel\mathcal{G}_2({}^{\texttt{h}}\mathbf{y}_{k}^{r,s-1},$ ${}^{\texttt{h}}\bfC_{k}^{v^{r,s-1}})\parallel/\parallel\mathcal{G}_2({}^{\texttt{h}}\mathbf{y}_{k}^{0,s-1},{}^{\texttt{h}}\bfC_{k}^{v^{0,s-1}})\parallel\leq TOL_2$ or $r>\texttt{I}_1$, then set $ {}^{\texttt{h}}\mathbf{y}_{k}^{0,s} = {}^{\texttt{h}}\mathbf{y}_{k}^{r,s-1}$, ${}^{\texttt{h}}\bfC_{k}^{v^{0,s}} = {}^{\texttt{h}}\bfC_{k}^{v^{r,s-1}}$, and $r=0$ and move to the next step; otherwise set $r\leftarrow r+1$ and go to Step 1.
}
    
\item{\emph{Step 4.} Having solved sub-problems (\ref{subproblem1}) and (\ref{subproblem2}), find ${}^{\texttt{h}}z^{s}_{k}\in{}^{\texttt{h}}\mathcal{Z}$ such that
\begin{align}
\mathcal{G}_3({}^{\texttt{h}}\mathbf{y}^{0,s}_{k},{}^{\texttt{h}}\bfC_{k}^{v^{0,s}},{}^{\texttt{h}}z_{k}^{s})=0. \label{subproblem3}
\end{align}
}
    
\item{\emph{Step 5.} If $\parallel\mathcal{G}_1({}^{\texttt{h}}\bfy_{k}^{0,s},{}^{\texttt{h}}\bfC_{k}^{v^{0,s}},{}^{\texttt{h}}z_{k}^{s})\parallel/\parallel\mathcal{G}_1({}^{\texttt{h}}\bfy_{k}^{0,0},{}^{\texttt{h}}\bfC_{k}^{v^{0,0}},{}^{\texttt{h}}z_{k}^{0})\parallel\leq TOL_1$,
     $\parallel\mathcal{G}_2({}^{\texttt{h}}\bfy_{k}^{0,s}, {}^{\texttt{h}}\bfC_{k}^{v^{0,s}})\parallel/\parallel\mathcal{G}_2({}^{\texttt{h}}\bfy_{k}^{0,0},{}^{\texttt{h}}\bfC_{k}^{v^{0,0}})\parallel\leq TOL_2$, and $\parallel\mathcal{G}_3({}^{\texttt{h}}\bfy_{k}^{0,s},{}^{\texttt{h}}\bfC_{k}^{v^{0,s}},{}^{\texttt{h}}z_{k}^{s})\parallel/\parallel\mathcal{G}_3({}^{\texttt{h}}\bfy_{k}^{0,0},{}^{\texttt{h}}\bfC_{k}^{v^{0,0}},{}^{\texttt{h}}z_{k}^{0})\parallel\leq TOL_3$ or $s>\texttt{I}_2$, then set ${}^{\texttt{h}}\mathbf{y}_{k} = {{}^{\texttt{h}}\mathbf{y}}_{k}^{0,s}$, ${}^{\texttt{h}}\bfC_{k}^{v} = {}^{\texttt{h}}\bfC_{k}^{v^{0,s}}$, ${}^{\texttt{h}}z_{k} = {{}^{\texttt{h}}z}_{k}^{s}$,  $s=0$ and move to the time step $t_{k+1}$; otherwise set $s\leftarrow s+1$ and go to Step 1.
}

\end{itemize}

\paragraph{The sub-problem (\ref{subproblem1})} In view of the fact that the internal variable ${}^{\texttt{h}}\bfC_{k}^{v^{r-1,s-1}}$ and phase field ${}^{\texttt{h}}z_{k}^{s-1}$ are kept fixed, the sub-problem (\ref{subproblem1}) is nothing more than a discretized finite elastostatics problem. While Newton's method, or any of its variants, can be successfully utilized to generate solutions for a variety of cases, it does not seem capable of delivering converged solutions for general specimen geometries and loading conditions. This is because in the context of finite deformations (as opposed to the classical context of small deformations), the nucleation of fracture at a material point can lead to sudden large changes in the deformation field in the neighborhood of that point, which is beyond the reach of Newton's method. As demonstrated in previous works \citep{Henao16,KFLP18}, one robust approach to address this issue is to use an implicit gradient flow method of solution. Accordingly, in this work, we make use of the implicit gradient flow scheme described in Subsection 4.4 of \citep{KFLP18} to solve (\ref{subproblem1}). We refer the interested reader to \cite*{KFLP18} for a complete description of the scheme and simply remark here that the scheme features quadratic rate of convergence when the solution is smooth and a linear one when the solution exhibits abrupt changes. 

\paragraph{The sub-problem (\ref{subproblem2})} The sub-problem (\ref{subproblem2}) corresponds to a nonlinear system of first-order ODEs wherein the non-convex constraint  $\det{}^{\texttt{h}}\bfC_{k}^{v}=1$ is built-in. Physically, this constraint describes that viscous dissipation in elastomers is an isochoric process. Because of the requirement of satisfying this non-convex constraint along the entire time domain, extreme care must be exercised in the choice of time-integration scheme; see, e.g., \cite{Simo92}. In this work, we make use of a fifth-order explicit Runge-Kutta integrator which has been shown capable of preserving the constraint $\det{}^{\texttt{h}}\bfC_{k}^{v}=1$ identically for all time steps \citep{Lawson66,KLP16,WLPM23}. The scheme reads

\begin{equation*}
{}^{\texttt{h}}\bfC_{k}^{v^{r,s-1}}=\frac{1}{\left(\det\mathbf{A}_{k-1}^{r,s-1}\right)^{1/3}}\mathbf{A}_{k-1}^{r,s-1}
\end{equation*}
with
\begin{equation}
\mathbf{A}_{k-1}^{r,s-1}={}^{\texttt{h}}\bfC^{v}_{k-1}+\frac{\Delta t_{k}}{90}\left(7\mathbf{G}_{1}+32\mathbf{G}_{3}+12\mathbf{G}_{4}+32\mathbf{G}_{5}+7\mathbf{G}_{6}\right),\label{A_definition}
\end{equation}
where

\begin{equation}
\begin{aligned}
		&\mathbf{G}_{1}= \mathbf{G}\left(\nabla{}^{\texttt{h}}\mathbf{y}_{k-1},{}^{\texttt{h}}\bfC_{k-1}^{v}\right), \\
		&\mathbf{G}_{2}= \mathbf{G}\left(\frac{1}{2}\nabla{}^{\texttt{h}}\mathbf{y}_{k-1}+\frac{1}{2}\nabla{}^{\texttt{h}}\mathbf{y}^{r,s-1}_{k},{}^{\texttt{h}}\bfC_{k-1}^{v}+\mathbf{G}_{1} \frac{\Delta t_{k}}{2}\right), \\
		&\mathbf{G}_{3}= \mathbf{G}\left(\frac{3}{4}\nabla{}^{\texttt{h}}\mathbf{y}_{k-1}+\frac{1}{4}\nabla{}^{\texttt{h}}\mathbf{y}^{r,s-1}_{k},{}^{\texttt{h}}\bfC_{k-1}^{v}+(3\mathbf{G}_{1}+\mathbf{G}_{2}) \frac{\Delta t_{k}}{16}\right), \\
		&\mathbf{G}_{4}= \mathbf{G}\left({\frac{1}{2}}\nabla{}^{\texttt{h}}\mathbf{y}_{k-1}+{\frac{1}{2}}\nabla{}^{\texttt{h}}\mathbf{y}^{r,s-1}_{k},{}^{\texttt{h}}\bfC_{k-1}^{v}+\mathbf{G}_{3} {\frac{\Delta t_{k}}{2}}\right), \\
		&\mathbf{G}_{5}= \mathbf{G}\left(\frac{1}{4}\nabla{}^{\texttt{h}}\mathbf{y}_{k-1}+\frac{3}{4}\nabla{}^{\texttt{h}}\mathbf{y}^{r,s-1}_{k},{}^{\texttt{h}}\bfC_{k-1}^{v}+3(-\mathbf{G}_{2}+2\mathbf{G}_{3}+3\mathbf{G}_{4}) \frac{\Delta t_{k}}{16}\right), \\
		&\mathbf{G}_{6}= \mathbf{G}\left(\nabla{}^{\texttt{h}}\mathbf{y}^{r,s-1}_{k},{}^{\texttt{h}}\bfC_{k-1}^{v}+(\mathbf{G}_{1}+4\mathbf{G}_{2}+6\mathbf{G}_{3}-12\mathbf{G}_{4}+8\mathbf{G}_{5}) \frac{\Delta t_{k}}{7}\right), 
\end{aligned}\label{G_definitions}
\end{equation}
$\Delta t_{k} = t_{k} - t_{k-1}$, and where we recall that the function $\mathbf{G}$ is defined by (\ref{BVP-Cv-theory-reg2}). We remark that the term $\tau\equiv\eta/(2\psi_{I^e_1}^{\mathrm{NEq}})$ in (\ref{BVP-Cv-theory-reg2}) describes the time scale over which viscous dissipation takes place. Accordingly, the value of $\Delta t_{k}$ in (\ref{A_definition})-(\ref{G_definitions}) must be chosen sufficiently smaller than $\tau$ in order to accurately resolve this dissipation process. 

\paragraph{The sub-problem (\ref{subproblem3})} Finally, given that the deformation field ${}^{\texttt{h}}\mathbf{y}^{0,s}_{k}$ and the internal variable ${}^{\texttt{h}}\bfC_{k}^{v^{0,s}}$ are kept fixed, the sub-problem (\ref{subproblem3}) is nothing more than a standard discretized elliptic problem for a scalar-valued field. As opposed to the sub-problem (\ref{subproblem1}), the sub-problem (\ref{subproblem3}) can be solved with a standard Newton's method. 

In all of the numerical results that are presented below, we have made use of the tolerance values $TOL_1=TOL_2=TOL_3=10^{-6}$ and a maximum number of iterations of $\texttt{I}_1=10$ and $\texttt{I}_2=50$. In regard to these choices, we remark that numerical tests have indicated that equations (\ref{subproblem1})-(\ref{subproblem3}) may not be satisfied up to the above-specified tolerances after $\texttt{I}_1=10$ and $\texttt{I}_2=50$ iterations, instead, hundreds of iterations  might be required. The tests have also indicated, however, that increasing the number of iterations beyond $\texttt{I}_1=10$ and $\texttt{I}_2=50$ does not significantly change the obtained solution, thus our choice of $\texttt{I}_1=10$ and $\texttt{I}_2=50$  to reduce computational cost. In order to prevent mesh-induced preference for nucleation and propagation of fracture \citep{Negri99,Negri03}, we have made use of unstructured meshes of size $\texttt{h}=\varepsilon/5$, which are sufficiently small. As alluded to above, the selection of the value for the penalty parameter $q$ in (\ref{stability-term}) is subtle as it depends on specifics of the initial-boundary-value problem being solved, in particular, on the compressibility of the elastomer. In general, however, our numerical tests have shown that choosing $q$ to be of the order of the initial equilibrium shear modulus $\mu^{{\rm Eq}}$ of the elastomer under consideration typically renders an optimally convergent formulation. Finally, the numerical tests have also shown that it suffices to set $\Delta t_{k} = 10^{-2}\tau$ for the time increment in order to properly resolve the time of viscous dissipation.

\section{Sample simulations and comparisons with experiments}\label{Sec: Simulations}

In the next three subsections, we illustrate via sample simulations the capability of the proposed phase-field theory (\ref{BVP-y-theory-reg})-(\ref{BVP-z-theory-reg}) to describe the nucleation and propagation of fracture in viscoelastic elastomers subjected to quasistatic mechanical loads. With the additional objective of also illustrating the predictive capabilities of the theory, the selected simulations pertain to three prototypical experiments dealing with nucleation of fracture in the bulk \citep{Mueller1968}, nucleation of fracture from a pre-existing crack \citep{Pharretal2012}, and propagation of fracture \citep{Greensmith55} in different types of elastomers.

\subsection{Fracture nucleation in the bulk: Uniaxial tension experiments on polyurethane rubber bands}\label{Sec: rubber bands}

We begin by illustrating the capability of the proposed phase-field theory to describe and predict fracture nucleation in the bulk under spatially uniform states of stress by means of comparisons with the classical experiments of \cite{Mueller1968}. 

\begin{figure}[H]
\centering
\includegraphics[width=0.8\linewidth]{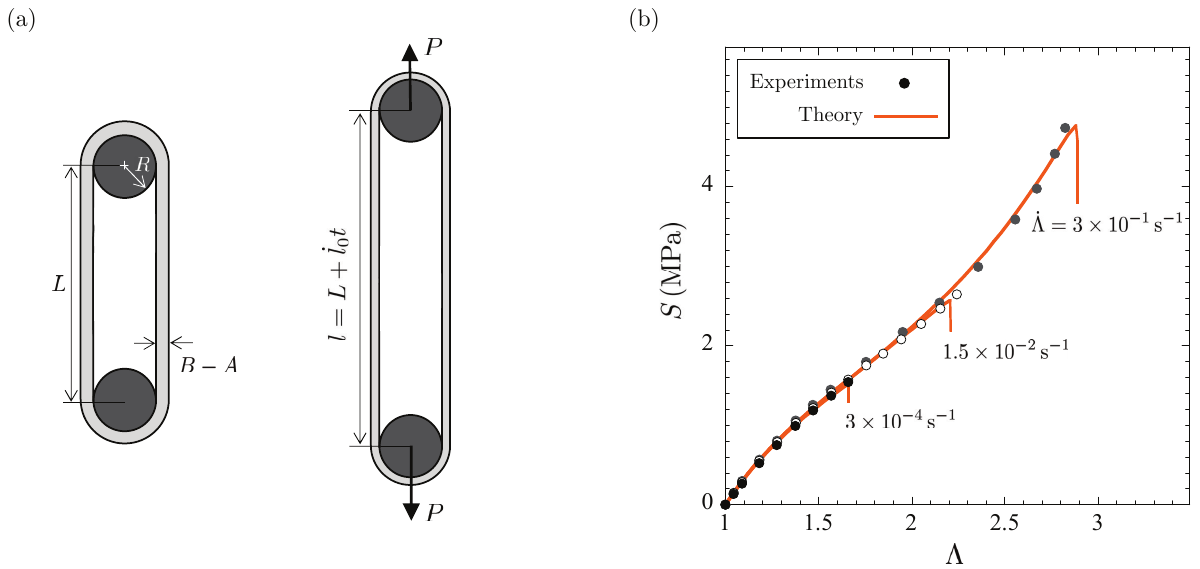}
\caption{{\small (a) Schematic of the ring specimens ($A=16.5$ mm, $B=19$ mm, $R=5$ mm, $L=\pi(A-R)=36.1$ mm) and loading in the experiments carried out by \cite{Mueller1968} on a polyurethane elastomer; the thickness of the specimens is $H=2.5$ mm. (b) Comparison between the response predicted by the phase-field theory (lines) and the experimental data (circles) of \cite{Mueller1968} for the stress $S=P/(2(B-A)H)$ as a function of the stretch $\Lambda=1+(l-L)/(\pi A)$ for three different constant stretch rates, $\dot{\Lambda}=\dot{l}_0/(\pi A)=3\times 10^{-4}, 1.5\times 10^{-2}, 3\times 10^{-1}$ s$^{-1}$.}}\label{Fig7}
\end{figure}

Following in the footsteps of Thor L. Smith, \cite{Mueller1968} carried out experiments on bands or ring specimens of inner radius $A=16.5$ mm, outer radius $B=19$ mm, and thickness $H=2.5$ mm that were stretched in tension by two circular pins, of radius $R=5$ mm, separated at ten different constant rates $\dot{l}_0\in[0.0157,15.7]$ mm/s; see Fig.~\ref{Fig7}(a). He reported results for the global stress $S=P/(2(B-A)H)$ as a function of the global stretch\footnote{For the specific geometry of the ring specimens used by \cite{Mueller1968}, viscoelastic calculations show that the local stretch in the gauge sections, say $\lambda_{g}$, is related to the global stretch $\Lambda$ by the linear relation $\lambda_{g}\approx\Lambda+0.32(\Lambda-1)$, irrespective of the rate of loading.} $\Lambda=1+(l-L)/(\pi A)$ until the specimens failed by severing, at some critical pair of global stress-stretch $(S_c,\Lambda_c)$, along one of the two gauge sections where the stress was spatially uniform and uniaxial. His most complete sets of results pertain to three different temperatures, $\Theta=-5, 20, 40$ $^\circ$C. Here, we focus on those at $\Theta=20$ $^\circ$C; see Figs. 15, 18, and 19 in his work.
%
%

\subsubsection{Calibration of the three material inputs entering the theory}

\begin{table}[H]\centering
\caption{Values of the material constants in the viscoelastic model (\ref{S-KLP})-(\ref{Evolution-KLP}) for the polyurethane elastomer studied by \cite{Mueller1968}.}
\begin{tabular}{lll}
\hline
$\mu_1= 0.9627$ MPa  & $\nu_1=0.244$ MPa    &  $\eta_0=200$ MPa s \\
$\mu_2=0$            & $\nu_2=0$            &  $\eta_\infty=0$ \\
$\kappa=1$ GPa       & $\beta_1=3.168$      &  $K_1=0$ \\
$\alpha_1= 0.946$    &                      &  $K_2=0$ \\
\hline
\end{tabular} \label{Table2}
\end{table}
\begin{table}[H]\centering
\caption{Values of the material constants in the uniaxial tensile strength model (\ref{Strength-Solithane}) and of the hydrostatic strength $\shs$ for the polyurethane elastomer studied by \cite{Mueller1968}.}
\begin{tabular}{ccccccc|c}
\hline
$a_0$ & $a_1$ (MPa) & $b_1$    & $a_2$  (MPa)  & $b_2$  & $c_0$  & $c_1$  &  $\shs$ (MPa) \\
\hline
$0$ & $4.3916$     & $0.3349$ & $9.9908\times 10^{-2}$       & $2.5298$ & $4.529$ & $20$ & $3$ \\
\hline
\end{tabular} \label{Table3}
\end{table}

Prior to the deployment of the phase-field theory (\ref{BVP-y-theory-reg})-(\ref{BVP-z-theory-reg}), we must first calibrate the material functions/constants that describe the viscoelasticity, the strength, and the fracture energy of the elastomer. 

Fitting the material constants in the viscoelastic model (\ref{S-KLP})-(\ref{Evolution-KLP}) to the experimental $S$ $vs.$ $\Lambda$ data for the slowest and the fastest stretch rates, $\dot{\Lambda}=3\times 10^{-4}$ and $3\times 10^{-1}$ s$^{-1}$, yields the values listed in Table \ref{Table2}. As seen from the comparisons presented in Fig.~\ref{Fig7}(b), the constitutive relation (\ref{S-KLP})-(\ref{Evolution-KLP}) with such material constants describes reasonably well the experimentally measured viscoelastic response of the elastomer.

Next, fitting the material constants in the uniaxial tensile strength model (\ref{Strength-Solithane}) to the experimental $S_c$ $vs.$ $\Lambda_c$ data for all ten stretch rates $\dot{\Lambda}\in[3\times 10^{-4}, 3\times 10^{-1}]$ s$^{-1}$ yields the values listed in Table \ref{Table3}. While \cite{Mueller1968} did not report strength data beyond uniaxial tension, the poker-chip experiments presented by \cite{Lindsey67} around the same time for the same polyurethane elastomer permit to estimate that $\shs\approx 3$ MPa. 

Finally, we consider a fracture energy of
\begin{equation*}
G_c=41\, {\rm N}/{\rm m},
\end{equation*}
as estimated from ``pure-shear'' experiments by \cite{Mueller1968} himself; see also \cite{Knauss71}.

\subsubsection{Computational aspects}

At this stage, it remains to select a sufficiently small value of the regularization length $\varepsilon$ and an appropriate FE mesh of sufficiently small size $\texttt{h}$. 

Given the width $B-A=2.5$ mm of the specimens and given the heuristic bound $G_c/\psi^{{\rm Eq}}_{\texttt{ts}}(I_1)\leq 0.1\,{\rm mm}$ on the material length scales associated with uniaxial tension, any regularization length $\varepsilon\leq 0.1$ mm is expected to be small enough. The results presented below correspond to $\varepsilon=0.02$ mm. Smaller and larger values were checked to lead to essentially the same results.

\begin{figure}[t!]
\centering
\includegraphics[width=0.55\linewidth]{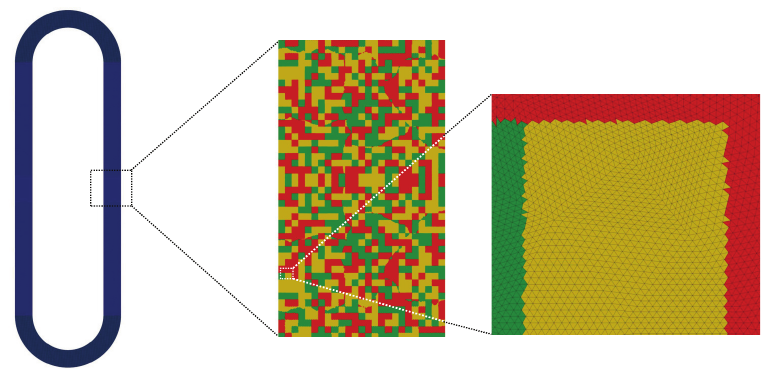}
\caption{{\small One of the FE meshes used to simulate the experiments of \cite{Mueller1968} and the subregions over which the strength material constants are stochastically perturbed.}}\label{Fig8}
\end{figure}

We treat the contact between the pins, assumed rigid, and the specimens to be semicircles where \emph{no} slip occurs. Additionally, since the stress field is uniaxial in the gauge sections away from the pins and it is in those sections that fracture occurs, it suffices to perform the simulations under conditions of plane stress and to make use of FE meshes that are refined only in the gauge sections. As noted at the end of Section \ref{Sec: Numerical Implementation}, the meshes there are chosen to be unstructured and of size $\texttt{h}=\varepsilon/5=0.004$ mm. Figure \ref{Fig8} shows one of the meshes used to carry out the simulations. 

Figure \ref{Fig8} also shows the subregions of size $5\varepsilon$ where the strength parameters $a_1$ and $a_2$ in Table \ref{Table3} are randomly perturbed by a $\pm5\%$ variation in value to account for the inherently stochastic nature of the strength. Consistent with previous works in the simpler context of finite elasticity \citep{KFLP18,KLP21,KKLP24}, this mild stochasticity is enough to break the symmetry of the problem and to lead to nucleation of fracture in just one of the two gauge sections, as observed in the experiments.

%
%

\subsubsection{Comparisons with the experiments}

\begin{figure}[H]
\centering
\includegraphics[width=0.99\linewidth]{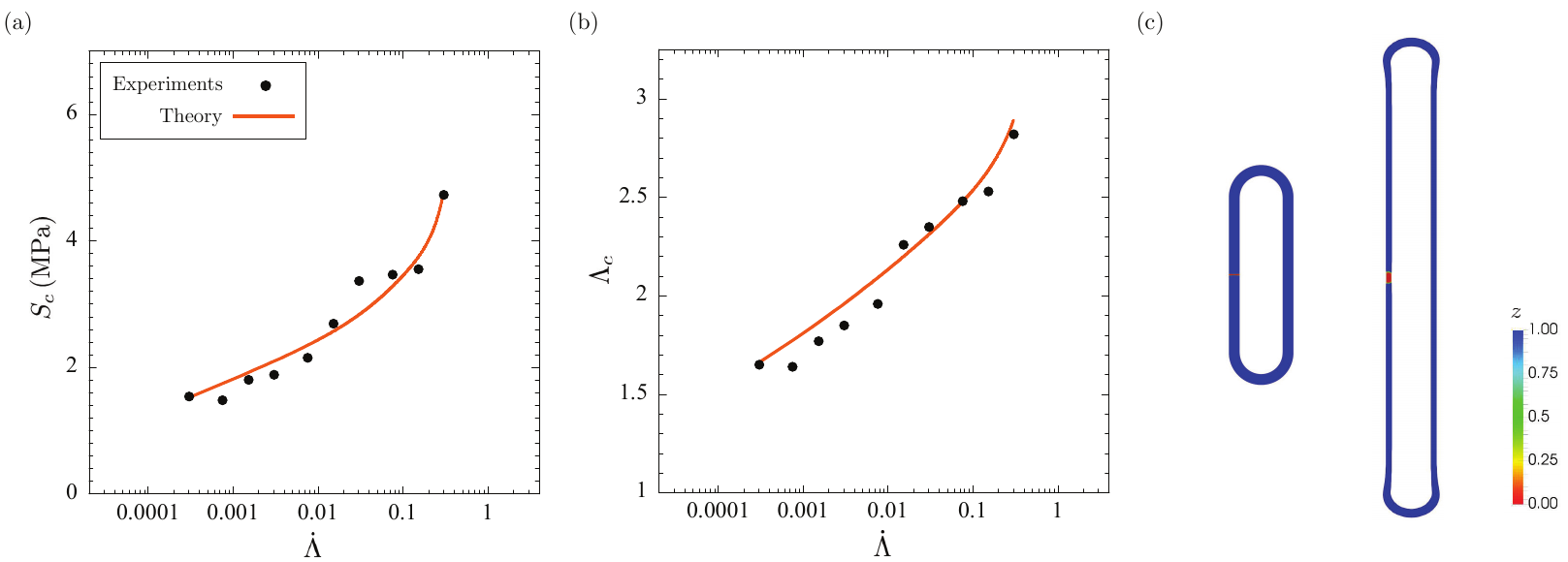}
\caption{{\small Comparisons between the values predicted by the phase-field theory (lines) and the experimental data (circles) of \cite{Mueller1968} for (a) the critical stress $S_c$ and (b) the critical stretch $\Lambda_c$ at which the specimens fracture. Both sets of results are plotted as functions of the stretch rate $\dot{\Lambda}$. (c) Contour plots of the phase-field $z(\bfX,t)$ over the undeformed and deformed configurations at $\Lambda=2.2$ of a specimen stretched at $\dot{\Lambda}=1.5\times 10^{-2}$ s$^{-1}$.
 }}\label{Fig9}
\end{figure}

Figure \ref{Fig9} compares the results predicted by the phase-field theory with the experimental data reported by \cite{Mueller1968} for (a) the critical stress $S_c$ and (b) the critical stretch $\Lambda_c$ at which the specimens fracture; both results are presented as functions of the stretch rate $\dot{\Lambda}$ at which the tests were performed. Furthermore, Fig.~\ref{Fig9}(c)  presents representative contour plots for the phase field $z(\bfX,t)$ over the undeformed and deformed configurations at the stretch $\Lambda=2.2$ for one of the tests carried out at the stretch rate $\dot{\Lambda}=1.5\times 10^{-2}$ s$^{-1}$.

It is immediately apparent from Fig.~\ref{Fig9}, as well as from Fig.~\ref{Fig7}(b), that the proposed phase-field theory predicts accurately the nucleation of fracture observed experimentally by \cite{Mueller1968} for the entire range of stretch rates $\dot{\Lambda}$ that he considered.
%
%

\subsection{Fracture nucleation from a pre-existing crack: ``Pure-shear'' experiments on VHB 4905}

Next, we illustrate the capability of the proposed phase-field theory to describe and predict fracture nucleation from a large pre-existing crack by means of comparisons with the  experiments of \cite{Pharretal2012}. These authors carried out ``pure-shear'' fracture experiments on VHB 4905 making use of specimens of length $L=152$ mm, height $H=10$ mm, and thickness $B=0.5$ mm with a pre-existing edge crack of length $A=20$ mm; see the schematic in Fig.~\ref{Fig2}. They reported results for the global stress $S=P/(BL)$ as a function of the global stretch $\Lambda=h/H$ for six constant global stretch rates, $\dot{\Lambda}=1.67\times 10^{-3}$, $1.67\times 10^{-2}$, $6.67\times 10^{-2}$, $1.67\times 10^{-1}$, $6.67\times 10^{-1}$, $1.67$ s$^{-1}$, until the pre-existing crack roughly started to grow at some critical global stretch $\Lambda_c$; see Figs.~2(a) and 3(b) in their work.

\subsubsection{Calibration of the three material inputs entering the theory}

\begin{table}[H]\centering
\caption{Values of the material constants in the viscoelastic model (\ref{S-KLP})-(\ref{Evolution-KLP}) for the acrylic elastomer VHB 4905 studied by \cite{Pharretal2012}.}
\begin{tabular}{lll}
\hline
$\mu_1= 13.96$ kPa  & $\nu_1=50.15$ kPa                &  $\eta_0=7007$ kPa s \\
$\mu_2=0.9255$ kPa  & $\nu_2=5.193\times 10^{-6}$ kPa  &  $\eta_\infty=14$ kPa s \\
$\kappa=50$ MPa      & $\beta_1=0.9660$                 &  $K_1=2833$ kPa s \\
$\alpha_1= 0.5104$  &  $\beta_2=7.107$                 &  $\gamma_1=3.467$ \\
$\alpha_2= 1.910$   &                                  &  $K_2=1.228$ kPa$^{-2}$ \\
                    &                                  &  $\gamma_2=0.0836$ \\
\hline
\end{tabular} \label{Table4}
\end{table}
As one of their comparisons with experiments, \cite{SLP23} employed the Griffith criticality (\ref{Gc-0}) condition to explain the same experiments of \cite{Pharretal2012} under consideration here. Their analysis showed that the viscoelastic model (\ref{S-KLP})-(\ref{Evolution-KLP}) with the materials constants listed in Table \ref{Table4} provides a reasonably accurate description of the experimentally measured viscoelastic response of VHB 4905. Their analysis also showed that the fracture energy of this elastomer is about

\begin{equation*}
G_c=634\, {\rm N}/{\rm m}.
\end{equation*}
The simulations presented below make use of both of these constitutive prescriptions. 

Now, save for the two experiments included in \citep{Mazza12,Chen17}, carried out at two different constant stretch rates ($4.5\times 10^{-3}$ s$^{-1}$ and $1.7\times 10^{-2}$ s$^{-1}$), no experimental study on the uniaxial strength of VHB 4905 appears to have been reported in the literature. We are also unaware of any experimental data on its hydrostatic strength. Fortunately, as discussed in Subsection \ref{Sec:Nucleation-crack} above, this lack of knowledge about the strength of VHB 4905 is \emph{not} a problem for the purposes of this subsection, as the growth of the pre-existing crack in ``pure-shear'' fracture experiments is governed by the Griffith criticality condition (\ref{Gc-0}), which, again, it primarily involves the viscoelasticity and the fracture energy of the elastomer. Accordingly, the simulations presented below are carried out with the strength materials constants listed in Table \ref{Table5}, which are consistent with the experiments reported in \citep{Mazza12,Chen17} and in \citep{BCLLP24} for a similar soft elastomer.

\begin{table}[H]\centering
\caption{Values of the material constants in the uniaxial tensile strength model (\ref{Strength-Solithane}) and of the hydrostatic strength $\shs$ used in the simulations for the acrylic elastomer VHB 4905.}
\begin{tabular}{cccccc|c}
\hline
$a_0$  & $a_1$ (kPa)               & $b_1$  & $a_2$  & $c_0$ & $c_1$   &  $\shs$ (kPa) \\
\hline
$0$    & $4.9362\times 10^{-7}$    & $6$ &     $0$   & $100$  & $120$  &   $500$ \\
\hline
\end{tabular} \label{Table5}
\end{table}

\subsubsection{Computational aspects}

In view of the height $H=10$ mm of the specimens and of the material length scales $G_c/\psi^{{\rm Eq}}_{\texttt{ts}}(I_1)\leq 1\,{\rm mm}$ associated with uniaxial tension, the results presented below correspond to a value of $\varepsilon=0.5$ mm for the regularization length. 

\begin{figure}[H]
\centering
\includegraphics[width=0.8\linewidth]{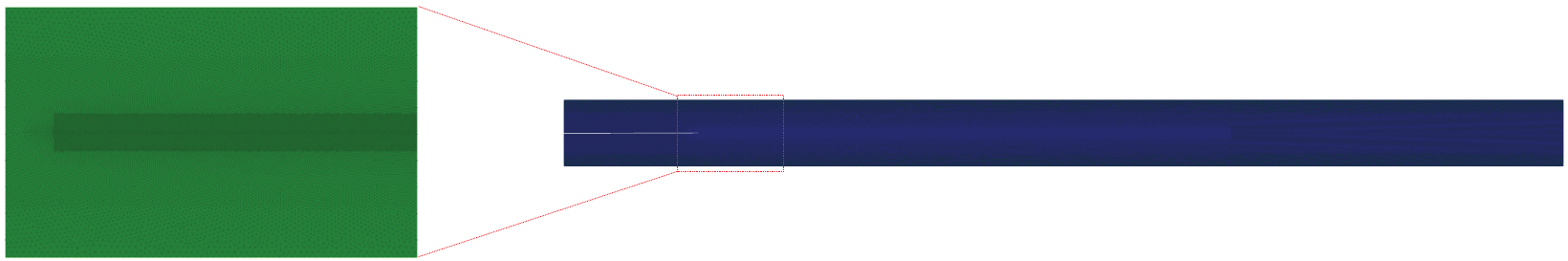}
\caption{{\small One of the FE meshes used to simulate the experiments of \cite{Pharretal2012}. A line has been drawn on the mesh to help visualize the location of the pre-existing crack.}}\label{Fig10}
\end{figure}

Moreover, given the small thickness $B=0.5$ mm of the specimens relative to their length $L=152$ mm and height $H=10$ mm, we carry out the simulations under conditions of plane stress. Additionally, since the region of crack growth is known \emph{a priori}, we make use of FE meshes that are refined only around and ahead of the crack front. In such a region, the size of the mesh is set to $\texttt{h}=\varepsilon/5=0.1$ mm. Figure \ref{Fig10} shows one of the meshes used to carry out the simulations.

%
%

\subsubsection{Comparisons with the experiments}

%
\begin{figure}[b!]
\centering
\includegraphics[width=0.99\linewidth]{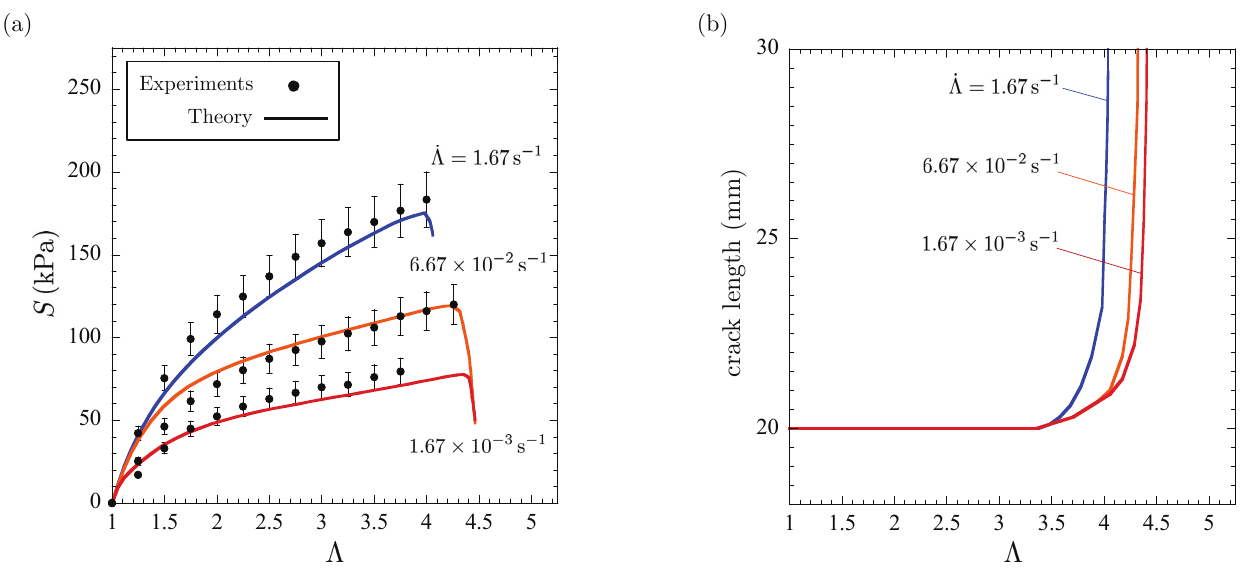}
\caption{{\small (a) Comparison between the response predicted by the phase-field theory (lines) and the experimental data (circles) of \cite{Pharretal2012} for the stress $S=P/(BL)$ as a function of the stretch $\Lambda=h/H$ for three different constant stretch rates, $\dot{\Lambda}=1.67\times 10^{-3}$, $6.67\times 10^{-2}$, $1.67$ s$^{-1}$. (b) The associated evolution of the crack length predicted by the phase-field theory as a function of the applied stretch. 
 }}\label{Fig11}
\end{figure}
%

%
\begin{figure}[t!]
\centering
\includegraphics[width=0.95\linewidth]{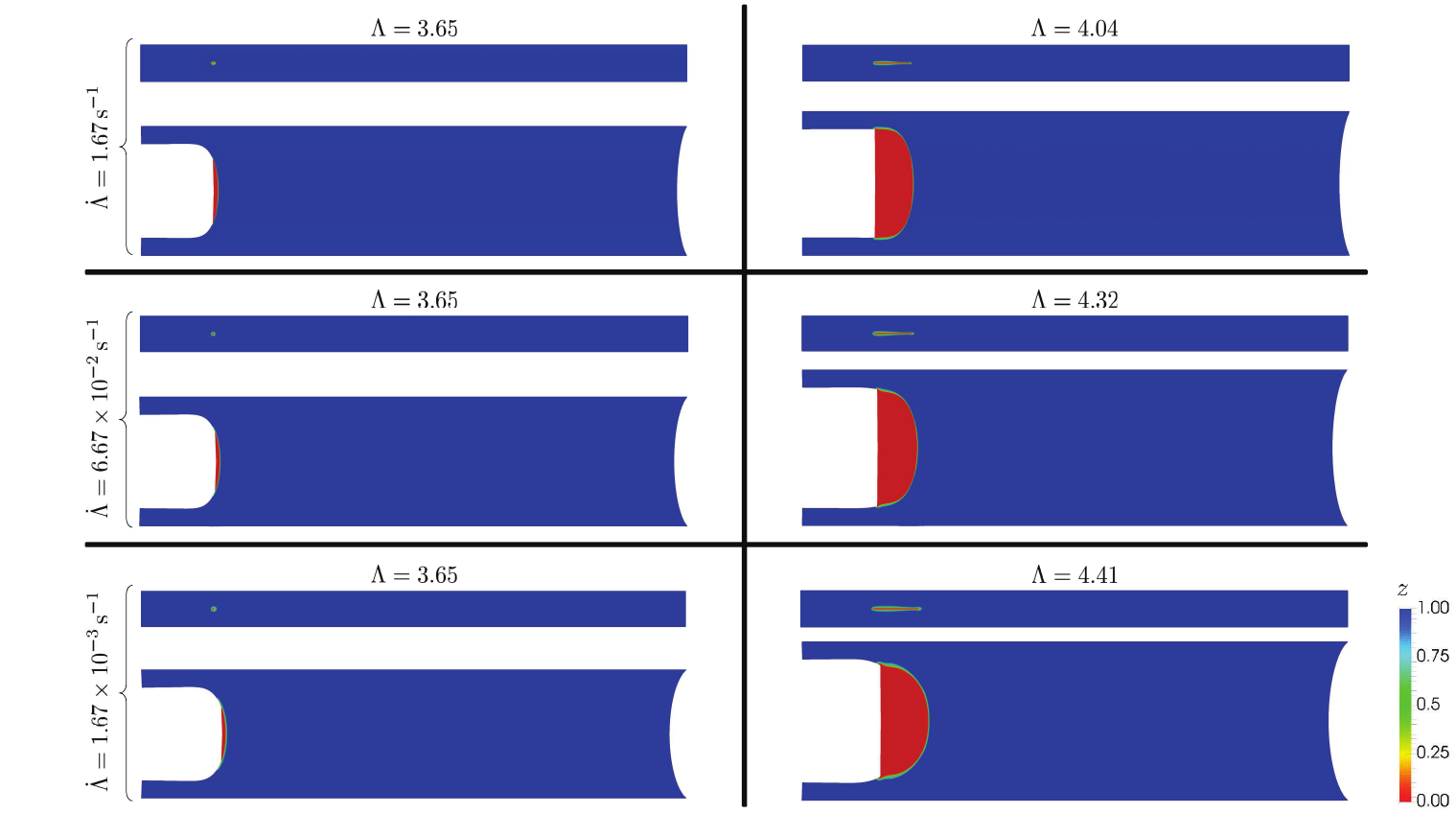}
\caption{{\small Contour plots of the phase field $z(\bfX,t)$ over the undeformed and deformed configurations of specimens stretched at $\dot{\Lambda}=1.67\times 10^{-3}$, $6.67\times 10^{-2}$, $1.67$ s$^{-1}$. The results pertain to two values of the applied stretch $\Lambda$ for each $\dot{\Lambda}$.}}\label{Fig12}
\end{figure}
%

%
\begin{figure}[t!]
\centering
\includegraphics[width=0.9\linewidth]{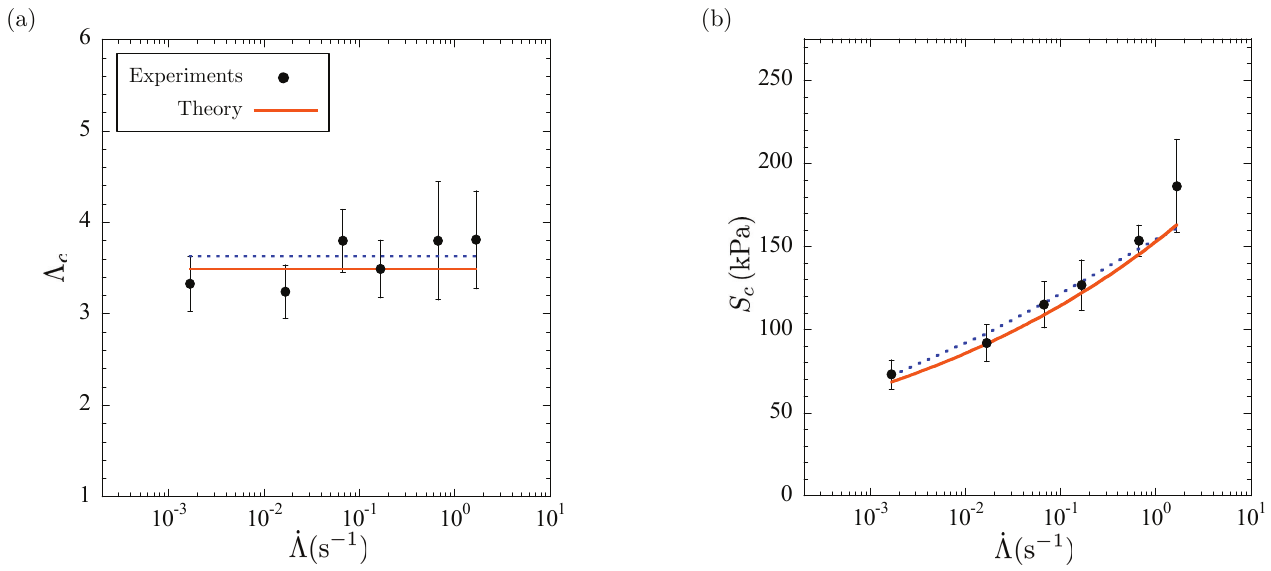}
\caption{{\small Comparisons between the values predicted by the phase-field theory (lines) and the experimental data (circles) of \cite{Pharretal2012} for (a) the critical stretch $\Lambda_c$ and (b) the critical stress $S_c$ at which fracture nucleates from the pre-existing crack. Both sets of results are plotted as functions of the stretch rate $\dot{\Lambda}$. For direct comparison, the sharp-fracture results generated from the Griffith critically condition (\ref{Gc-0}), as reported in \citep{SLP23}, are also included in the plots (dotted lines).}}\label{Fig13}
\end{figure}
%

Figure \ref{Fig11}(a) compares the stress-stretch response predicted by the phase-field theory with the experimental data reported by \cite{Pharretal2012} for three different stretch rates, $\dot{\Lambda}=1.67\times 10^{-3}$, $6.67\times 10^{-2}$, $1.67$ s$^{-1}$. The main observation from this figure is that the theoretical predictions are in good agreement with the experimental observations. 

In particular, the results in Fig.~\ref{Fig11}(a) show that the theoretical predictions for the critical stretch $\Lambda_c$ and the corresponding critical stress $S_c$ at which the crack starts to grow agree with the experiments and, by the same token, with the Griffith critically condition (\ref{Gc-0}), which predicts onset of crack growth at $\Lambda_c=3.63$ irrespective of the loading rate \citep{SLP23}. These comparisons are presented in a different and more complete form by Fig. \ref{Fig13}, where $\Lambda_c$ and $S_c$ are plotted as functions of all the six stretch rates $\dot{\Lambda}$ at which the experiments were carried out.

A few words about the identification of the critical value $\Lambda_c$ (and hence also of $S_c$) from the simulations are in order. Numerical experiments show that the crack length can be estimated by identifying the phase-field crack tip as the material point at which the equilibrium free energy function $\psi^{{\rm Eq}}(I_1,J)$ attains its maximum value along the axis of symmetry of the specimen. The critical stretch $\Lambda_c$ is then defined as the value of the applied stretch $\Lambda$ at which the crack tip starts to move away from its initial location. By way of an example, Fig.~\ref{Fig11}(b) presents results for the evolution of the crack length as a function of the applied stretch $\Lambda$ for the stretch rates $\dot{\Lambda}=1.67\times 10^{-3}$, $6.67\times 10^{-2}$, $1.67$ s$^{-1}$.  Figure \ref{Fig12} presents corresponding contour plots of the phase field over the undeformed and deformed configurations at two different stretches, $\Lambda=3.65$ and $4.41$ for $\dot{\Lambda}=1.67\times 10^{-3}$ s$^{-1}$, $\Lambda=3.65$ and $4.32$ for $\dot{\Lambda}=6.67\times 10^{-2}$ s$^{-1}$, and $\Lambda=3.65$ and $4.04$ for $\dot{\Lambda}=1.67$ s$^{-1}$.

\subsection{Propagation of fracture: Trousers experiments on a canonical viscoelastic elastomer}

Finally, we turn to illustrating the capability of the proposed phase-field theory to describe fracture propagation by means of comparisons with the classical experiments of \cite{Greensmith55}. 

Specifically, \cite{Greensmith55} carried out trousers fracture experiments on various natural rubbers and SBR elastomers making use of specimens of lengths $L\in[100,150]$ mm, height $H=40$ mm, and thicknesses $B\in[1,2]$ mm with a pre-existing edge crack of length of about $A=50$ mm; see the schematic in Fig.~\ref{Fig5}. They reported results for the normalized critical force $2P_c/B$ at which the crack propagated either in a steady state manner (for the case of the SBR elastomers) or by a stick-slip mechanism (for the case of the natural rubbers) for constant rates $\dot{l}\in[5\times 10^{-3},500]$ mm/s of separation between the grips. The results pertain to four SBR elastomers (vulcanizates A, B, C, D) and three natural rubbers (vulcanizates E, F, and G) at several temperatures in the range $\Theta\in[-20, 90]$~$^\circ$C; see Figs.~4 through 7 in their work and also Fig.~\ref{Fig5} above. They also reported the initial shear modulus, as well as the critical nominal stress and associated stretch at which the various elastomers fractured when subjected to uniaxial tension at $\Theta=20$~$^\circ$C, but unspecified stretch rate; see Table I in their work. Unfortunately, these results are insufficient to calibrate the material constants in (\ref{BVP-y-theory-reg})-(\ref{BVP-z-theory-reg}) to carry out a quantitative comparison. Nevertheless, we can carry out a qualitative comparison.

\subsubsection{The dimensions of the specimen and the three material inputs entering the theory}

For definiteness, consistent with the specimens tested by \cite{Greensmith55}, we carry out the simulations on a specimen of length $L=100$ mm, height $H=40$ mm, thickness $B=1$ mm, with a pre-existing edge crack of length $A=50$ mm.

%
\begin{figure}[b!]
\centering
\includegraphics[width=0.9\linewidth]{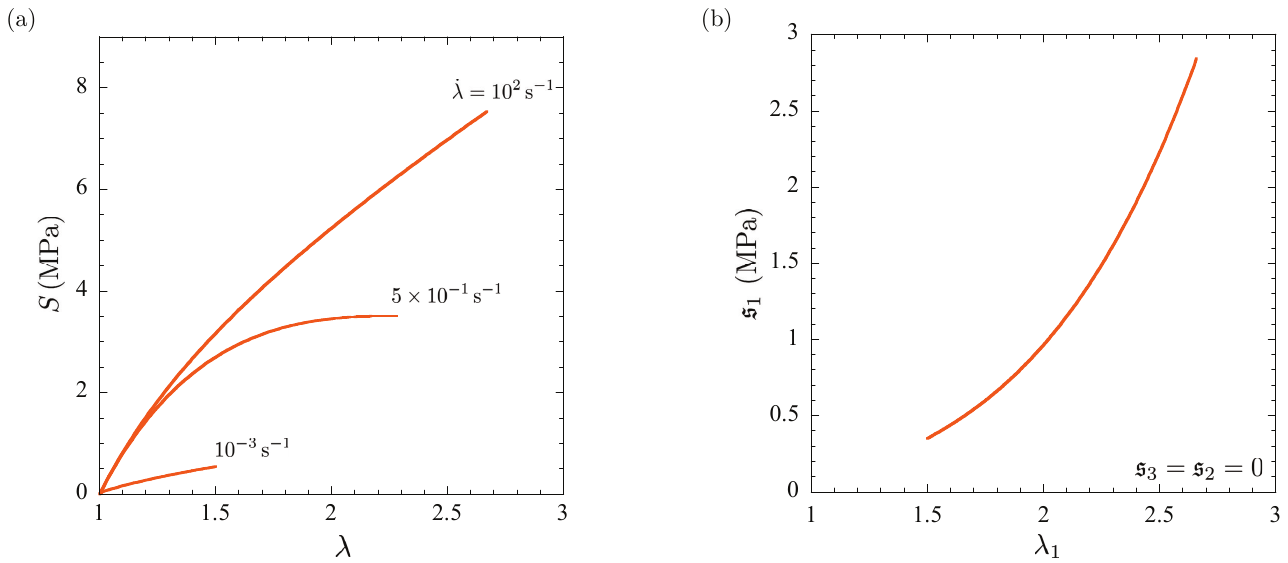}
\caption{{\small (a) Uniaxial stress-stretch response at various constant stretch rates described by the viscoelastic model (\ref{S-KLP})-(\ref{Evolution-KLP}) with the material constants listed in Table \ref{Table6}. (b) The uniaxial tensile strength model (\ref{Strength-Solithane}) with the material constants listed in Table \ref{Table7}.}}\label{Fig14}
\end{figure}
%

To keep the results as basic as possible, we consider the case of a canonical elastomer whose viscoelastic behavior is described by a Gaussian equilibrium elasticity, a Gaussian non-equilibrium elasticity, and a constant viscosity. This amounts to setting $\mu_2=\nu_2=0$ and $\eta_{\infty}=K_1=0$, $K_2=0$ in the viscoelastic model (\ref{S-KLP})-(\ref{Evolution-KLP}). As for the remaining viscoelastic material constants, consistent with the initial shear moduli reported by \cite{Greensmith55},  we make use of those listed in Table \ref{Table6}. Figure \ref{Fig14}(a) shows the uniaxial tensile stress-stretch response of such a canonical elastomer for three different constant stretch rates, $\dot{\lambda}=10^{-3}, 5\times10^{-1}, 10^{2}$ s$^{-1}$.

\begin{table}[H]\centering
\caption{Values of the material constants in the viscoelastic model (\ref{S-KLP})-(\ref{Evolution-KLP}) used for the trousers simulations.}
\begin{tabular}{lll}
\hline
$\mu_1= 0.5$ MPa  & $\nu_1=2.5$ MPa    &  $\eta_0=10$ MPa s \\
$\mu_2=0$            & $\nu_2=0$    &  $\eta_\infty=0$ \\
$\kappa=0.5$ GPa       & $\beta_1=1$      &  $K_1=0$ \\
$\alpha_1= 1$   &       &  $K_2=0$ \\
\hline
\end{tabular} \label{Table6}
\end{table}

Based on the uniaxial tensile strength data that \cite{Greensmith55} reported in Table I of their work, and based on additional uniaxial tensile strength data that \cite{Greensmith60a} presented in a subsequent contribution, as well as the recent poker-chip experiments reported in \citep{Euchler20}, we make use of the material constants listed in Table \ref{Table7} for the uniaxial tensile strength $\sts(I_1)$ and the hydrostatic strength $\shs$ of the canonical elastomer. Figure \ref{Fig14}(b) provides a plot of the former.
\begin{table}[H]\centering
\caption{Values of the material constants in the uniaxial tensile strength model (\ref{Strength-Solithane}) and of the hydrostatic strength $\shs$ used for the trousers simulations.}
\begin{tabular}{cccccc|c}
\hline
$a_0$ & $a_1$ (MPa) & $b_1$    & $a_2$    & $c_0$ & $c_1$   &  $\shs$ (MPa) \\
\hline
$0$ & $0.0292$     & $4.1002$ & $0$  & $3.5833$ & 7.8129  & $2$ \\
\hline
\end{tabular} \label{Table7}
\end{table}

Finally, based on a number of experimental results on common elastomers reported in the literature \citep{GT82,Gent96}, we make use of the fracture energy
\begin{equation*}
G_c=200\, {\rm N}/{\rm m}.
\end{equation*}

\subsubsection{Computational aspects}

Given the thickness $B=1$ mm of the specimens and given the heuristic bound $G_c/\psi^{{\rm Eq}}_{\texttt{ts}}(I_1)\leq 0.25\,{\rm mm}$ on the material length scales associated with uniaxial tension, the results presented below correspond to regularization length $\varepsilon=0.25$ mm. 

In actual trousers experiments, the specimens are mounted in the testing machine by bending in opposite directions the two legs of the trousers until they are brought to lie in the same plane, at which point their bottom ends are firmly gripped and the specimen is ready to be loaded. For this reason, it proves convenient not to use the initial configuration as the reference configuration to carry out the simulations, but to use, instead, the configuration as initially mounted in the testing machine; see the schematic in Fig.~\ref{Fig5}.

%
\begin{figure}[b!]
\centering
\includegraphics[width=0.95\linewidth]{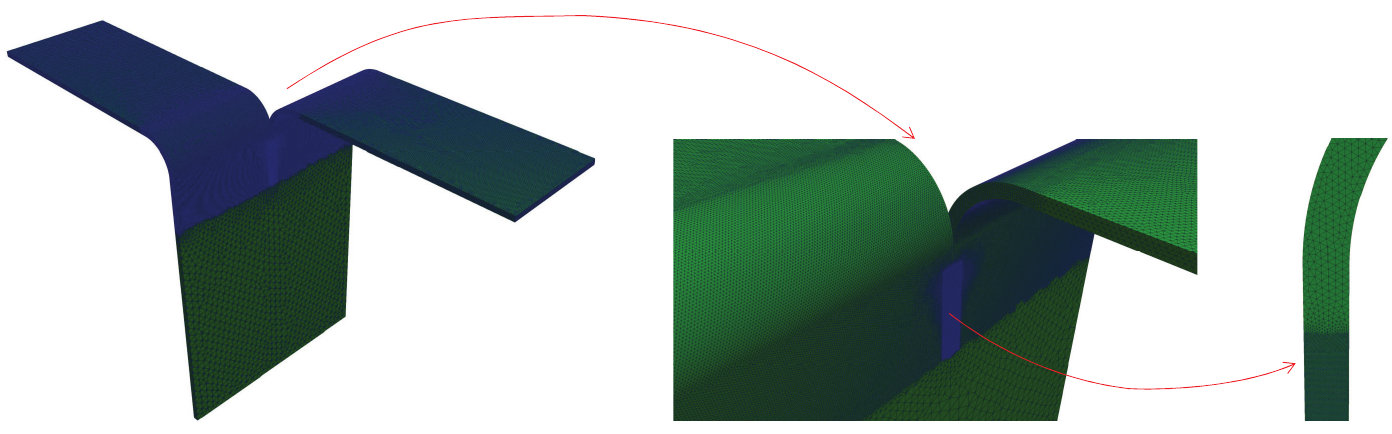}
\caption{{\small One of the FE meshes used to simulate the trousers experiments.}}\label{Fig15}
\end{figure}
%

When mounted in the testing machine, following in the footsteps of \cite{SLP23c}, we take the transition from the legs to the un-cracked part of the specimen to be circular fillets of inner radius $R_f$ plus an additional straight segment of length $A_b$.
Note that in this reference configuration the initial grip separation is $l_0=2(A-A_b)+B-(\pi-2)R_f$. The simulations presented below correspond to $R_f=5B=5$ mm, $A_b=2$ mm, and hence $l_0=91.29$ mm. Additionally, since the region of crack growth is known \emph{a priori}, we make use of FE meshes that are refined around and ahead of the crack front. In such a region, the size of the mesh is set to $\texttt{h}=\varepsilon/5=0.05$ mm. Figure \ref{Fig15} shows one of the meshes used to carry out the simulations.

\subsubsection{Results and qualitative comparisons with experimental observations}

\begin{figure}[H]
\centering
\includegraphics[width=0.9\linewidth]{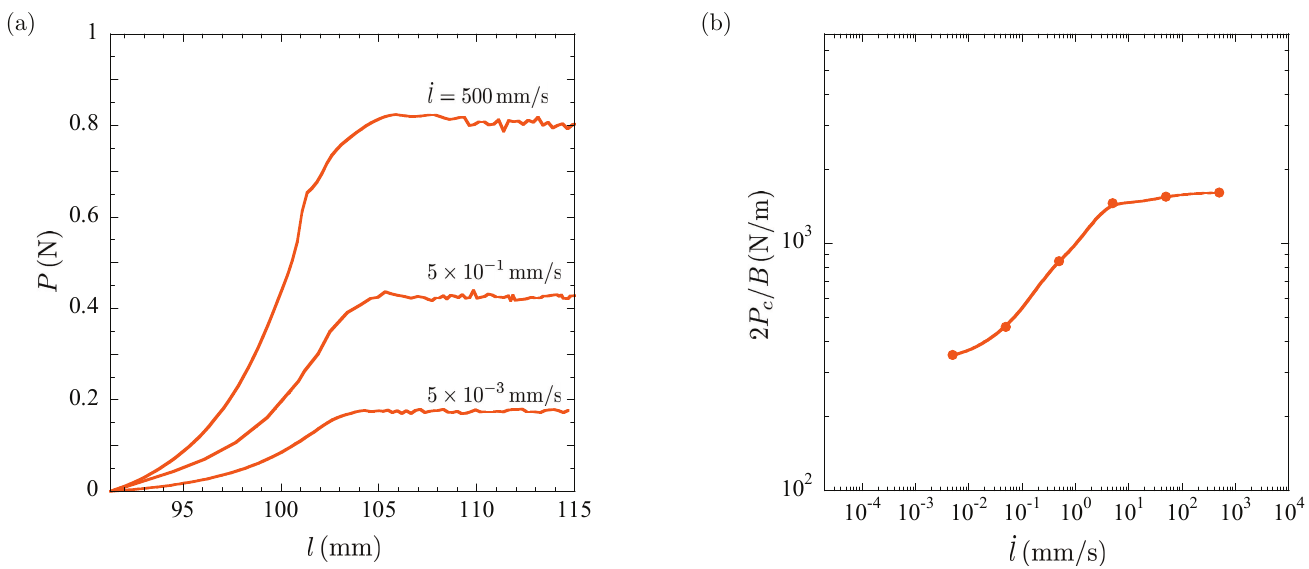}
\caption{{\small (a) Results predicted by the phase-field theory for the force $P$ as a function of the applied grip-to-grip separation $l$ for three different constant grip separation rates, $\dot{l}=5\times10^{-3}, 5\times10^{-1}, 500$ mm/s. (b) Values predicted by the phase-field theory for the normalized critical force $2P_c/B$ at which the crack propagates at constant speed for different constant rates $\dot{l}$ of separation between the grips.}}\label{Fig16}
\end{figure}

\begin{figure}[H]
\centering
\includegraphics[width=0.97\linewidth]{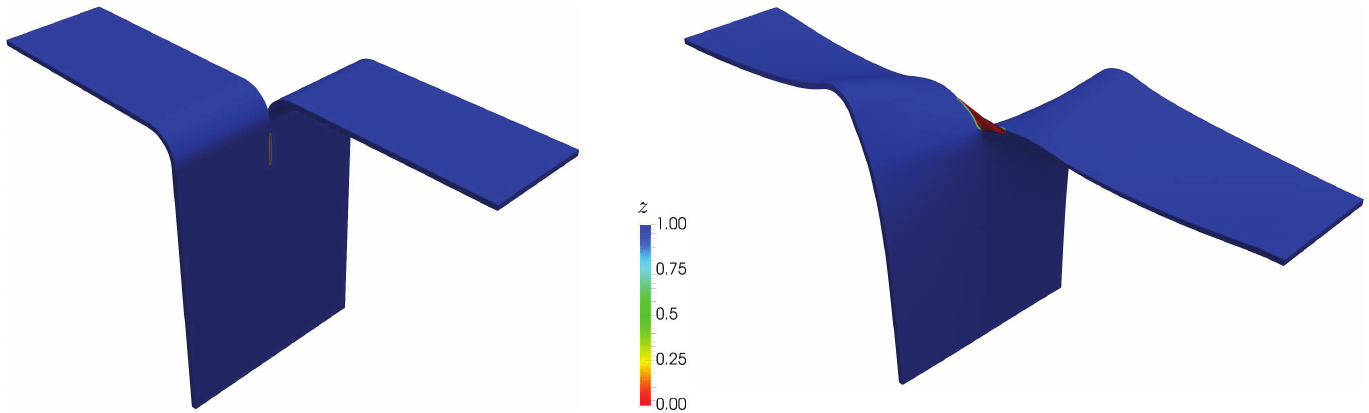}
\caption{{\small Contour plot of the phase-field $z(\bfX,t)$ over the undeformed and deformed configurations at the grip-to-grip separation $l=114$ mm for the specimen teared at $\dot{l}=500$ mm/s.}}\label{Fig17}
\end{figure}

Figure \ref{Fig16}(a) presents results, plotted as functions of the applied grip-to-grip separation $l$, for the force $P$ predicted by the phase-field theory for trousers tests carried out at three different constant rates of separation between the grips, $\dot{l}=5\times10^{-3}, 5\times10^{-1}, 500$ mm/s. Figure \ref{Fig17} presents corresponding contour plots for the phase field $z(\bfX,t)$ over the undeformed and deformed configurations at the grip-to-grip separation $l=114$ mm for one of the tests, that carried out at $\dot{l}=500$ mm/s.

Consistent with the Griffith criticality condition (\ref{Gc-0-Propagation}), after a transient response, the results in Fig.~\ref{Fig16}(a) show that the force $P$ reaches a plateau during which the crack propagates at a constant speed. What is more, consistent with the Griffith criticality condition (\ref{Gc-0-Propagation}), the plateau value $P_c$ of the force increases with the rate of loading $\dot{l}$. 

%
%

%
%

From the type of results presented in Fig.~\ref{Fig16}(a), we can readily determine the values predicted by the phase-field theory for the normalized critical force $2P_c/B$ associated with each constant rate $\dot{l}$ of grip separation. Figure \ref{Fig16}(b) presents such predictions. Similar to the results presented in the preceding two subsections, the main observation from Fig.~\ref{Fig16}(b) is that the proposed phase-field theory predicts accurately the propagation of fracture described by the Griffith criticality condition (\ref{Gc-0-Propagation}). The results are also in qualitative agreement with those found experimentally since \cite{Greensmith55}; see, e.g., Fig.~\ref{Fig5} above.

\section{Summary and final comments}\label{Sec: Final Comments}

In this paper, with direct guidance from the substantial body of experimental evidence that has been amassed since the 1930s until present times, we have put forth a macroscopic theory that describes, explains, and predicts the nucleation and propagation of fracture in viscoelastic elastomers subjected to arbitrary quasistatic mechanical loads. 

The theory amounts to a threefold generalization of the phase-field theory initiated by \cite*{KFLP18} for elastic brittle materials that accounts for: $i$) the mechanics of deformation; $ii$) the mechanics of strength; and $iii$) the mechanics of toughness inherent to viscoelastic elastomers. 

While the mechanics of deformation of viscoelastic elastomers is by now a mature research topic, the same is \emph{not} true about their mechanics of strength and toughness, particularly so for the former. Generalizing the definition introduced in \citep{KLP20,KBFLP20} for elastic brittle materials, the proposed theory makes use of the following definition of strength: the strength of a viscoleastic elastomer is the set of all pairs of critical stresses and deformation gradients $(\bfS,\bfF)$ at which the elastomer fractures when it is subjected to a state of \emph{spatially uniform}, but otherwise arbitrary, history of stress $\{\bfS(t),t\in[0,T]\}$. Such a set of critical pairs defines a hypersurface $\mathcal{F}\left(\bfS,\bfF\right)=0$ in stress-deformation space, which is referred to as the strength surface of the elastomer.

As for the mechanics of toughness, the proposed theory makes use of the Griffith criticality condition recently brought to light by \cite{SLP23}, which states that the underlying Griffith energy competition in viscoelastic elastomers is one where only the ``equilibrium'' part of the stored elastic energy --- as opposed to the entire stored elastic energy and possibly part of the dissipated viscous energy --- competes with the fracture energy. 

From an applications point of view, the proposed theory amounts to solving the initial-boundary-value problem (\ref{BVP-y-theory-reg})-(\ref{BVP-z-theory-reg}), comprised of two nonlinear PDEs coupled with a nonlinear ODE, for the deformation field $\bfy(\bfX,t)$, the tensorial internal variable $\bfC^v(\bfX,t)$, and the phase field $z(\bfX,t)$. In order to generate solutions for these equations, we have developed and implemented a robust scheme that makes use of a non-conforming Crouzeix-Raviart FE discretization of space and a high-order accurate explicit Runge-Kutta FD discretization of time.

The sample simulations and comparisons with experiments that we have presented in Section \ref{Sec: Simulations} have served to illustrate the use of the proposed theory (including the calibration of the material functions/constants entering the theory and the proper selection of the various computational parameters), as well as its capabilities to describe and predict the nucleation and propagation of fracture in representative elastomers of practical relevance under a range of commonplace --- yet theoretically/computationally challenging --- loading conditions. Such validation results provide ample motivation to continue the study of the proposed phase-field theory as a complete framework for the description of fracture nucleation and propagation in viscoelastic elastomers at large. Paramount to such a study is having measurements of the strength of elastomers under different stress triaxialities and loading conditions. Such measurements, however, are largely missing in the literature. We hope that the analysis presented in this work will be a launching point for many investigations to follow on this topic in both experimental mechanics from a top-down or macroscopic point of view, as well as in polymer science from a bottom-up or microscopic point of view. Other topics for future work, particularly relevant for emerging applications, include accounting for: $i$) non-monotonic loading, in particular, fatigue loading; $ii$) additional dissipation mechanisms beyond viscous dissipation, such as strain-induced crystallization; $iii$) additional external stimuli beyond mechanical forces, such as heat, electric/magnetic fields, light, or moisture; and $iv$) inertial effects.

\section*{Acknowledgements}

This work was supported by the 3M Company and the National Science Foundation through the Grant DMS--2308169. This support is gratefully acknowledged.

\bibliographystyle{elsarticle-harv}
\bibliography{References}

\end{document}